\documentclass[twocolumn,natbib]{svjour3}          

\smartqed  

\usepackage{amssymb, color, xcolor}
\usepackage[colorlinks=true,linkcolor=blue,
            urlcolor=black,
            citecolor=blue]{hyperref}

\usepackage{amsmath}
\usepackage{graphicx}

\usepackage{bbding}
\usepackage[normalem]{ulem}
\usepackage{supertabular,booktabs}

\begin{document}\sloppy 


\title{Diffuse Radio Emission from Galaxy Clusters\thanks{
}}


\author{R.~J.~van~Weeren \and F.~de~Gasperin  \and H.~Akamatsu \and M.~Br\"uggen \and L.~Feretti \and H.~Kang \and A.~Stroe 
         \and F.~Zandanel}


\institute{R.~J. van Weeren \at
              Leiden Observatory, Leiden University, PO Box 9513, 2300 RA Leiden, The Netherlands \\
              \email{rvweeren@strw.leidenuniv.nl}           
           \and
           F. de Gasperin and M. Br\"uggen \at
              Hamburger Sternwarte, University of Hamburg, Gojenbergsweg 112, 21029 Hamburg, Germany
           \and 
           H. Akamatsu \at 
           SRON Netherlands Institute for Space Research, Sorbonnelaan 2, 3584 CA Utrecht, The Netherlands               
           \and 
           L. Feretti \at
           INAF - Istituto di Radioastronomia, Via Gobetti 101, I–40129 Bologna, Italy
           \and
           H. Kang \at
           Department of Earth Sciences, Pusan National University, Busan 46241, Republic of Korea           
  \and 
           A. Stroe \at
            Harvard-Smithsonian Center for Astrophysics, 60 Garden Street, Cambridge, MA 02138, USA \&  European Southern Observatory, Karl-Schwarzschild-Str. 2, 85748, Garching, Germany         
           \and
           F. Zandanel \at 
             GRAPPA, University of Amsterdam, Science Park 904, 1098XH, Amsterdam, The Netherlands            
}

\date{Received: date / Accepted: date}

\maketitle

\begin{abstract}
In a growing number of galaxy clusters diffuse extended radio sources have been found. These sources are not directly associated with individual cluster galaxies. The radio emission reveal the presence of cosmic rays and magnetic fields in the intracluster medium (ICM). We classify diffuse cluster radio sources into radio halos, cluster radio shocks (relics), and revived AGN fossil plasma sources. Radio halo sources can be further divided into giant halos, mini-halos, and possible ``intermediate'' sources. Halos are generally positioned at cluster center and their brightness approximately follows the distribution of the thermal ICM. Cluster radio shocks (relics) are polarized sources mostly found in the cluster's periphery. They trace merger induced shock waves. Revived fossil plasma sources are characterized by their radio steep-spectra and often irregular morphologies. In this review we give an overview of the properties of diffuse cluster radio sources, with an emphasis on recent observational results. We discuss the resulting implications for the underlying physical acceleration processes that operate in the ICM, the role of relativistic fossil plasma, and the properties of ICM shocks and magnetic fields. We also compile an updated list of diffuse cluster radio sources which will be available on-line (\url{http://galaxyclusters.com}). We end this review with a discussion on the detection of diffuse radio emission  from the cosmic web.


\end{abstract}

\section{Introduction}
\label{sec:intro}
Galaxy clusters are the largest virialized objects in our Universe, with masses up to $\sim10^{15}$~M$_{\odot}$. Elongated filaments of galaxies, located between clusters, form even larger unbound structures, making up the cosmic web. Galaxy clusters are located at the nodes of filaments, like ``spiders'' in the cosmic web. 

\begin{figure*}[htbp]
\centering
\includegraphics[width=0.32\textwidth]{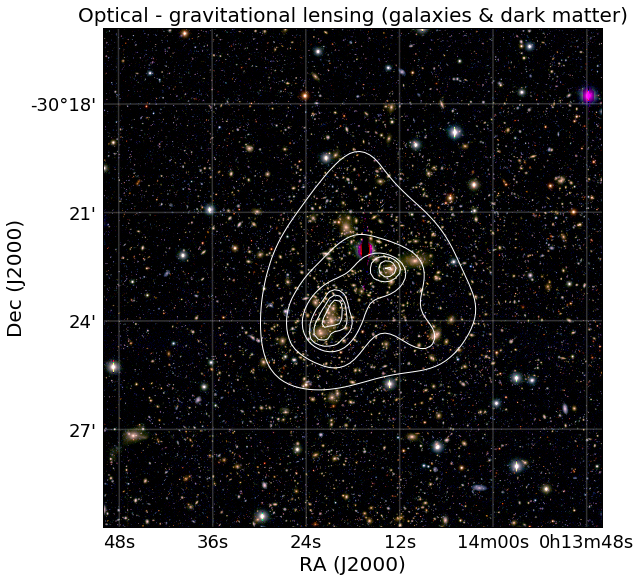}
\includegraphics[width=0.32\textwidth]{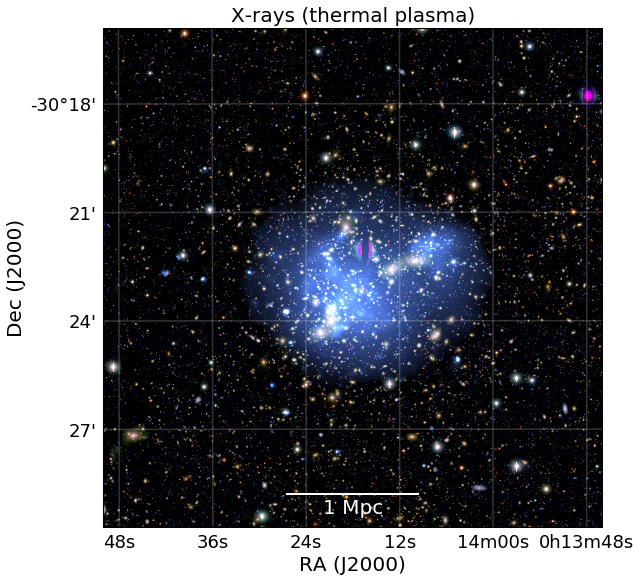}
\includegraphics[width=0.32\textwidth]{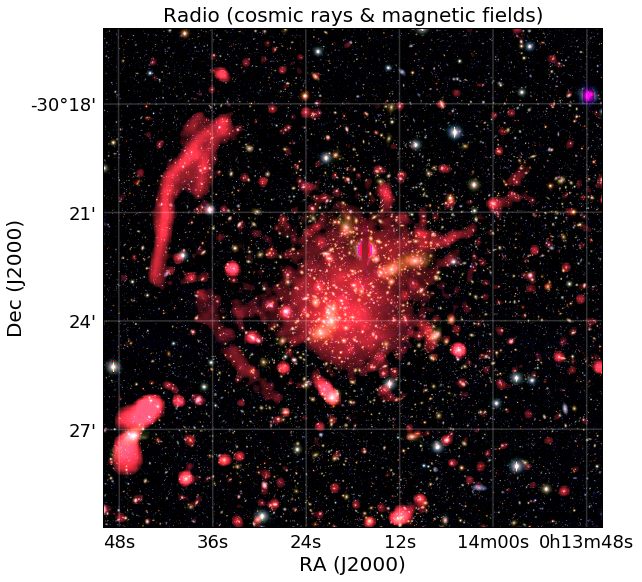}
\caption{The galaxy cluster Abell\,2744. The \textit{left} panel shows an optical \citep[Subaru BR\textit{z};][]{2016ApJ...817...24M} view of the cluster. White linearly spaced contours represent the mass surface density ($\kappa$) derived from a weak lensing study ($\kappa = {\Sigma}/{\Sigma_{\rm{cr}}}$, with $\Sigma_{(\rm{cr})}$ the (critical) mass surface density density) overlaid from \cite{2011MNRAS.417..333M,2017ApJ...837...97L}. 
In the \textit{middle} panel the X-ray emission from the thermal ICM (Chandra 0.5--2.0~keV band) is displayed in blue. In the \textit{right} panel a 1--4~GHz Very Large Array (VLA) image is shown in red, tracing cosmic rays and magnetic fields.
For more details about the images see \cite{2017ApJ...845...81P}.}
\label{fig:a2744overview}   
\end{figure*}

Clusters contain up to several thousands of galaxies. However, the galaxies comprise only a few percent of a cluster's total mass. Most of the baryonic mass of clusters is contained in a hot ($10^{7}$--$10^{8}$~K) ionized intracluster medium (ICM), held together by the clusters’  gravitational pull. This dilute magnetized plasma ($\sim 10^{-3}$ particles~cm$^{-3}$) emits thermal Bremsstrahlung at X-ray wavelengths, permeating the cluster's volume \citep[e.g.,][]{1976MNRAS.175P..29M,1977ApJ...211L..63S,1982ARA&A..20..547F}, see Figure~\ref{fig:a2744overview}. The ICM makes up $\sim$15\% of a cluster's mass budget. Most of the mass, $\sim80\%$, is in the form of dark matter \citep[e.g.,][]{1984Natur.311..517B,1995MNRAS.273...72W,1999ApJ...511...65J,1999MNRAS.305..631A,2003MNRAS.340..989S,2006ApJ...640..691V}.

Elongated filaments of galaxies span the regions between clusters. The so-called warm-hot intergalactic medium (WHIM) pervades these galaxy filaments \citep{1999ApJ...514....1C}. Compared to the ICM, the intergalactic medium of galaxy filaments (WHIM) has a significantly lower density ($\lesssim 10^{-4}$ particles cm$^{-3}$) and cooler temperature ($10^{5}$--$10^{7}$~K). About half of the Universe’s baryons reside in this WHIM \citep[e.g.,][]{1999ApJ...514....1C,2001ApJ...552..473D,2015Natur.528..105E}. Galaxy filaments are expected to be surrounded by strong accretion shocks, where the plasma is first shock-heated \citep{1972A&A....20..189S}. However, studying the WHIM and associated shocks is difficult due to a lack of sensitive observational tools. 
Galaxy clusters form by accretion from the WHIM and through a sequence of mergers of clusters and groups \citep[e.g.,][]{1970ApJ...162..815P,1974ApJ...187..425P,2005RvMP...77..207V,2012ARA&A..50..353K}. Cluster mergers are very energetic events, releasing energies up to $\sim10^{64}$~ergs on a few Gyr timescale. This energy is dissipated through low-Mach number shocks and turbulence, heating the ICM \citep[e.g.,][]{2007PhR...443....1M}. Clusters can thus be divided as either ``relaxed'' (undisturbed) or ``merging'' (disturbed) systems, depending on their dynamical (merging) state.

Galaxy clusters often host a number of active galactic nuclei (AGN) that emit radio synchrotron emission (i.e., radio galaxies) \citep[e.g.,][]{1984PhR...111..373D,2002pers.book.....D,2016A&ARv..24...10T}. The sizes of these sources range from a few kpc to about $\sim$1~Mpc, extending well beyond the host galaxy.
A major difference with radio galaxies that are located outside clusters (and groups) is that the jets and lobes of cluster radio galaxies often show signs of interaction with the ICM \citep[e.g.,][]{1980ARA&A..18..165M,1998Sci...280..400B,2015IAUS..313..321J}. 
These interactions result in morphologies that range from wide-angle (WAT), narrow angle (NAT), to ``head-tail'' radio sources.

 Gas in the central regions of many relaxed clusters has a radiative cooling time that is much shorter than the Hubble time. In the absence of a heating source, a cooling flow is expected to develop, whereby the temperature in the central region of the cluster drops and gas flows inwards \citep[e.g.,][]{1994ARA&A..32..277F,2006PhR...427....1P,2012ARA&A..50..455F,2012NJPh...14e5023M}. X-ray observations do show these temperature drops in some cluster cores (``cool core'' clusters), but there is much less cool gas than what would be expected from the short radiative cooling time \citep{2001A&A...365L..99K,2001A&A...365L.104P,2003ApJ...590..207P}. Therefore, some source of heating must balance the radiative losses. 
Radio galaxies, associated with the brightest cluster galaxy (BCG), 
have been identified as the main source of energy input into the ICM. X-ray observations show numerous cavities in cool core clusters, coincident with the lobes of the central radio galaxy. Here the radio plasma has displaced the X-ray emitting gas, creating a low-density bubble which rises buoyantly and expands, distributing energy to the surrounding ICM \citep[e.g.,][]{2002MNRAS.332..729C}. This process is commonly referred to as ``radio-mode'' feedback, although it is still being debated what the precise mechanism is that transfers the energy to the ICM.

\subsection{Extended synchrotron radio emission from galaxy clusters}
Radio observations have shown that the ICM can also contain a non-thermal component of cosmic rays (CR, see  Figure~\ref{fig:a2744overview}) which is not directly associated with cluster radio galaxies \citep[e.g.,][]{1959Natur.183.1663L,1970MNRAS.151....1W}. These GeV CR electrons (i.e., Lorentz factors of $\gamma > 10^{3}$) emit synchrotron radiation in the presence of $\sim \mu$Gauss ICM magnetic fields. During the last decade significant progress has been made in our understanding of this non-thermal component, through observations, theoretical, and numerical work. There is now compelling evidence that ICM shocks waves, and likely also turbulence, are able to (re-)accelerate particle to relativistic energies creating this non-thermal CR component of the ICM. 

The presence of extended synchrotron emission also indicates the existence of large-scale ICM magnetic fields with a strength of the order of 0.1--10~$\mu$Gauss \citep[e.g.,][]{2012SSRv..166..187B,2001ApJ...547L.111C,2015aska.confE..92J}. Cluster magnetic fields play an important role in particle acceleration processes. Additionally, magnetic fields inhibit transport processes like heat conduction, spatial mixing of gas, and the propagation of cosmic rays \citep[e.g.,][]{2017MNRAS.465.4500P,2010ApJ...713.1332R}. However, few details are known about the precise properties of these fields since they are difficult to measure \citep[e.g.,][]{2004IJMPD..13.1549G}.

The synchrotron emitting CR electrons should scatter photons from the cosmic microwave background (CMB) to X-ray energies, resulting in a hard tail on top of the thermal X-ray spectrum of clusters \citep{1979ApJ...227..364R,1994ApJ...429..554R,2000ApJ...533...73S}.
So far, no conclusive detection of this inverse-Compton (IC) radiation has been made \citep[e.g.,][]{2000ApJ...534L...7F,2001ApJ...552L..97F,2004ApJ...606..825R,2004A&A...414L..41R,2004ASSL..309..125F,2008SSRv..134...71R,2008A&A...479...27E,2009ApJ...696.1700W,2014ApJ...792...48W}.
However, even a non-detection of IC X-ray emission, in combination with radio observations, is useful to set lower limits on the ICM magnetic fields strength \citep[e.g.,][]{2009PASJ...61.1293S,2010ApJ...715.1143F,2015PASJ...67..113I}. Similarly, CR protons can interact hadronically with the protons of the ICM and generate pions that can then decay into gamma-rays \citep[c.f.,][]{1980ApJ...239L..93D,1999APh....12..169B,2007IJMPA..22..681B}. Gamma-ray observations are particularly important to understand the dynamical role of CR protons in clusters, and the role of secondary electrons, also coming from pion decays, in generating the extended radio emission.

\subsection{This review}
Galaxy clusters provide a unique environment to study the physics of particle acceleration in collisionless, high-$
\beta$, turbulent plasmas, where $\beta$ is the ratio of the thermal pressure to the magnetic pressure\footnote{$\beta = \frac{8\pi n T}{B^2}\sim 100$ for the ICM, taking $T=5$~keV, $B=3$~$\mu$Gauss, and $n=5 \times 10^{-3}$~cm$^{-3}$}, and at low Mach numbers shocks. Furthermore, diffuse radio emission from clusters can be used as a signpost of ICM shocks and turbulence, which are often difficult to detect and characterize at other wavelengths. Since shocks and turbulence trace the dynamical state of the ICM, radio observations also provide us with a probe of the cluster's evolutionary stage, important for our understanding of structure formation in the Universe. Finally, diffuse radio emission can be used as a complementary method to discover clusters that were missed by X-ray, SZ, or optical surveys \citep{2011ApJ...727L..25B,2012AA...546A.124V,2014AA...565A..13M,2017AA...597A..15D}.

In this paper we review the observational properties of diffuse extended cluster radio emission. Previous observational reviews on this subject were presented by \cite{2002IAUS..199..133F,2002ASSL..272..197G,2003ASPC..301..143F,2008SSRv..134...93F,2012A&ARv..20...54F}. Here we provide an update, encompassing recent results that have helped to improve our understanding of these sources. For a more theoretical review we refer the reader to \cite{2014IJMPD..2330007B}.
Observational progress in this field has been made through a combination of high-resolution multi-frequency studies, the availability of deep low-frequency observations, an increasing number of polarimetric studies, the compilation of larger cluster samples with deep radio data, and high-frequency detections. The joint analyses of radio data and observations at other wavelengths, in particular in the X-ray and Gamma-ray bands, has also played an important role.

The outline of this paper is as follows. In Section~\ref{sec:synchrotron} we briefly discuss synchrotron radiation and particle acceleration mechanisms. The classification of diffuse cluster radio sources is discussed in Section~\ref{sec:classification}. A review of  cluster magnetic fields is given in Section~\ref{sec:magneticfields}. Overviews of radio halos, including mini-halos, and cluster radio shocks and revived fossil plasma sources are presented in Sections~\ref{sec:halos} and~\ref{sec:relics}. In Section~\ref{sec:radiocosmicweb} we end this review with a discussion on the detection of diffuse radio emission outside cluster environments.

\section{Synchrotron radiation and radio spectra}
\label{sec:synchrotron}

In this section we briefly discuss some relevant theory about the synchrotron spectra of CR electrons. For a more detailed treatment of synchrotron radiation we refer the reader to the references provided in \cite{2012A&ARv..20...54F}.
A standard assumption is that the ICM CR population can be described by a power law energy ($E$) distribution
\begin{equation}
n(E)\mathrm{d} E \propto E^{-p} \mathrm{d} E \mbox{ .}
\end{equation}
The index of the energy (or momentum) distribution $p$ is directly related to the radio spectral index\footnote{$F_\nu \propto \nu^{\alpha}$, where $\alpha$ is the spectral index}
\begin{equation}
p=1-2\alpha \mbox{ .}
\label{eg:palpha}
\end{equation}

Diffuse cluster radio emission typically has a steep spectral index , i.e., $\alpha \lesssim -1$. The spectral shape is related to the physics of the acceleration mechanism and the electron synchrotron and IC energy losses. 
The characteristic lifetime ($t_{\mathrm{age}}$) of the synchrotron emitting electrons ($\gamma\sim 10^4$; GeV energy) due to these energy losses is 
\begin{equation}
t_{\mathrm{age}} \mbox{ [yr]} \approx 3.2 \times 10^{10} \frac{B^{1/2}}{B^2 + B^{2}_{\mathrm{CMB}}} \left[(1+z)\nu\right]^{-1/2} \mbox{ ,} 
\label{eq:tage}
\end{equation} 
where $B$ the magnetic field strength, $z$ the source redshift, 
 $B_{\mathrm{CMB}}$ the equivalent magnetic 
field strength of the CMB ($B_{\mathrm{CMB}} \mbox{ }[\mu \rm{Gauss}] \approx 3.25\left(1+z\right)^2$), and $\nu$ the observing frequency in MHz. In clusters, we have $t_{\mathrm{age}} \lesssim 10^{8}$~yrs. 
 The typical diffusion length-scale in the ICM of a GeV electron, using the  Bohm approximation, is of the order of 10~pc \citep[e.g.,][]{2002NewA....7..249B}. {Plasma motions can increase the distance over which GeV electrons travel, but this distance is still expected to remain well below a Mpc.} This means that Mpc-scale diffuse radio sources cannot trace CR electrons that are accelerated at a single location in the ICM. Instead, they need to be (re-)accelerated or produced in-situ \citep{1977ApJ...212....1J}, providing important constraints on the possible acceleration/production mechanisms.

\begin{figure}[htbp]
\centering
\includegraphics[width=0.49\textwidth]{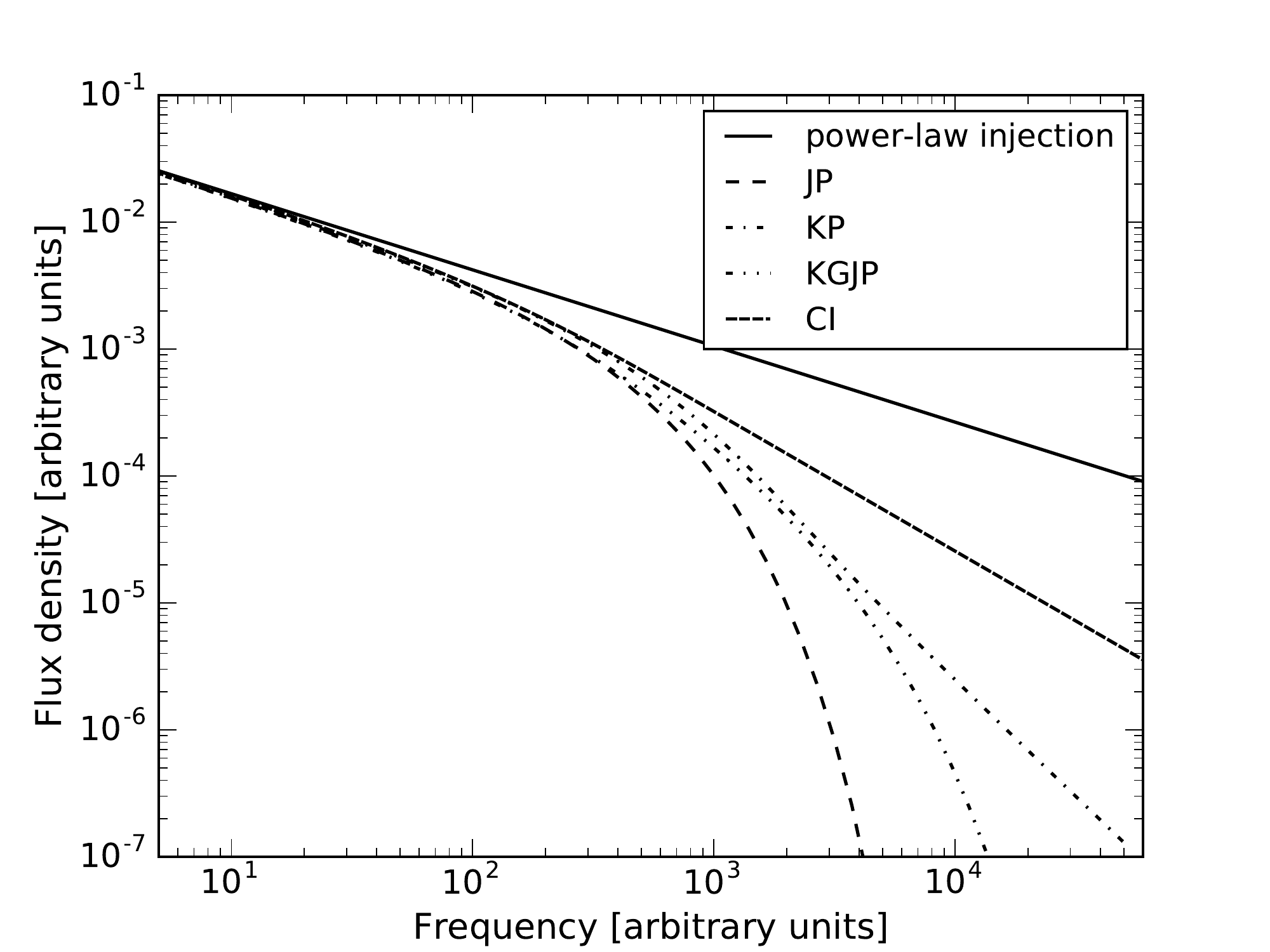}
\caption{An overview of radio spectral shapes. All spectral models have $\alpha_{\rm{inj}} = -0.6$. The power-law spectrum depicts the spectral shape before any energy losses.}
\label{fig:syncspectra}   
\end{figure}

Due to the energy losses, the initial power-law spectrum steepens beyond a break frequency, whose position is related to the time since acceleration. The power-law spectrum is commonly refereed to as the injection spectrum, characterized by an injection spectral index ($\alpha_{\rm{inj}}$). For the JP (Jaffe-Perola) synchrotron spectrum \citep{1973A&A....26..423J}, one assumes that there is a continuous isotropization of the electron pitch angles (i.e., angle between the magnetic field and the electron velocity) on a timescale that is shorter than $t_{\mathrm{age}}$. A JP spectrum describes a synchrotron spectrum from a single burst of acceleration and then aging. The KP (Kardashev-Pacholczyk) model \citep{1962SvA.....6..317K,1970ranp.book.....P} also represents such a spectrum, but without the isotropization of the pitches angles. A collection of spectral shapes is displayed in Figure~\ref{fig:syncspectra}.

Since it is usually difficult to spatially isolate electrons that all have the same spectral age, there are also composite models. These models sum JP (or KP) spectra with different amounts of spectral aging. The CI (continuous injection) composite model \citep{1970ranp.book.....P} describes the integrated spectrum of a source with continuous particle injection. For the KGJP/KGKP (Komissarov-Gubanov) model \citep{1994A&A...285...27K}, the particles are only injected for a finite amount of time before the injection in the source stops.

\subsection{Particle acceleration mechanisms}

There are several physical mechanisms to accelerate particles in the ICM and produce the synchrotron emitting CR electrons. We briefly give an overview of these processes below. Further details will be discussed in later sections where relevant. 

\begin{itemize}
\item \textit{First order Fermi acceleration} (Fermi-I): This process, also known as diffusive shock acceleration (DSA), plays an important role in various astrophysical environments \citep{1977DoSSR.234R1306K,1977ICRC...11..132A,1978MNRAS.182..147B,1978MNRAS.182..443B,1978ApJ...221L..29B,1983RPPh...46..973D,1987PhR...154....1B,1991SSRv...58..259J,2001RPPh...64..429M}. For DSA, particles are accelerated at a shock with the acceleration taking place diffusively. In this process, particles cross back and forward across the shock front as they scatter from magnetic inhomogeneities in the shock down and upstream region. At each crossing, particles gain additional energy, forming a power-law energy distribution of CR.

\item \textit{Second order Fermi acceleration} (Fermi-II): This is a stochastic process where particles scatter from magnetic inhomogeneities, for example from magneto-hydrodynamical (MHD) turbulence \citep{1987AA...182...21S,1993JPlPh..49...63S,2001MNRAS.320..365B,2001ApJ...557..560P}. Particles can either gain or loose energy when scattering. When the motions are random, the probability for a head-on collision, where energy is gained, is slightly larger. Because of its random nature, second order Fermi acceleration is an inefficient process.


\item \textit{Adiabatic compression:} 
A shock wave can adiabatically compress a bubble/lobe/cocoon of (old) relativistic radio plasma from an AGN. Due to the compression, the CR  electrons in the cocoon regain energy boosting the radio synchrotron emission \citep{2001A&A...366...26E,2002MNRAS.331.1011E}.

\item \textit{Secondary models:} Another mechanism to produce CR electrons is via a secondary process, meaning that the CR electrons are produced as secondary particles (decay products). In the hadronic model, collisions between relativistic protons and the thermal ions produce secondary CR electrons  \citep{1980ApJ...239L..93D,1999APh....12..169B,2000A&A...362..151D,2001ApJ...562..233M,2010ApJ...722..737K,2010MNRAS.407.1565D,2011A&A...527A..99E}. Since CR protons have a very long lifetime compared to CR electrons, they will accumulate over the lifetime of a cluster once they are accelerated. Possible mechanisms to produce CR protons are first order Fermi acceleration at shocks, AGN activity, and galactic outflows (supernovae, winds). 

\end{itemize}

\section{Classification}
\label{sec:classification}

\begin{figure*}[htbp]
\centering
\includegraphics[width=0.49\textwidth]{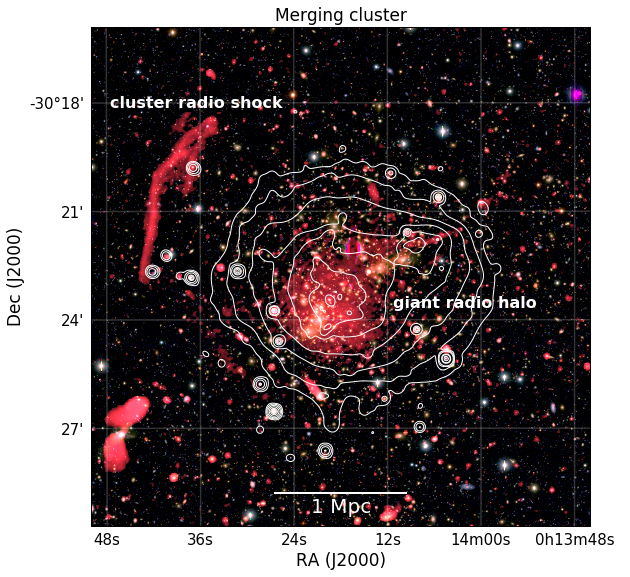}
\includegraphics[width=0.49\textwidth]{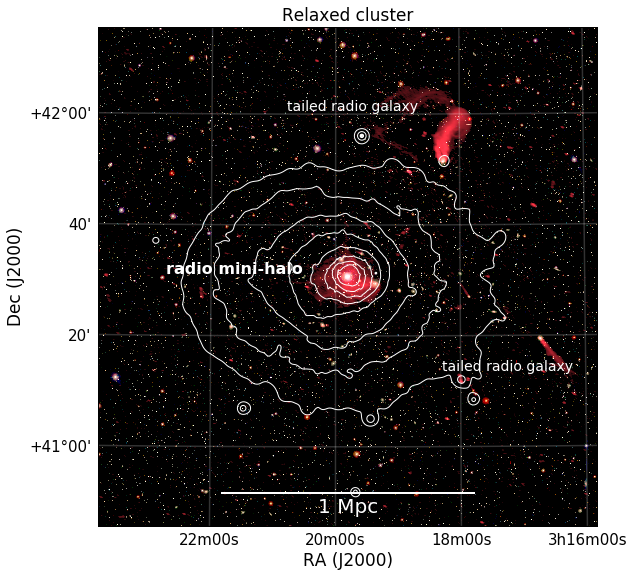}
\caption{\textit{Left panel:} VLA 1--4~GHz image of the merging galaxy cluster Abell\,2744 with different source classes labeled (see also Figure~\ref{fig:a2744overview}). Chandra X-ray contours are shown in white. This cluster hosts a luminous giant radio halo and a cluster radio shock (relic). X-ray surface brightness contour are drawn proportional to $[1,4,16,64,\ldots]$. \textit{Right panel:} VLA 230--470 MHz image of the relaxed cool core Perseus cluster from \cite{2017MNRAS.469.3872G}. XMM-Newton X-ray contours in the 0.4--1.3~keV band are overlaid in white with the same contour spacing as in the left panel. The Perseus cluster hosts a radio mini-halo as well as two prominent tailed radio galaxies.}
\label{fig:a2744tax}   
\end{figure*}

Diffuse cluster radio sources have historically been divided into three main classes, relics, halos, and mini-halos \citep{1996IAUS..175..333F}. In addition, radio filaments were proposed to trace the large-scale filaments of the cosmic web, outside of clusters. Note that the term filament has also sometimes been used to describe radio relics (or relic-type structures) in clusters. We will discuss radio emission outside the cluster environment in Section~\ref{sec:radiocosmicweb}.

\emph{Radio halos} are centrally located diffuse sources in merging clusters. They do not have any optical counterparts. \emph{Mini-halos} have smaller sizes and are located in relaxed cool core clusters which also host a powerful radio galaxy associated with the BCG. 
\emph{Radio relics} have been defined as extended sources that show high levels of polarization ($\gtrsim 10\%$ at GHz frequencies) and are located in the cluster periphery. Similar to radio halos, they not show optical counterparts. Relics were further subdivided \citep{2004rcfg.proc..335K} into large \emph{Radio Gischt}, large Mpc-size sources that trace particles accelerated at shocks via Fermi-I  processes; \emph{Radio Phoenices}, AGN fossil plasma compressed and revived by merger shocks; and \emph{AGN Relics}, fossil radio plasma that is passively evolving from an AGN that has been switched off. For radio relics, the boundaries between the different categories is not always very obvious and the term relics itself is somewhat unfortunate because large relics could be ``young'' sources with on-going (re-)acceleration.

Here we propose to classify cluster emission into three broad classes: 
\begin{itemize}
 \item[$\bullet$] \uline{Radio halos} are extended  sources that roughly follow the ICM baryonic mass distribution. This class includes \uline{giant radio halos} and \uline{mini-halos}, see Figure~\ref{fig:a2744tax}. This class would also contain possible ``intermediate'' or ``hybrid'' radio halos,  with properties falling somewhere in between those of classical giant radio halos and mini-halos.  Another property of the halo class is that these sources are not localized, in the sense that particle (re-)acceleration/production occurs throughout a significant volume of the cluster and is not associated with a particular shock which location can be pint-pointed. In terms of a physical interpretation, these ``global'' sources should trace Fermi-II processes and/or secondary electrons.
 
 \item[$\bullet$] \uline{Cluster radio shocks (radio relics)} are extended diffuse sources tracing particles that are (re-)accelerated at ICM shock waves (Figure~\ref{fig:a2744tax}). They have commonly been referred to as radio relics.
 This radio shock classification is somewhat similar to the that of Gischt, but it does not necessarily require DSA or Fermi-I type acceleration. In that sense, cluster radio shocks are an observationally defined class, unrelated to the details of the actual acceleration mechanism. However, based on our current understanding of these sources, we do anticipate that in most cases cluster radio shocks are associated with Fermi-I acceleration processes. It is not required that cluster radio shocks are located in the cluster periphery, although for large cluster radio shocks that will typically be the case. Due to their nature, the large majority of these sources are expected to show a high degree of polarization. Sources  previously classified as large radio relics, Gischt, and double relics, fall in the cluster radio shock category. Unlike radio halos, cluster radio shocks can be associated to a specific cluster region where a shock wave is present, or where a shock wave recently passed. {A drawback of the radio shock classification is that the detection of shocks in the ICM is observationally challenging. Therefore, the classification will remain uncertain for some sources. However, for a number of sources the presence of a shock at their location has been confirmed by X-ray observations (see Section~\ref{sec:radioxrayshock}) which we argue warrants the creation of a radio shock class. In this review we will use the term radio shock for sources previously classified as large radio relics, Gischt, and double relics. It is important to keep in mind that for a number of sources the presence of a shock remains to be confirmed.}

\item[$\bullet$] \uline{Revived AGN fossil plasma sources, phoenices, and GReET}
In this class we group sources that trace AGN radio plasma that has somehow been re-energized through processes in the ICM, unrelated to the radio galaxy itself. Low-frequency observations are starting to reveal more and more of these type of sources. However, their precise origin and connection to cluster radio shocks and possibly also halos is still uncertain. The main observational property that the sources have in common is the AGN origin of the plasma and their ultra-steep radio spectra due to their losses.
For this review we decided to keep the radio phoenix classification \citep{2004rcfg.proc..335K}. Often these phoenices display irregular filamentary morphologies. They have relatively small sizes of at most several hundreds of kpc. Gently re-energized tails \citep[GReETs;][]{2017SciA....3E1634D} are tails of radio galaxies that are somehow revived, showing unexpected spectral flattening, opposite from the general steepening trend caused by electron energy losses. With the new and upgraded
low-frequency radio telescopes that have become operational, we expect that the nature of these revived fossil plasma sources will become more clear over the next decade.

\end{itemize}

Fossil radio plasma plays and important role in some of the models for the origin of radio halos and cluster radio shocks. In these models fossil plasma is re-accelerated via first and second order Fermi processes. This implies that when clusters are observed at low enough frequencies, both halos and cluster radio shocks will blend with regions of old AGN radio plasma, complicating the classification.

The classification can also be hindered by projection effects. For example, a cluster radio shock observed in front of the cluster center might mimic halo-type emission if the signal to noise of the image is not very high. However, these are observation related difficulties, which can in principle be resolved with better data. On the website \url{http://galaxyclusters.com} we provide an up to date list of the currently known diffuse cluster radio sources and their classification. An up-to-date list of clusters with (candidate) diffuse radio emission at the time of writing (September 2018) is shown in Table~\ref{tab:clusterlist}.

\section{Cluster magnetic fields} 
\label{sec:magneticfields}
\subsection{Global}
Magnetic fields permeate galaxy clusters and the intergalactic medium on Mpc-scales. These fields play key roles in particle acceleration and on the process of large scale structure formation, having effects on turbulence, cloud collapse, large-scale motions, heat and momentum transport, convection, viscous dissipation, etc. 
In particular, cluster magnetic fields inhibit transport processes like heat conduction, spatial mixing of gas, and propagation of cosmic rays.
The origin of the fields that are currently observed remains largely uncertain. A commonly accepted hypothesis is that they result from the amplification of much weaker pre-existing {\it seed}  fields via shock/compression and/or turbulence/dynamo amplification during merger events and structure formation, and different magnetic field scales survive as the result of turbulent motions \citep[e.g.,][]{2013ApJ...770...47K}. The origin of {\it seed} fields is is unknown. They could be either {\it primordial},
i.e., generated in the early Universe prior to recombination, or
produced locally at later epochs of the Universe, in early stars
and/or (proto)galaxies, and then injected in the interstellar and
intergalactic medium \citep{2006AN....327..395R}. For a review about magnetic field amplification in clusters we refer the reader to \cite{2018SSRv..214..122D}.

Magnetic fields are difficult to measure. Some estimates have relied on the idea that the energies in cosmic rays and magnetic fields in the radio emitting regions are the same \citep[``equipartition'';][]{2005AN....326..414B}.  In this way, magnetic field values in the range 0.1--10~$\mu$Gauss are obtained. However, this method is inherently uncertain due to the many assumptions that are required. Cosmological simulations of clusters predict $\mu$Gauss-level magnetic field strengths in the cluster centers and
a decrease of the magnetic field strength with radius in the outer regions \citep{1999A&A...348..351D,2001A&A...378..777D,2002A&A...387..383D,2014MNRAS.445.3706V,2018MNRAS.474.1672V}. These values are roughly consistent with equipartition magnetic field strengths  estimates of the order of a $\mu$Gauss.

The most promising technique to derive a more detailed view of the magnetic fields in clusters is via the analysis of the Faraday rotation of radio galaxies located inside and behind the cluster \citep[e.g.,][]{2004JKAS...37..337C,2004IJMPD..13.1549G}. Faraday rotation changes the intrinsic polarization angle ($\chi_0$). The Faraday depth ($\phi$) is related to the properties of the plasma that cause the Faraday rotation \citep{1966MNRAS.133...67B,2005A&A...441.1217B}
\begin{equation}
\phi (\mathbf{r}) = 0.81 \int_{\mathrm{source}}^{\mathrm{telescope}} n_{\mathrm{e}}\mathbf{B} \cdot \, d\mathbf{r} \mbox{  } [\mbox{rad m}^{-2}  ] \mbox{ ,} 
\end{equation}
where $n_{\mathrm{e}}$ is the electron density in units of cm$^{-3}$, $\mathbf{B}$ the magnetic field in units of $\mu$Gauss, and $d\mathbf{r}$ is an infinitesimal path length in along the line of sight in units of parsec. The rotation measure (RM) is defined as
\begin{equation}
\mathrm{RM} = \frac{d\chi(\lambda^2)}{d\lambda^2} \mbox{ ,}
\end{equation}
where $\lambda$ is the observing wavelength. The Faraday depth equals the RM if there is only one source along the line of sight (and there is no internal Faraday rotation). This means that the RM does not depend on the observing wavelength. Also, all polarized emission comes from a single Faraday depth $\phi$ and the measured polarization angle ($\chi$) is given by
\begin{equation}
\chi =   \chi_0 + \phi \lambda^2 \mbox{ .}
\label{eq:chi0}
\end{equation}
From RM measurements, the strength and structure of cluster magnetic fields can be constrained by semi-analytical approaches, numerical techniques or RM synthesis \citep{2005A&A...441.1217B}.  To this aim, a
spherically symmetric model ($\beta$-model) is generally assumed for the thermal gas. Moreover, one needs to assume that the interaction between the ICM and the radio galaxy plasma does not affect the measured RM. It is still being debated to what extent this assumption holds. Deviations of the Faraday rotation from the simple $\lambda^2$--law (Equation~\ref{eq:chi0}) have been detected \citep[e.g.,][]{2009AA...503..707B}, likely implying either that the
magnetized screen is non--uniform and/or that the ICM thermal plasma is mixed with the relativistic plasma.

\subsubsection{Results from RM studies}
\label{sec:rm}
\begin{figure}[htbp]
\centering
\includegraphics[width=0.49\textwidth]{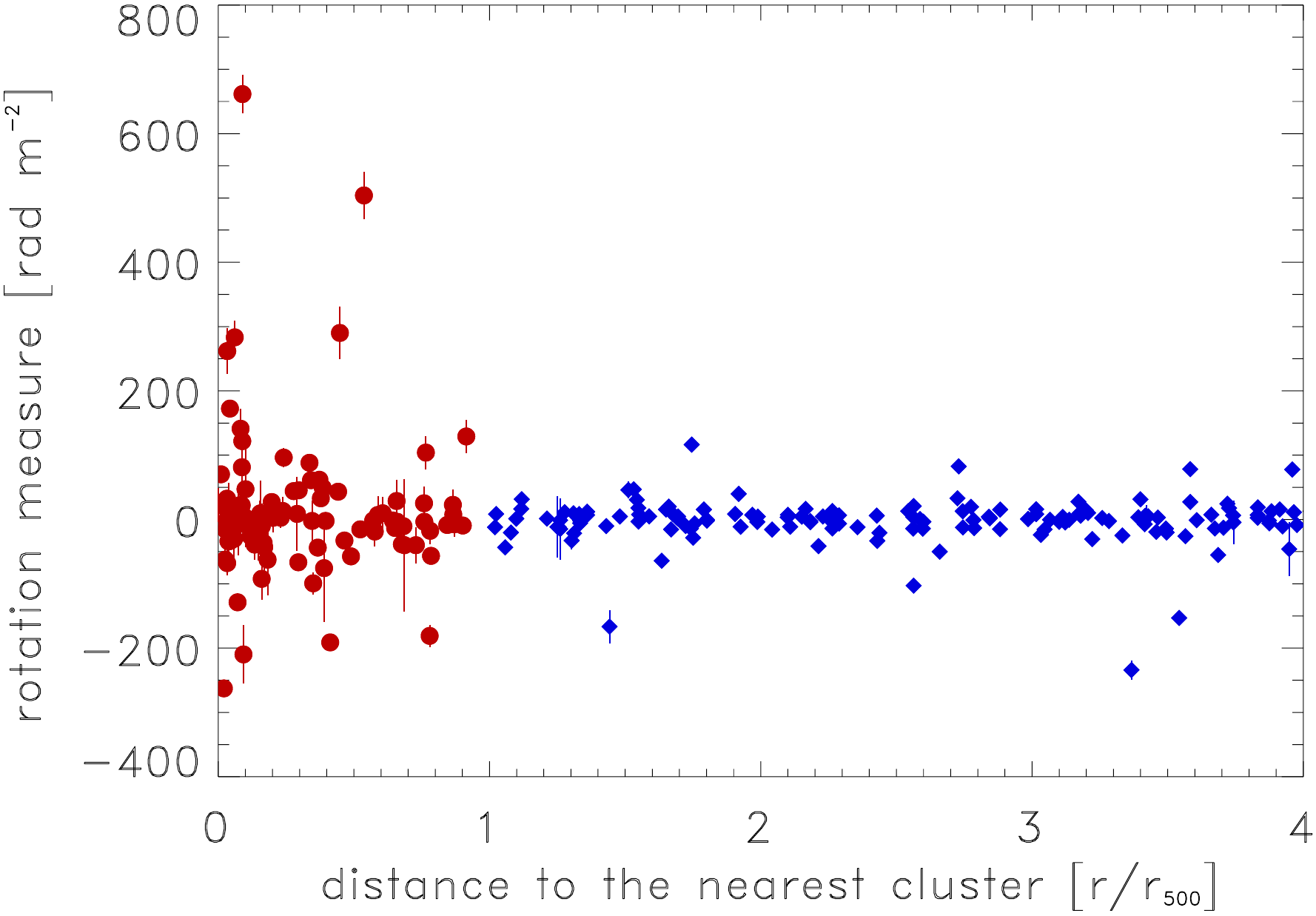}
\caption{Rotation measure as a function of cluster centric radius (scaled by R$_{500}$)  for a sample of X-ray selected clusters. The figure is taken from \cite{2016A&A...596A..22B}. Red circles are for rotation measures inside $R_{500}$, those outside are marked with blue diamonds.}
\label{fig:RMimpact}   
\end{figure}

The presence of magnetic field in clusters is demonstrated by
statistical studies. The comparison between the RMs of polarized
extragalactic radio sources in the line of sight of galaxy clusters and RM measurements made outside of the projected cluster regions shows excess of the standard deviations of RM values in the cluster areas \citep[c.f.,][]{2001ApJ...547L.111C,2016A&A...596A..22B}, see Figure~\ref{fig:RMimpact}. This is consistent with ubiquitous cluster magnetic fields of a few $\mu$Gauss strength, coherent cells of about 10 kpc, and a magnetic field energy density of a few per mille of the thermal energy density.

Information about the magnetic field in individual clusters through RM studies has been obtained so far for about 30 objects, including both merging and relaxed clusters.  The best studied cluster is Coma, whose magnetic field has been obtained with RM information on 7 radio galaxies in the cluster central region \citep{2010A&A...513A..30B}, and 7 additional radio galaxies in the peripheral Coma southwest 
region, where the NGC 4839 infalling group and the cluster radio shock are located \citep{2013MNRAS.433.3208B}. A single-cell model is not appropriate to describe the observed data, which are generally consistent with a turbulent field following a Kolmogorov power-law spectrum. From energy considerations, i.e., to avoid that the magnetic pressure exceeds the thermal pressure in the outer cluster regions, it is inferred that the magnetic field profile scales with the gas density $n_{th}$ as $B \propto n_{th}^{\eta}$.
The value of the index $\eta$ reflects the magnetic field formation and amplification.  It is expected that $\eta$=2/3 in the case of adiabatic compression during a spherical collapse due to gravity. In this case, the field lines are frozen into the plasma and compression of the plasma results in compression of the flux lines (as a consequence of magnetic flux conservation). A value $\eta$=1/2 is instead expected if the energy in the magnetic field scales as the energy in the thermal plasma.  Other values of $\eta$ may be obtained by specific combinations of compression orientation and magnetic field
orientation.

\begin{figure}[htbp]
\centering
\includegraphics[width=0.49\textwidth]{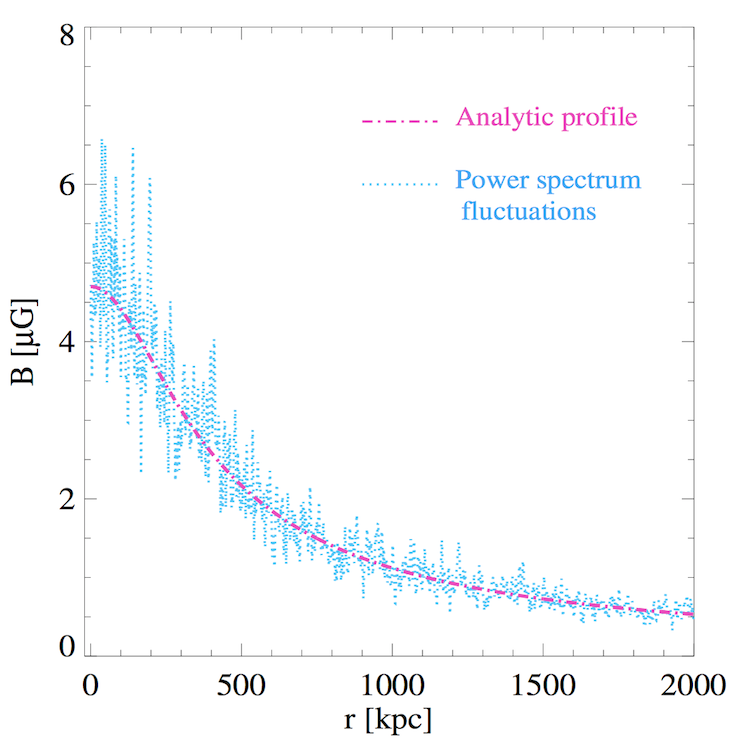}
\caption{The best fitting radial magnetic field strength profile (magenta line) for the Coma cluster from \cite{2010A&A...513A..30B}. Simulated power spectrum fluctuations on the profile are shown in blue.}
\label{fig:comaBfield}   
\end{figure}

The Coma cluster magnetic field is well represented by a Kolmogorov power spectrum with minimum scale of $\sim$2~kpc and maximum scale of $\sim$34~kpc.  The central field strength is  4.7~$\mu$Gauss and the radial slope is $\propto n_{th}^{0.7}$ \citep{2010A&A...513A..30B}, see Figure~\ref{fig:comaBfield}. The magnetic field of the southwest peripheral region is found to be $\sim$2~$\mu$Gauss, i.e., higher than that derived from the extrapolation of the radial profile obtained for the cluster center; a boost of magnetic field of $\sim$ a factor of 3 is required. The magnetic field amplification does not appear to be limited to the cluster radio shock region, but it must occur throughout the whole southwestern cluster sector, including the NGC~4839 group \citep{2013MNRAS.433.3208B}.

In the clusters analyzed so far, it is derived that cool core clusters have central magnetic field intensities of the order of a few 10~$\mu$Gauss, while merging clusters are characterized by intensities of a few $\mu$Gauss. The fields are turbulent, with spatial scales in the range 5--500~kpc, and coherence lengths of a few 10~kpc.  The values of the profile index $\eta$ are in the range 0.4--1, therefore no firm conclusion can be drawn on the radial trend of the magnetic field. Recently, \cite{2017A&A...603A.122G} found a correlation between the central electron density and mean central magnetic field strength ($\eta$=0.47) using data for 9 clusters. No correlation seems to be present between the mean central magnetic field and the cluster temperature.  In conclusion, good  information about the central magnetic field intensity in clusters has been obtained, whereas the magnetic field structure (profile, coherence scale, minimum and maximum scales, power spectrum, link to cluster properties) is still poorly known.  

\subsubsection{Statistical studies from fractional polarization}

From the analysis of the fractional polarization of radio sources in a sample of X-ray luminous clusters from the NVSS, a clear trend of the fractional polarization increasing with the distance from the cluster center has been derived \citep{2011A&A...530A..24B}. The low fractional
polarization in sources closer to the cluster center is interpreted as the result of higher beam depolarization, occurring in the ICM because of fluctuations within the observing beam and higher magnetic field and gas densities in these regions. Results are consistent with fields
of a few $\mu$Gauss, regardless of the presence or not of  radio halos.  A marginally significant difference between relaxed and merging clusters has been found.

\subsubsection{Lower limits from IC emission}

CR electrons present in the ICM should scatter photons from the CMB, creating a hard power-law of X-ray emission, on top of the thermal Bremsstrahlung from the ICM \citep{1979ApJ...227..364R,1994ApJ...429..554R,2000ApJ...533...73S}. Despite several claims made over the last decades, it seems that there is no conclusive evidence yet for this IC emission from the diffuse CR component of the ICM \citep[e.g.,][]{2000ApJ...534L...7F,2001ApJ...552L..97F,2004ApJ...606..825R,2004A&A...414L..41R,2004ASSL..309..125F,2008SSRv..134...71R,2008A&A...479...27E,2009ApJ...696.1700W,2014ApJ...792...48W,2009ApJ...690..367A,2009A&A...493...13M,2010PASJ...62..115K,2012ApJ...748...67W,2015ApJ...800..139G}. {The difficultly associated with the detection of IC emission is related to the requirement of accurately modeling the contributions of the instrumental and astronomical backgrounds.}

Following \cite{2001ApJ...557..560P,2016ApJ...823...94R}, the
monochromatic IC X-ray and synchrotron radio flux ratio ($R_{\rm obs}$) can be written as  
\begin{equation} \label{eq:flux_ratio}
\begin{aligned}
R_{\rm obs} \equiv \frac{f_{\rm IC}(kT)}{f_{\rm sync}(\nu)} & = 1.86 \times 10^{-8}
  \left( \frac{\rm photons}{\rm cm^{2}\, s\, keV\, Jy} \right) \\ &
  \times \left (\frac{kT}{20\, {\rm
      keV}} \right)^{-\Gamma}\left( \frac{\nu}{\rm{GHz}} \right)^{\Gamma - 1} \\ &
  \times \left( \frac{T_{\rm CMB}}{2.8 {\rm K}} \right)^{\Gamma + 2} \left(\frac{B}{\mu
    {\rm Gauss}} \right)^{-\Gamma}c(p),
\end{aligned}
\end{equation}
where $\Gamma = (p + 1)/2$,  $p$ is the power-law slope of the electron energy
distribution $N(E) \propto E^{-p}$ (see  Equation~\ref{eg:palpha} for the relation between radio spectral index $\alpha$ and $p$), $f_{\rm IC}(kT)$ is the IC flux
density at energy $kT$,
$f_{\rm sync}(\nu)$ is the synchrotron flux density at
frequency $\nu$, $T_{\rm CMB}$ is the CMB temperature at the
cluster's redshift, and $c(p)$ is a normalization factor that is a function of $p$. For typical values of $p$, $10 < c(p) < 1000$, see \cite{1979rpa..book.....R}. The function $c(p)$, for values of $2 \lesssim p \lesssim 5$ 
can be approximated as $c(p) \approx e^{1.42 p - 0.51}$.  With Equation~\ref{eq:flux_ratio} and this approximation the expression for the magnetic field strength becomes
\begin{equation} \label{eq:bfield}
\begin{aligned}
&B = \left( \frac{20 {\rm keV}}{kT} \right) \left( \frac{\nu}{{\rm
      GHz}} \right)^{(p-1)/(p+1)} e^{\frac{2.84(p-r)}{p+1}} \mu \rm{Gauss}, \\ &
r = 0.7 \ln \left[ \frac{R_{\rm obs}(kT, \nu)}{1.11 \times 10^{-8}}
\right] .
\end{aligned}
\end{equation}
In the above derivations a power-law distribution of electrons down to low energies is assumed. If this assumption does not hold \citep[e.g.,][]{2015A&A...582A..20B}, for example because there is flattening of the spectrum at low frequencies, the magnetic field values will be overestimated. 

By deriving upper limits on the IC X-ray emission and combining that with radio flux density measurements of radio halos, lower limits on the global ICM magnetic field strength can be computed. For radio halos, it is generally challenging to obtain stringent lower limits. The reason is that radio halos are typically faint. In addition, the IC emission is co-spatial with the thermal ICM, making it harder to separate the components. Furthermore, bright radio galaxies located in the cluster center can also produce non-thermal X-ray emission. The obtained lower magnetic field strength limits are therefore less constraining than the ones obtained for radio shocks (see Section~\ref{sec:relicsbfield}). The lower limits that have been computed for radio halo hosting clusters range around $0.1-0.5$~$\mu$Gauss. For example, for the Coma cluster \cite{2004A&A...414L..41R} found $B > 0.2-0.4$~$\mu$Gauss and \cite{2009ApJ...696.1700W} reported $B > 0.15$~$\mu$Gauss. For the Bullet cluster a limit of $B > 0.2$~$\mu$Gauss was determined \citep{2014ApJ...792...48W}. Magnetic field strength limits for the cluster Abell\,2163 are $B>0.2$~$\mu$Gauss and  $B>0.1$~$\mu$Gauss \citep{2009PASJ...61.1293S,2014A&A...562A..60O}. A recent overview of constraints on the volume-average magnetic field for radio halo and relic hosting clusters is given by \cite{2015A&A...582A..20B}.

\subsection{Magnetic fields at cluster radio shocks}
\label{sec:relicsbfield}

Similar to radio halos, measurements of IC X-ray emission can be used to determine magnetic field strength at the location of cluster radio shocks  \citep{1979ApJ...227..364R,1994ApJ...429..554R,2000ApJ...533...73S,2016ApJ...823...94R}, but so far no undisputed detections have been made. With deep X-ray observations, mostly from the XMM-Newton and Suzaku satellites, interesting lower limits on the magnetic field strength have been determined. \cite{2010ApJ...715.1143F} placed a lower limit of $3$~$\mu$Gauss on the northwest cluster radio shock region in Abell\,3667, consistent with an earlier reported lower limit of $1.6$~$\mu$Gauss  by \cite{2009PASJ...61..339N}. \cite{2015PASJ...67..113I} reported a lower limit of $1.6$~$\mu$Gauss for the Toothbrush Cluster. For the radio shock in the cluster RXC~J1053.7+5453, the lower limits was found to be $0.7$~$\mu$Gauss \citep{2017PASJ...69...88I}.

Another method to constrain the magnetic field strength at the location of cluster radio shocks is to use the source's width. Here the assumption is that the source's width is determined the characteristic timescale of electron energy losses (synchrotron and IC) and the shock downstream velocity. Using this method, values of either $\sim$1 or $\sim$5~$\mu$Gauss were found for the Sausage Cluster \citep{2010Sci...330..347V}. However, recent work by \cite{2018ApJ...852...65R} suggests that there are more factors affecting the downstream radio brightness profiles making the interpretation more complicated, for example, due to the presence of filamentary structures in the radio shock and a distribution of magnetic fields strengths \citep[see also][]{2018ApJ...865...24D}.
Taking some of these complications into account, \cite{2018ApJ...852...65R} concluded that  the magnetic field strength is less than 5~$\mu$Gauss for the Toothbrush cluster. 

\subsection{Future prospects}
{Surveys at frequencies of $\gtrsim 1$ GHz, such ongoing VLA Sky Survey at 2--4~GHz \citep[VLASS;][]{2016AAS...22732409L,2016AAS...22732408M}, and future surveys carried out with MeerKat \citep{2009arXiv0910.2935B,2009IEEEP..97.1522J}, ASKAP \citep{2011PASA...28..215N,2010AAS...21547013G}, and WSRT-APERTIF \citep{2008AIPC.1035..265V,2018AAS...23135404A} will provide larger samples of polarized radio sources that can be utilized for ICM magnetic field studies. In the more distant future, the SKA will provide even larger samples. This will enable the detailed characterization of magnetic fields in some individual (nearby) clusters, employing background and cluster sources \citep{2009MNRAS.400..646K,2015aska.confE..95B,2015aska.confE..92J,2016JApA...37...42R}.}

{Another important avenue to further pursue are hard X-ray observations to directly measure the IC emission from the CRe in the ICM \citep[e.g.,][]{2015A&A...582A..20B}. This will enable direct measurements of the ICM magnetic field strength at the location of radio shocks and halos.}

\section{Radio halos}
\label{sec:halos}

\subsection{Giant radio halos}

\begin{figure*}[htbp]
\centering
\includegraphics[width=0.49\textwidth]{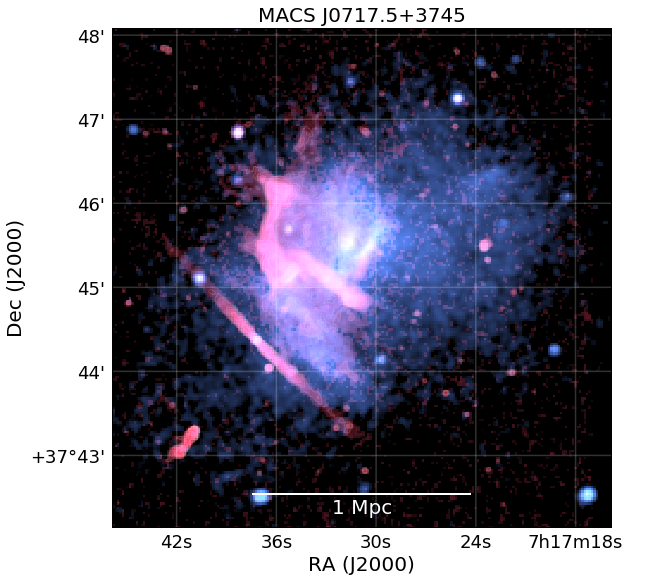}
\includegraphics[width=0.49\textwidth]{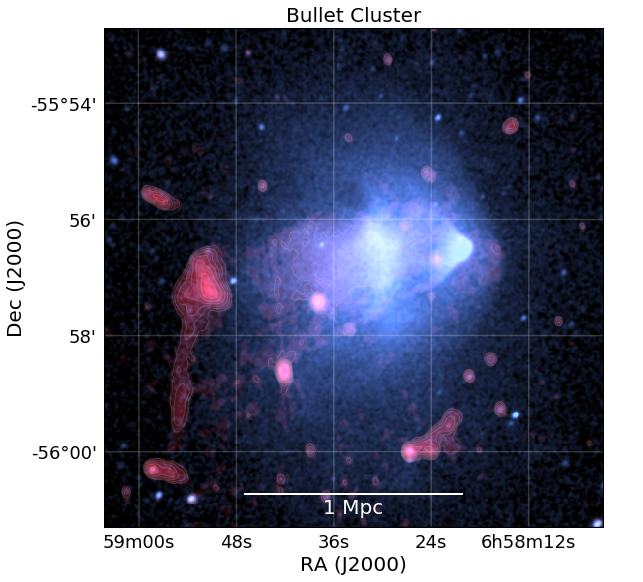}
\includegraphics[width=0.49\textwidth]{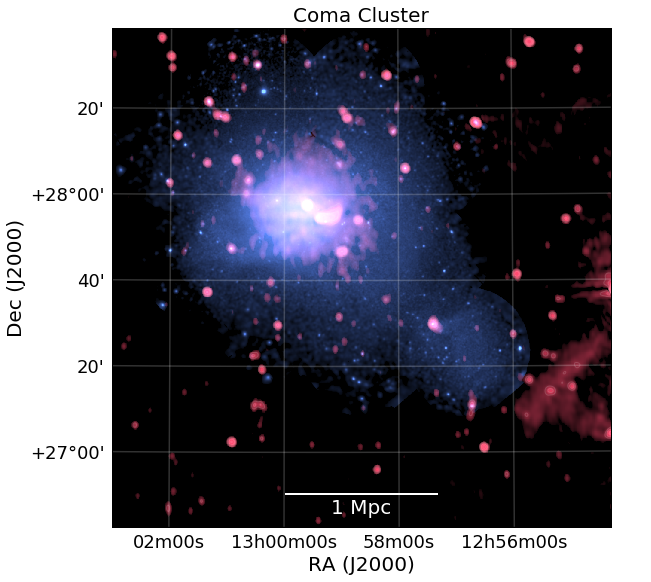}
\includegraphics[width=0.49\textwidth]{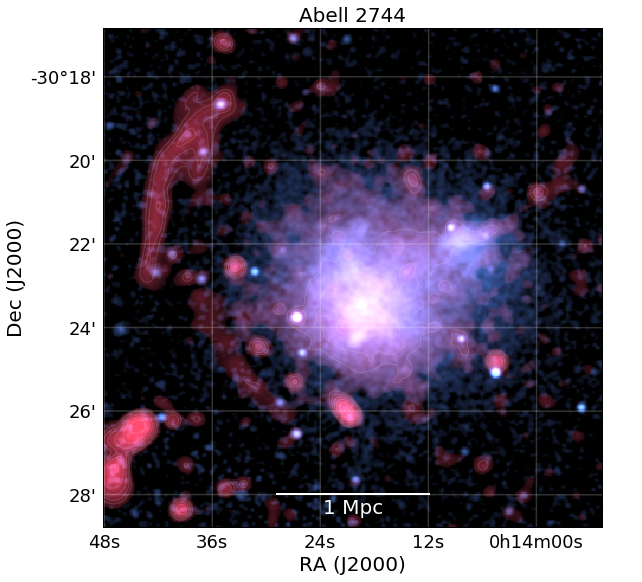}
\includegraphics[width=0.49\textwidth]{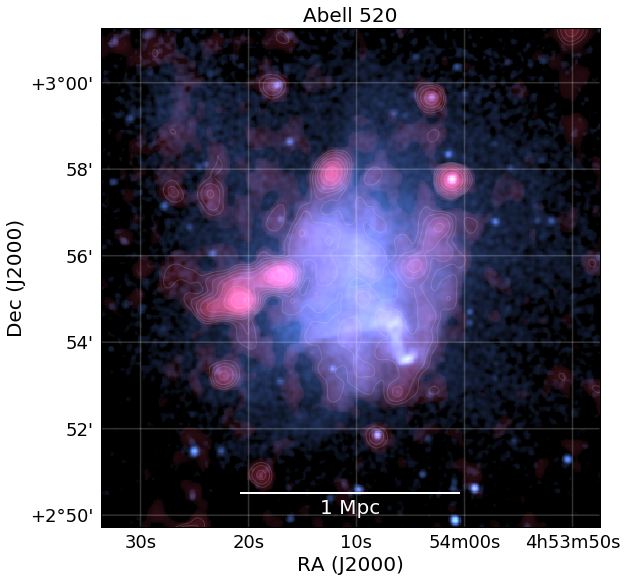}
\includegraphics[width=0.49\textwidth]{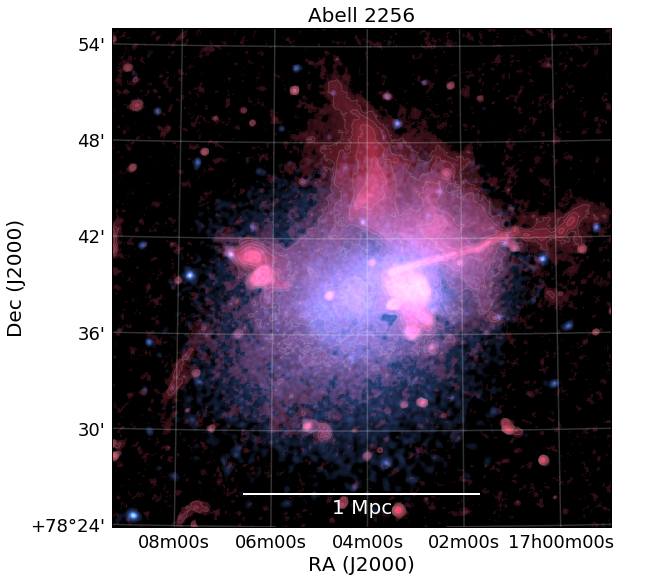}
\caption{Examples of clusters hosting giant radio halos. The radio emission is shown in red and the X-ray emission in blue. MACS\,J0717.5+3745: VLA 1--6 GHz and Chandra 0.5--2.0~keV \citep{2017ApJ...835..197V}. Bullet cluster: ATCA  1.1–-3.1~GHz and Chandra 0.5--2.0~keV \citep{2015MNRAS.449.1486S,2017ApJ...843...76A}. Coma cluster: WSRT 352~MHz and XMM-Newton 0.4--1.3~keV \citep{2011MNRAS.412....2B}. Abell\,2744: VLA 1--4~GHz and Chandra 0.5--2.0~keV \citep{2017ApJ...845...81P}. Abell\,520: VLA~1.4~GHz and Chandra 0.5--2.0~keV \citep{2018ApJ...856..162W,2017ApJ...843...76A}. Abell\,2256: LOFAR 120--170~MHz and XMM-Newton  0.4--1.3~keV (van Weeren et al. in prep).}
\label{fig:haloexamples}
\end{figure*}

Radio halos are diffuse extended sources that roughly follow the brightness distribution of the ICM. Giant Mpc-size radio halos are mostly found in massive dynamically disturbed clusters \citep{1999NewA....4..141G,2001ApJ...553L..15B,2010ApJ...721L..82C}.  The prototypical example is the radio halo found in the Coma cluster  \citep[e.g.,][]{1959Natur.183.1663L,1970MNRAS.151....1W,1993ApJ...406..399G,2003AA...397...53T,2011MNRAS.412....2B}. In Table~\ref{tab:clusterlist} we list the currently known giant radio halos and candidates. Some examples of clusters hosting giant radio halos are shown in Figure~\ref{fig:haloexamples}.

Giant radio halos have typical sizes of about 1--2~Mpc. The most distant radio halo is found in El Gordo at $z=0.87$ \citep{2012ApJ...748....7M,2014ApJ...786...49L,2016MNRAS.463.1534B}. The 1.4~GHz radio powers of observed halos range between about $10^{23}$ and $10^{26}$~W~Hz$^{-1}$, with the most powerful radio halo ($P_{\rm 1.4 GHz} = 1.6 \times 10^{26}$~W~Hz$^{-1}$) being present in the quadruple merging cluster MACS\,J0717.5+3745 \citep{2009AA...503..707B,2009A&A...505..991V}. The radio halo with the lowest power known to date ($P_{\rm 1.4 GHz} = 3.1 \times 10^{23}$~W~Hz$^{-1}$) is found in ZwCl\,0634.1+4747 \citep{2018AA...609A..61C}.
Other noteworthy examples are the double radio halos in the pre-merging cluster pairs Abell\,399--401 \citep{2010AA...509A..86M} and Abell\,1758N--1758S \citep{2018MNRAS.478..885B}.

Currently there are about 65~confirmed radio halos. Initially, most halos were found via the NVSS\footnote{NRAO VLA Sky Survey} \citep{1998AJ....115.1693C} and  WENSS\footnote{Westerbork Northern Sky Survey} \citep{1997A&AS..124..259R} surveys  \citep[e.g.,][]{1999NewA....4..141G,2001ApJ...548..639K,2009ApJ...697.1341R,2011AA...533A..35V,2017MNRAS.467..936G}. More recently, halos have been uncovered with targeted GMRT campaigns\footnote{Giant Metrewave Radio Telescope} \citep{2008AA...484..327V,2007AA...463..937V,2013AA...557A..99K,2015AA...579A..92K,2018arXiv180609579K}, and via low-frequency surveys such as  GLEAM\footnote{GaLactic and Extragalactic All-sky MWA Survey} \citep{2015PASA...32...25W,2017MNRAS.464.1146H} and LoTSS\footnote{The LOFAR Two-metre Sky Survey} \citep{2017A&A...598A.104S,2018arXiv181107926S}.
In addition, radio halo searches have been carried out with the VLA\footnote{Very Large Array}, ATCA\footnote{Australia Telescope Compact Array}, MWA\footnote{Murchison Widefield Array}, KAT-7\footnote{Seven-dish MeerKAT precursor array}, and LOFAR\footnote{LOw-Frequency ARray} \citep{2009AA...507.1257G,2016MNRAS.459.2525S,2016AA...595A.116M,2018AA...611A..94M,2016MNRAS.456.1259B,2018AA...609A..61C,2018arXiv181107929W,2018arXiv181108410S}.

\subsubsection{Morphology}
Radio halos typically have a smooth and regular morphology with the radio emission approximately following the distribution of the thermal ICM. This is supported by quantitative studies which find a point-to-point correlation between the radio and X-ray brightness distributions \citep{2001A&A...369..441G,2001AA...373..106F,2005AA...440..867G,2011MNRAS.412....2B,2018ApJ...852...65R}), although there are some exceptions. One example is the Bullet cluster, where no clear correlation is found \citep{2014MNRAS.440.2901S}. 

A few radio halos with more irregular shapes have been uncovered \citep[e.g.,][]{2009ApJ...704L..54G,2009AA...507.1257G,2011AA...530L...5G}. One striking example is MACS\,J0717.5+3745, where a significant amount of small scale structure is present within the radio halo \citep{2017ApJ...835..197V}. Although, it is not yet clear whether these structures really belong to the radio halo or if they are projected on top of it.
Two other peculiar cases are the ``over-luminous'' halos in the low luminosity X-ray cluster Abell\,1213 \citep{2009ApJ...704L..54G} and 0217+70 \citep{2011ApJ...727L..25B}. \cite{2011AA...530L...5G} discussed the interesting possibility that over-luminous halos represent a new class. However, better data is required to further investigate this possibility since none of these ``peculiar'' halos  have been studied in great detail, making the classification and interpretation more uncertain. For example, the peculiar ``halo'' in A523 has also been classified as a possible radio shock by \cite{2011AA...533A..35V}.

\subsubsection{Radio spectra}

The spectral properties of radio halos can provide important information about their origin. Therefore, considerable amount of work has gone into measuring the spectral properties of  halos.

A complication is that reliable flux density measurements of extended low signal to noise ratio sources are often not trivial to obtain. Reported uncertainties on flux density measurements in the literature often take into account the (1) map noise, assuming the noise is Gaussian distributed and not varying spatially across the radio halo, (2) flux-scale uncertainty, usually somewhere between 2 and 20\%, and (3) uncertainty in the subtraction of flux from discrete sources embedded in the diffuse emission. Correctly assessing latter effect can be hard, in particular at low frequencies when extended emission from radio galaxies (i.e., their tails and lobes) becomes more prominent and partly blends with the halo emission. 
Errors from incomplete uv-coverage and deconvolution are usually not included in the uncertainties. However, in principle they can be determined but this requires some amount of work. The uncertainties related to calibration errors, for example coming from model incompleteness or ionosphere, are often not fully taken into account. Calibration errors affect discrete source subtraction, the map noise distribution, deconvolution, and can lead to flux ``absorption''. For the above reasons, the reported uncertainties on radio halo flux-density measurements and spectral index maps in the literature can usually be thought of as lower limits on the true uncertainty.

\subsubsection{Integrated spectra}
\label{sec:halointspectra}
Most radio halos have integrated spectral indices in the range $-1.4<\alpha<-1.1$ \citep[e.g.,][]{2009AA...507.1257G}.

\begin{figure}[htbp]
\centering
\includegraphics[width=0.49\textwidth]{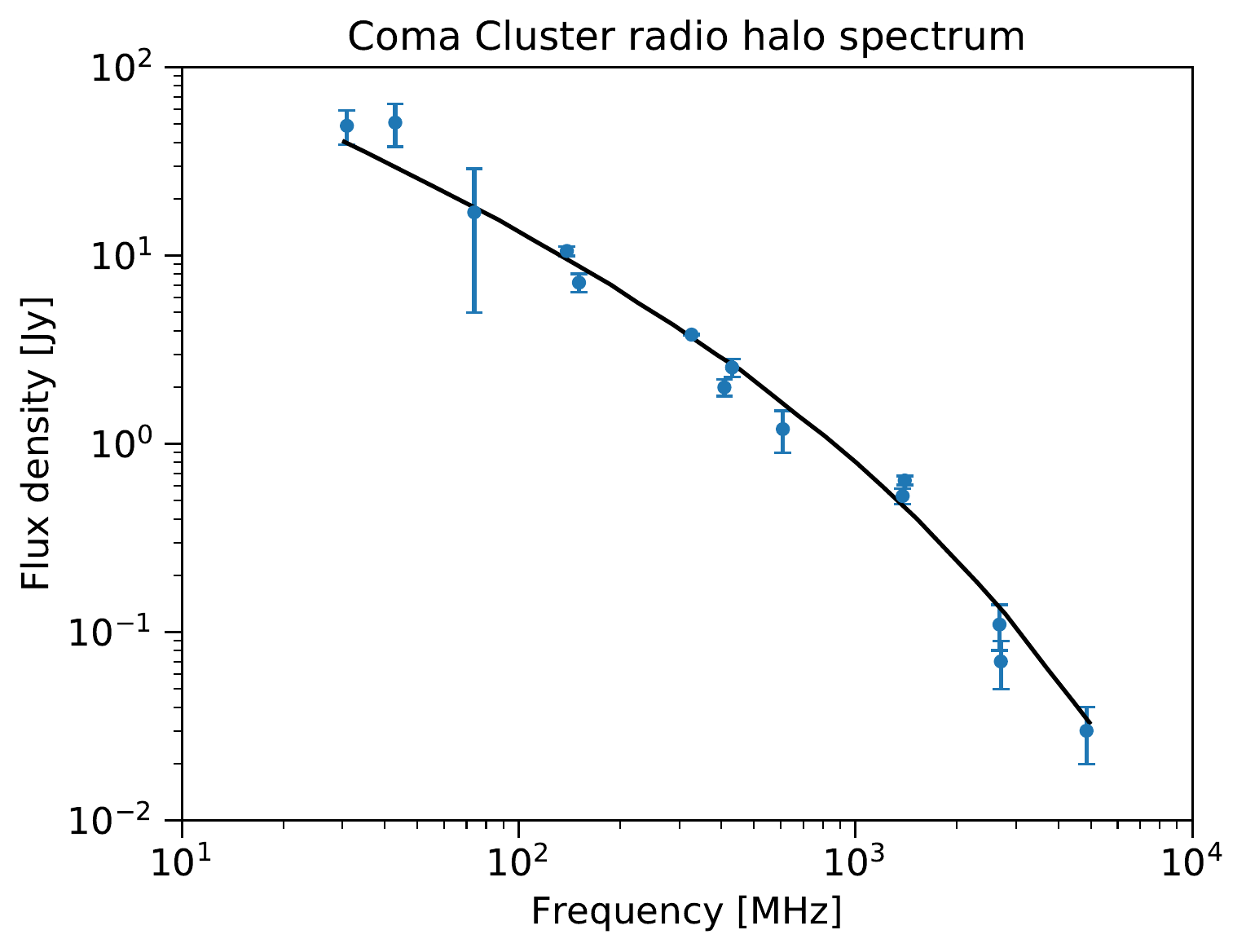}
\caption{The integrated spectrum of the radio halo in the Coma cluster.  The black line shows an in-situ acceleration model fit. The measurements and fit are taken from \cite{2010PhDT.......259P} and references therein.}
\label{fig:comahspec}
\end{figure}

The spectral information of most radio halos is based on measurements at just two frequencies.  Recently, two systematic campaigns have been carried out with the GMRT to follow-up clusters at lower frequencies to obtain spectra \citep{2013AA...551A.141M,2013AA...551A..24V}. Flux density measurements at more than three frequencies that also cover a large spectral baseline are rare. Therefore, deviations from power-law spectral shapes are difficult to detect. The best example of a radio halo with an observed spectral steepening, displayed in Figure~\ref{fig:comahspec}, is the Coma cluster \citep{2003AA...397...53T}. Importantly, it has also been shown that most of this steepening is not due to the Sunyaev-Zel'dovich effect (SZ) decrement \citep{2013A&A...558A..52B}. Other halos with well sampled spectra include the Toothbrush and Bullet cluster which show power-law spectral shapes  \citep{2000ApJ...544..686L,2012AA...546A.124V,2014MNRAS.440.2901S}.

There is some evidence that the integrated spectra of radio halos show a correlation with the global ICM  temperature of clusters, where hotter clusters host halos with flatter spectra \citep{2004NewAR..48.1137F,2009AA...507.1257G}.
However, \cite{2010ApJ...718..939K} pointed out that comparing  the average values of ICM temperatures and of spectral indices can give inconclusive results.

\subsubsection{Resolved spectra}
\begin{figure*}[htbp]
\centering
\includegraphics[width=0.49\textwidth]{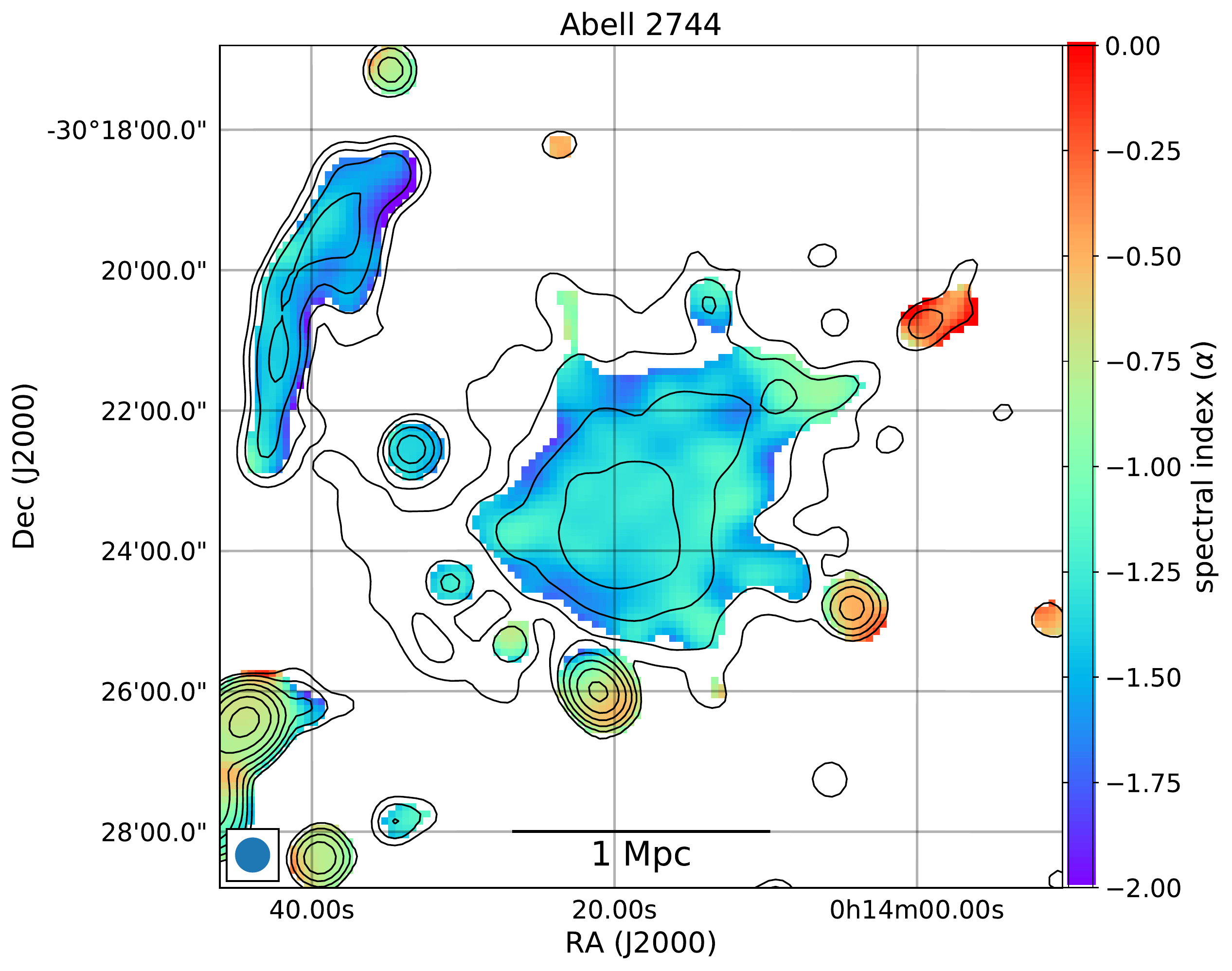}
\includegraphics[width=0.49\textwidth]{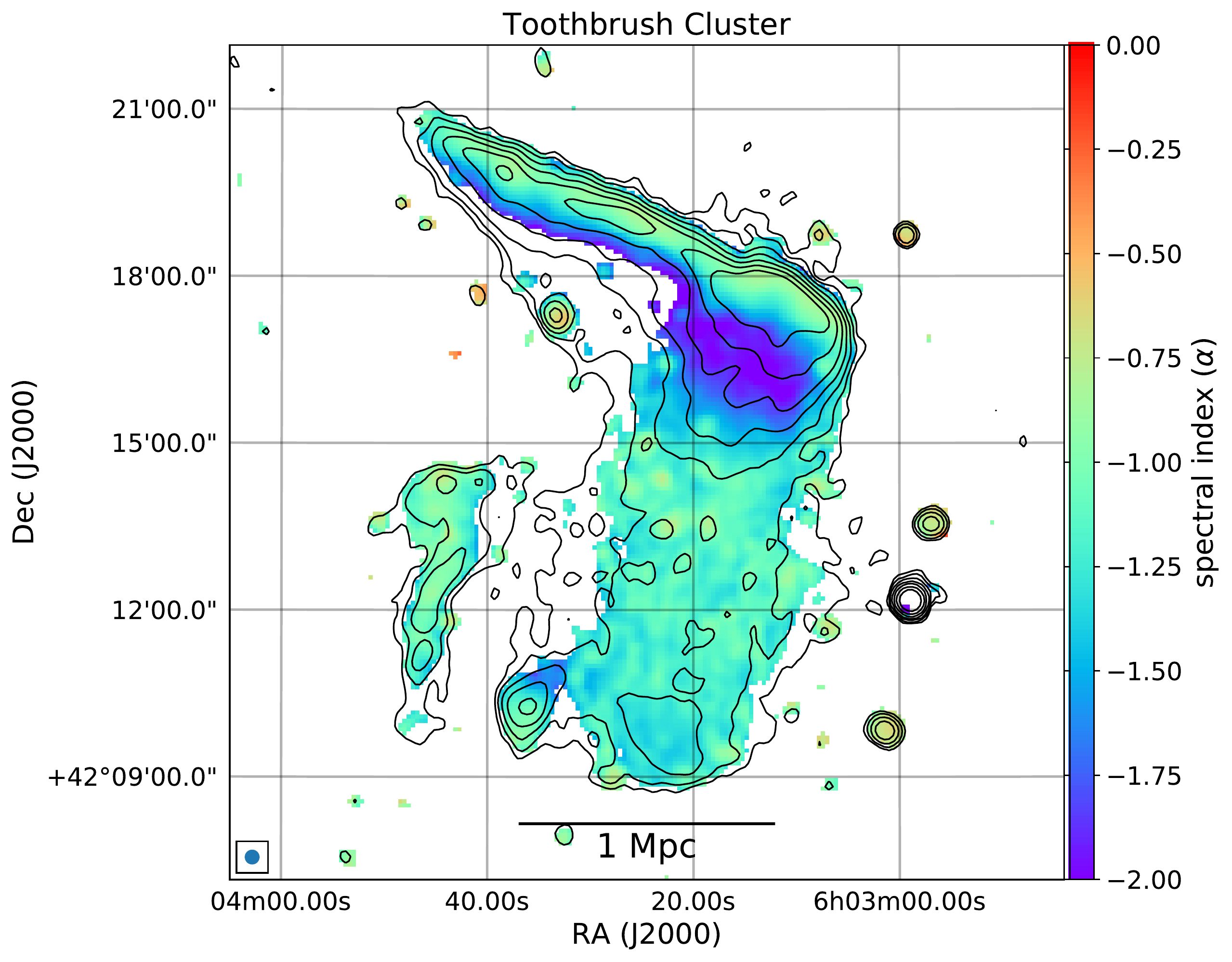}
\caption{\textit{Left panel:} Spectral index map of the radio halo in Abell\,2744 between 1.5 and 3.0~GHz obtained with the VLA \citep{2017ApJ...845...81P}. The 1.5~GHz radio contours are overlaid in black at levels of $[1,4,16,\ldots] \times 4 \sigma_{\rm rms}$, where $\sigma_{\rm rms}$ is the map noise. Besides a radio halo, the image also displays a large radio shock to the northwest of the cluster central region. \textit{Right panel:} Spectral index map of the radio halo in the Toothbrush cluster between 150~MHz and 1.5~~GHz using LOFAR and the VLA \citep{2018ApJ...852...65R}. Contours are from the 150~MHz LOFAR image and drawn at the same levels as in the left panel. North of the radio halo, 
a luminous 2~Mpc radio shock is also present.}
\label{fig:A2744spixhalo}   
\end{figure*}

The first detailed study of the spatial distribution of the radio spectral index across a radio halo was carried out by \cite{1993ApJ...406..399G}. They found a smooth spectral index distribution for the Coma cluster radio halo, with evidence for radial spectral steepening. For Abell\,665 and Abell\,2163 hints of radial spectral steepening where also found in undisturbed cluster regions \citep{2004AA...423..111F}. A caveat of these studies is that they were not done with matched uv-coverage, which could lead to errors in the derived spectral index distributions. Some other studies of radio halo spectral index distributions are \cite{2005AA...440..867G,2007AA...467..943O,2009AA...507..639P,2010ApJ...718..939K,2014MNRAS.440.2901S,2017ApJ...845...81P}. Two examples radio halo spectral index maps, for the massive merging clusters Abell\,2744 and the Toothbrush, are shown in Figure~\ref{fig:A2744spixhalo}. It shows that the spectral index is rather uniform across these radio halos.

A spatial correlation between radio spectral index and ICM temperature ($T$) for Abell\,2744 was reported by \cite{2007AA...467..943O}, with flatter spectral index regions corresponding to higher temperatures. However, using deeper VLA and Chandra data this result was not confirmed \citep{2017ApJ...845...81P}. Similarly, no clear evidence for such a correlation was founding in Abell\,520 \citep{2014AA...561A..52V}, the Toothbrush Cluster \citep{2016ApJ...818..204V}, the Bullet cluster \citep{2014MNRAS.440.2901S}, and Abell\,2256 \citep{2010ApJ...718..939K}. The current results therefore indicate there is no strong $T-\alpha$ correlation present, although more studies are necessary. It has been noted that even in the presence of an underlying $T-\alpha$ correlation, projection effects might also significantly reduce the detectability \citep{2010ApJ...718..939K}.

\subsubsection{Ultra-steep spectrum radio halos}
\label{sec:halouss}
Some halos have been found that have ultra-steep spectra, up to $\alpha\sim -2$. Radio halos with $\lesssim-1.6$ have been called ultra-steep spectrum radio halos (USSRH). The existence of USSRH is expected if the integrated spectra of radio halos include a cutoff. When we measure the spectral index close to the cutoff frequency ($\nu_{\rm b}$) it becomes very steep. Any radio halo can thus appear as an USSRH as along as we observe it close to (or beyond) the cutoff frequency. It is expected that only the most luminous radio halos, corresponding to the most energetic merger events, have cutoff frequencies of $\gtrsim 1$~GHz. In the turbulent re-acceleration model, the location of the cutoff frequency approximately scales as  \citep{2010A&A...509A..68C}, 
\begin{equation}
\nu_{\rm b} \propto M^{4/3} \mbox{ ,}
\end{equation}
where $M$ is the mass of the main cluster. In connection with major merger events
\begin{equation}
\nu_{\rm b} \propto \left( 1 + \Delta M/M \right)^{3}  \mbox{ ,}
\end{equation}
where $\Delta M$ the mass the merging subcluster. Because of these scalings, it is expected that more USSRH radio halos, corresponding to less energetic merger events, can be uncovered with sensitive observations at low frequencies.

The prime example of a USSRH is found in Abell\,521 \citep{2008Natur.455..944B,2009ApJ...699.1288D}, 
Other clusters with USSRH or candidate USSRH are Abell\,697 \citep{2010AA...517A..43M,2011AA...533A..35V,2013AA...551A.141M}, Abell\,2256 \citep{2008AA...489...69B}, Abell\,2255 \citep{1997AA...317..432F,2009AA...507..639P}, Abell\,1132 \citep{2018MNRAS.473.3536W}, MACS\,J0416.1--2403 \citep{2015sf2a.conf..247P}, MACS\,J1149.5+2223 \citep{2012MNRAS.426...40B}, 
Abell\,1300 \citep{1999MNRAS.302..571R,2013AA...551A..24V}, and
PSZ1\,G171.96--40.64 \citep{2013ApJ...766...18G}. 
It should be noted that a number of these USSRH still need to be confirmed. The reason is that reliable spectral index measurements are difficult to obtain because of differences in uv-coverage, sensitivity, resolution, and absolute flux calibration. This situation will improve with the new and upgraded radio telescopes that have become operational, in particular at low frequencies.
One example of a candidate radio halo with an ultra-steep spectrum was Abell\,1914 \citep{2003AA...400..465B}. Recent LOFAR and GMRT observations suggest that the most of the diffuse emission in this cluster does not come from a halo but instead from a radio phoenix \citep{2018arXiv181108430M}.

\subsubsection{Polarization}
Radio halos are found to be generally unpolarized. 
This likely is caused by the limited angular resolution of current observations, resulting in beam depolarization. This effect is significant when the beam size becomes larger than the angular scale of coherent magnetic field regions.  Even at high-angular resolution, magnetic field reversals and resulting Faraday rotation will reduce the amount of observed polarized flux.

For three clusters, Abell\,2255, MACS\,J0717.5+3745, and  Abell\,523 significant polarization has been reported \citep{2005AA...430L...5G,2009AA...503..707B,2016MNRAS.456.2829G}, but it is not yet fully clear whether this emission is truly from the radio halos, or from polarized cluster radio shocks projected on-top or near the radio halo emission \citep{2011AA...525A.104P,2017ApJ...835..197V}.

\cite{2013A&A...554A.102G} modeled the radio halo polarization signal at 1.4~GHz and inferred that radio halos should be intrinsically polarized. The fractional polarization at the cluster centers is about 15--35\%, varying from cluster to cluster, and increasing with radial distance. However, the polarized signal is generally undetectable if it is observed with the low sensitivity and  resolution of current radio interferometers. The \cite{2013A&A...554A.102G} results are based on MHD simulations by \cite{2011ApJ...739...77X,2012ApJ...759...40X} which are probably not  accurate enough yet to resolve the full dynamo amplification. Whether this will affect the predicted fractional polarization levels is not yet clear, see \cite{2018SSRv..214..122D}. If the polarization properties of radio halos can be obtained from future observations it would provide very valuable information on the ICM magnetic field structure.

\subsubsection{Samples and scaling relations, merger connection}

Statistical studies of how the radio halo properties relate to the ICM provide important information on the origin of the non-thermal CR component.

\begin{figure*}[htbp]
\centering
\includegraphics[width=0.49\textwidth]{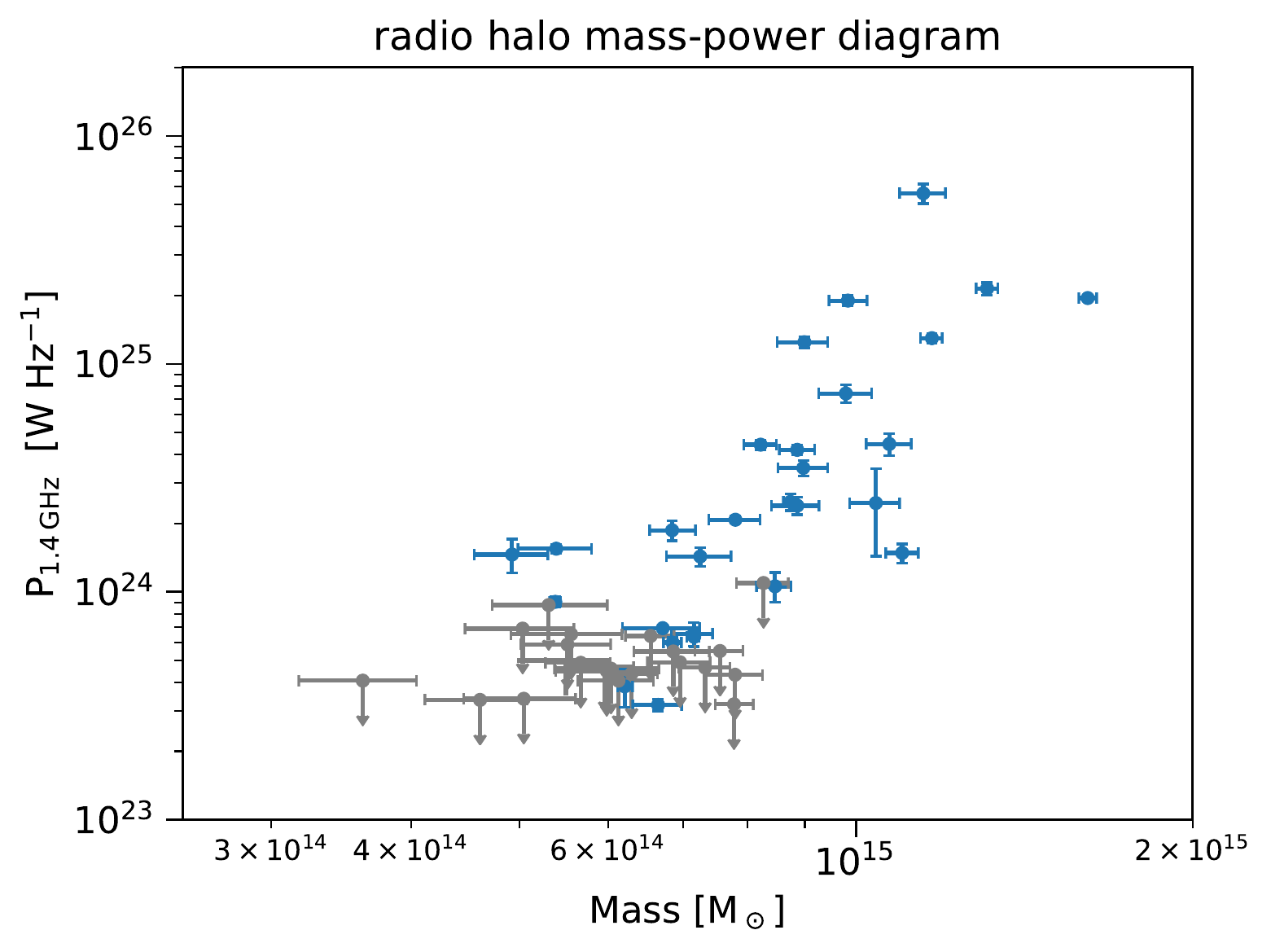}
\includegraphics[width=0.49\textwidth]{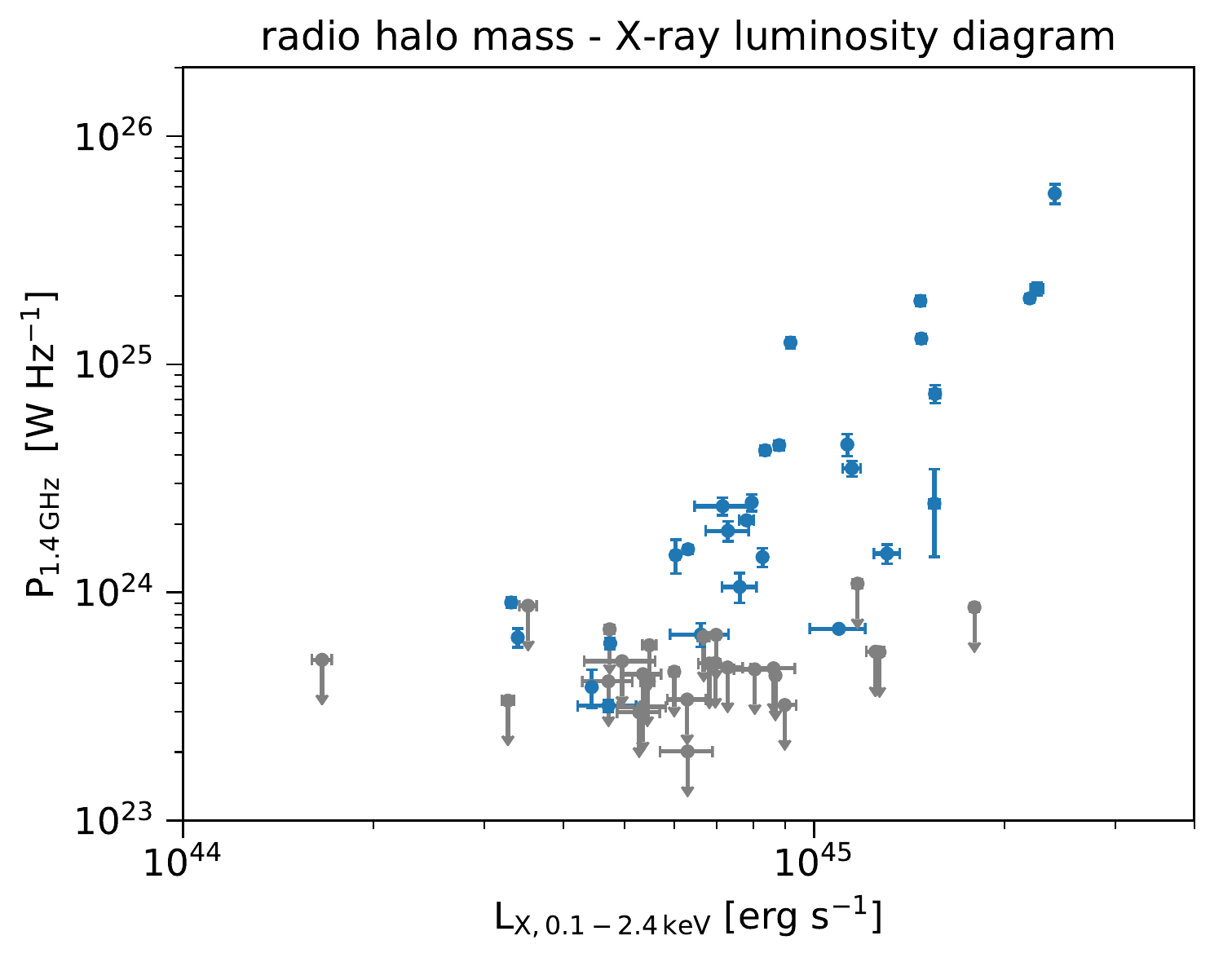}
\caption{Radio halos in the mass (\textit{left panel}) and $L_{\rm{X}}$ (\textit{right panel}) -- radio power diagrams. Radio halos are taken from \cite{2013ApJ...777..141C,2015AA...579A..92K,2018AA...609A..61C} and references therein. Cluster masses are taken from the Planck PSZ2 catalog \citep{2016A&A...594A..27P}.}
\label{fig:haloscorr}
\end{figure*}

It is well known \citep[e.g.,][]{2000ApJ...544..686L,2002A&A...396...83E,2003ASPC..301..143F,2015ApJ...813...77Y} that the radio power (luminosity) of giant halos correlates with the cluster X-ray luminosity ($L_{\rm{X}}$), and thus cluster mass.
For observational reasons, the radio power at 1.4~GHz ($P_{\rm{1.4 GHz}}$) is commonly used to study scaling relations. The X-ray luminosity is often reported in the 0.1--2.4~keV ROSAT band. Figure~\ref{fig:haloscorr} shows a compilation of radio halos and upper limits on a mass-$P_{\rm{1.4 GHz}}$ and $L_{\rm{X}}$-$P_{\rm{1.4 GHz}}$ diagram. Detailed investigations of the scaling relations between radio power and X-ray luminosity (or mass), based on the turbulent re-acceleration model, were performed by \cite{2006MNRAS.369.1577C,2007MNRAS.378.1565C,2008A&A...480..687C}. These models were also used to predict the resulting statistics for upcoming  radio surveys \citep{2010A&A...509A..68C,2010A&A...517A..10C,2012A&A...548A.100C}. More recently, the integrated Sunyaev-Zel'dovich Effect signal (i.e., the Compton $Y_{\rm{SZ}}$ parameter) has been used as a proxy of cluster mass \citep{2012MNRAS.421L.112B,2013ApJ...777..141C,2014MNRAS.437.2163S}. The advantage from using this proxy stems from the fact that $Y_{\rm{SZ}}$ should be less affected by the dynamical state of a cluster, providing less scatter compared to $L_{\rm{X}}$ \citep[e.g.,][]{2005ApJ...623L..63M,2008ApJ...680...17W}.

To determine radio halo power or upper limits for statistical studies, it is important to derive these quantities in a homogeneous way and minimize the dependence on map noise or uv-coverage. This argues against using a certain contour level, often $3\sigma_{\rm rms}$ has been used, to define the radio halo flux density integration area. Assumptions have to be made on the brightness distribution to determine upper limits for non-detections \citep{2007ApJ...670L...5B,2009AA...499..679M,2011MNRAS.417L...1R}. For example, \cite{2017MNRAS.470.3465B} used an exponential radial profile of the form
\begin{equation}
I(r) = I_0e^{-r/r_{e}} \mbox{ ,}
\label{eq:expradialprofile}
\end{equation} 
with added brightness fluctuations, with the characteristic sizes ($r_{e}$, e-folding radius) determined from previously found correlations between power and size \citep{2007MNRAS.378.1565C,2009AA...499..679M}. In addition, ellipsoidal profiles were employed for clusters with very elongated X-ray brightness distributions. The effects of uv-coverage, visibility weighting, mosaicking (for observations that combine several pointings), and deconvolution can be quantified by injection of mock radio halos into the uv-data \citep{2007ApJ...670L...5B,2017arXiv170604930J}.

Radio halos are rather common in massive clusters. An early study by \cite{1999NewA....4..141G} showed that about 6\%--9\% of $L_{\rm{X}} < 5\times 10^{44}$~erg~s$^{-1}$ clusters host halos at the limit of the NVSS survey, while this number increases to 27\%--44\% above this luminosity.
Extensive work, mainly using the GMRT, provided further improvements on the statistics, showing that the occurrence fraction  for clusters with $L_{\rm{X}} > 5\times 10^{44}$~erg~s$^{-1}$ is about 30\% \citep{2007AA...463..937V,2008AA...484..327V,2013ApJ...777..141C,2015AA...579A..92K}. For a mass-selected sample ($M >6\times 10^{14}$~M$_{\odot}$), \cite{2015A&A...580A..97C} found evidence for a drop in the halo occurrence fraction for lower mass clusters. For clusters with $M >8\times 10^{14}$~M$_{\odot}$ this fraction is $\approx 60\%-80\%$, dropping to $\approx 20\%-30\%$ below this mass.

An important result from observations is that giant radio halos are predominately found in merger clusters, as indicated by a disturbed ICM and/or other indicators of the cluster's dynamical state, e.g., the velocity distribution of cluster member galaxies, presence of multiple BCGs, and galaxy distribution. Early work already established evidence that radio halos were related to cluster merger events as determined from X-ray observations  \citep[e.g.,][]{2000cucg.confE..37F,2001ApJ...553L..15B,2001A&A...378..408S,2002HiA....12..519S,2002IAUS..199..133F,2002ASSL..272..197G,2002ASSL..272..133B}. This conclusion is also supported by optical studies \citep{2003A&A...399..813F,2004A&A...416..839B,2006A&A...449..461B,2006A&A...455...45G,2007A&A...467...37B,2008A&A...491..379G,2008A&A...487...33B,2009A&A...495...15B,2010A&A...517A..65G,2011A&A...536A..89G,2012A&A...540A..43B,2012A&A...547A..44B,2014MNRAS.442.2216B,2016MNRAS.456.2829G,2016ApJ...831..110G}. A common method is to use the cluster's X-ray morphology as an indicator of the cluster's dynamical state, such as the centroid shift, power ratio, and concentration parameter \citep{2001ApJ...553L..15B,2010ApJ...721L..82C}. 
Almost all giant ($\gtrsim 1$~Mpc) radio halos so far have been found in dynamically disturbed clusters. Recent studies also confirm this general picture \citep{2013ApJ...777..141C,2015AA...579A..92K,2015A&A...580A..97C}, but see Section~\ref{sec:halounification} for some  exceptions. 

Further support for the relation between cluster mergers and the presence of radio halos was presented by \cite{2009A&A...507..661B}. They found that there is a  radio bi-modality between merging and relaxed clusters. Merging clusters host radio halos, with the radio power increasing with $L_{\rm X}$. Relaxed clusters do not show the presence of halos, with upper limits located well below the expected correlation. Similarly, \cite{2011A&A...532A.123R,2011ApJ...740L..28B} find that the occurrence of halos is related to the cluster's evolutionary stage. 
Early work by \cite{2012MNRAS.421L.112B} reported a lack of a radio bimodality in the Y--P plane. However, this was not confirmed by \cite{2013ApJ...777..141C}. On the other hand, X-ray selected cluster samples are biased towards selecting cool core clusters, which generally do not host giant radio halos, and hence the occurrence fraction of radio halos in SZ-selected samples is expected to be higher \citep{2014MNRAS.437.2163S,2017ApJ...843...76A}. Recently, \cite{2018AA...609A..61C} found two radio halos that occupy the region below the mass-$P_{\rm{1.4 GHz}}$ correlation. These two underluminous radio halos do not have steep spectra and could be generated during minor mergers where turbulence has been dissipated in smaller volumes, or be ``off-state'' radio halos originating from hadronic collisions in the ICM.

Some merging clusters that host cluster double radio shocks (see Section~\ref{sec:doublerelics}), do not show the presence of a radio halo \citep{2017MNRAS.470.3465B}. This absence of a radio halo might be related to early or late phase mergers, and
the timescale of halo formation and disappearance. Although, these results are not yet statistically significant given the small sample size.

\cite{2016A&A...593A..81C} investigated whether giant radio halos can probe the merging rate of galaxy clusters. They suggested that merger events generating radio halos are characterized by larger mass ratios. Another possible explanation is that  radio halos may be generated in all mergers but their lifetime is shorter than the timescale of the merger-induced disturbance. The lack of radio halos in some merging clusters can also be caused by the lack of sufficiently deep observations. One prime example is Abell\,2146 \citep{2011MNRAS.417L...1R} where no diffuse emission was found in GMRT observations. However, recent deep VLA and LOFAR observations revealed the presence of a radio halo in this cluster \citep{2018MNRAS.475.2743H,2018arXiv181109708H}.

\subsubsection{Origin of radio halos}
\label{sec:originhalo}

The origin of radio halos have been historically debated between two models: the hadronic and turbulent re-acceleration models. In the hadronic model, radio emitting electrons are produced in the hadronic interaction between CR protons and ICM protons 
\citep{1980ApJ...239L..93D,1999APh....12..169B,2000A&A...362..151D,2001ApJ...562..233M,2008MNRAS.385.1211P,2010ApJ...722..737K,2011A&A...527A..99E}. In the re-acceleration model, a population of seed electrons \citep[e.g.,][]{2017MNRAS.465.4800P} is re-accelerated during powerful states of ICM turbulence \citep{2001MNRAS.320..365B,2001ApJ...557..560P,2013MNRAS.429.3564D,2014MNRAS.443.3564D}, as a consequence of a cluster merger event. While indirect arguments against the hadronic model can be drawn from the integrated radio spectral \citep{2008Natur.455..944B} and spatial characteristics of halos, and from radio--X-ray scaling relations \citep[for a review see][]{2014IJMPD..2330007B}, only gamma-ray observations, which will be discussed in more detail below (Section~\ref{sec:halogamma}), of the Coma cluster directly determined that radio halos cannot be of hadronic origin. {The spatial distribution of spectral indices across radio halos,  which can go from being very uniform to more patchy, might provide further tests for turbulent re-acceleration model. Furthermore, additional high-frequency ($\gtrsim 5$~GHz) observations of known radio halos would enable a search for possible spectral cutoffs. Such cutoffs are expected in the framework of the turbulent re-acceleration model, but have so far rarely been observed (see Sections~\ref{sec:halointspectra} and~\ref{sec:halouss}). Such measurements would be quite challenging though, requiring single dish observations to avoid resolving out diffuse emission.}

Nowadays, turbulent re-acceleration is thought to be the main mechanism responsible for generating radio halos, even if other mechanisms as magnetic reconnection have been proposed \citep[e.g.,][]{2016MNRAS.458.2584B}. However, one of the main open questions for the re-acceleration model is the source of the seed electrons. There are several possibilities, with secondary electrons coming from proton-proton interactions being an obvious candidate \citep{2005MNRAS.363.1173B,2011MNRAS.410..127B}. The seed electrons could also have been previously accelerated at cluster merger and accretion shocks. A third possibility is that the seed electrons are related to galaxy outflows and AGN activity. 
The latter, in particular, is becoming more and more evident thanks to the recent low-frequency observation of re-energized tails \citep[][see Section~\ref{sec:fossil}]{2017SciA....3E1634D} and fossil plasma sources \citep[e.g.,][]{2016MNRAS.459..277S}. While it is difficult to determine the possible contribution of these primary sources of seed electrons, gamma-ray observations can be used to study the contribution of secondary electrons. Another important open question in this context is the connection with the generation mechanism for mini-halos that will be discussed in Section~\ref{sec:halounification}.

\cite{2017ApJ...843L..29E} used the amplitude of density fluctuations in the ICM as a proxy for the turbulent velocity. Importantly, they inferred that radio halo hosting clusters have one average 
and a factor of two higher turbulent velocities. However, this indirect method relies on number of  assumptions making the result somewhat open to interpretation. Direct measurements of ICM turbulence have so far only been performed for the Perseus cluster with the Hitomi satellite \citep{2016Natur.535..117H,2018PASJ...70....9H},  finding a line-of-sight velocity dispersion of $164 \pm 10$~km~s$^{-1}$. {Future measurements with \emph{XRISM} (X-ray Imaging and Spectroscopy Mission) and \emph{Athena} \citep{2013arXiv1306.2307N,2016SPIE.9905E..2FB} of the turbulent motions in halo and non-halo hosting clusters will provide crucial tests for the turbulent re-acceleration model.}

\subsubsection{Gamma-ray upper limits}
\label{sec:halogamma}
Gamma-rays in clusters of galaxies are expected from neutral pion decays coming from proton-proton interactions \citep[for more details see][]{2004JKAS...37..307R,2007IJMPA..22..681B,2011PhRvD..84l3509P}. As mentioned earlier, CR protons can be injected in clusters by structure formation shocks and galaxy outflows, and can accumulate there for cosmological times. The quest for the detection of these gamma-rays have been going on for about two decades now \citep{2003ApJ...588..155R,2004NewAR..48..481R,2009A&A...502..437A,2010ApJ...717L..71A,2010ApJ...710..634A,2012ApJ...757..123A,2012A&A...547A.102H,2013A&A...560A..64H,2014ApJ...787...18A,2014MNRAS.440..663Z,2014A&A...567A..93P,2014ApJ...795L..21G,2016ApJ...819..149A,2016PhRvD..93j3525L,2017ApJS..228....8B}. Unfortunately, the detection of diffuse gamma-ray emission connected with the ICM has been so far elusive. There is no conclusive evidence for an observation yet. 

Nevertheless, gamma-ray observations have been very important in the last few years for three reasons: to put a direct limit on the CR content in clusters, to test the hadronic nature of radio halos and mini-halos, and to test the contribution of secondary electrons in re-acceleration models. The number of works on this topic are numerous, thanks to the observations of imaging atmospheric Cherenkov telescopes and of gamma-ray satellites, and the most relevant ones have been cited in the previous paragraph.

Of particular importance for this review are the observations of Coma and Perseus clusters (results for the Perseus cluster will be discussed in Section~\ref{sec:minihalogamma}), and of larger combined samples of nearby massive and X-ray luminous clusters. The combined likelihood analysis of the \emph{Fermi}-Large Area Telescope \citep[LAT;][]{2009ApJ...697.1071A} satellite of 50 HIFLUGCS clusters have been a milestone in constraining the amount of CR protons in merging clusters to be below a few percent \citep{2014ApJ...787...18A}. However, the most constraining object is the Coma cluster due to its high mass, closeness and radio-halo brightness. In fact, thanks to the \emph{Fermi}-LAT observations, we are now able to exclude the hadronic origin of the prototypical radio halo of Coma independently from the exact magnetic field value in the cluster  \citep{2012MNRAS.426..956B,2017MNRAS.472.1506B}, a long standing issue in the field \citep[e.g.,][]{2011ApJ...728...53J}. In particular, the CR-to-thermal energy in Coma is limited to be $\lesssim 10$\%, almost independently (within a factor or two) from the specific model considered, i.e., re-acceleration or hadronic, and from the magnetic field \citep{2017MNRAS.472.1506B}. Additionally, the \emph{Fermi}-LAT observations of Coma are starting to test re-acceleration models. These first gamma-ray constraints on re-acceleration are obtained under the assumption that only CR protons and their secondaries are present in the ICM \citep{2017MNRAS.472.1506B}. While we obviously know that this is not the case (see the discussion in the previous Sec.~\ref{sec:originhalo}), it is possible that CR protons and their secondaries give the dominant seed contribution.

\subsubsection{Radio halo-shock edges}
\begin{figure*}[htbp]
\centering
\includegraphics[width=0.49\textwidth]{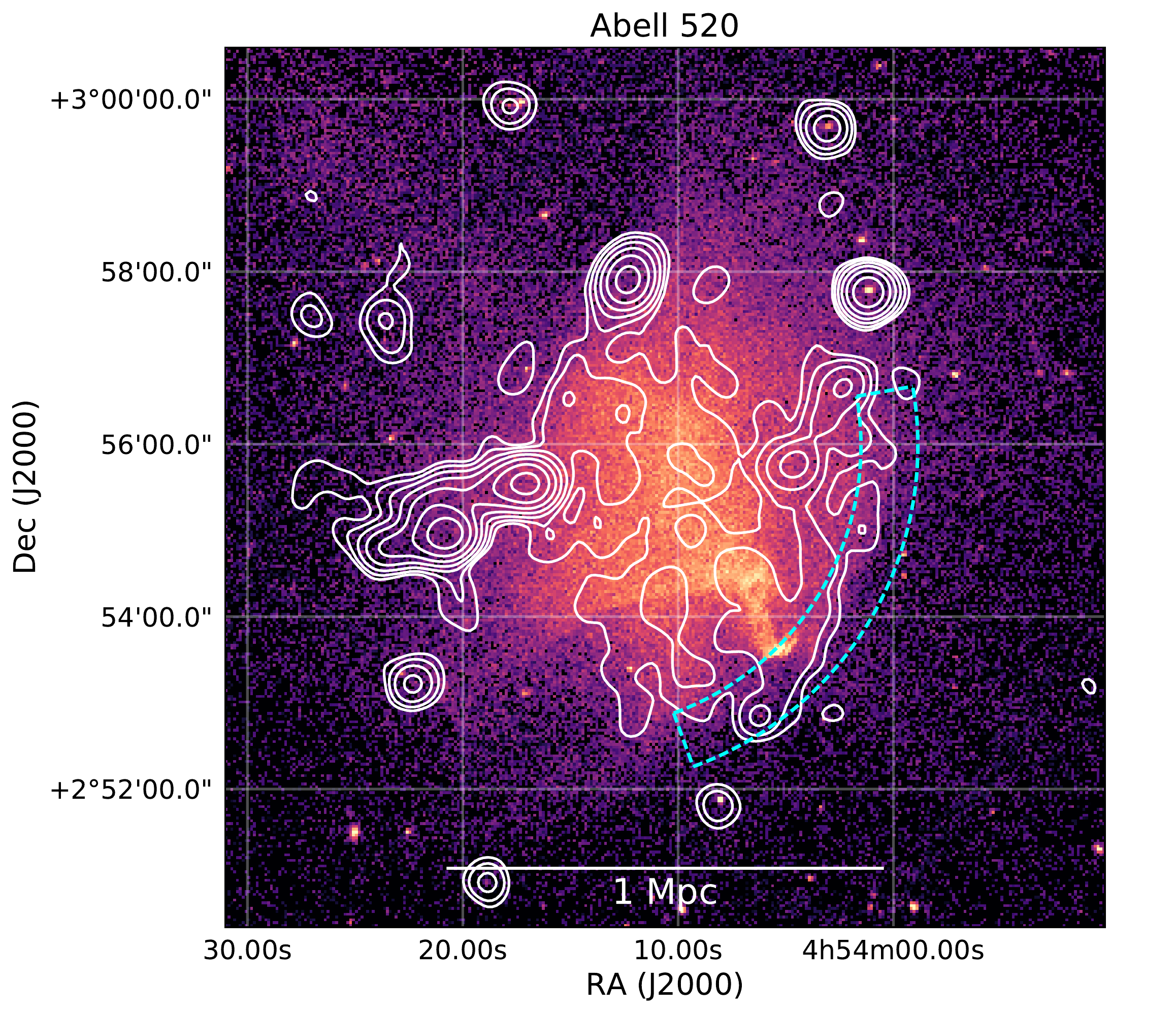}
\includegraphics[width=0.49\textwidth]{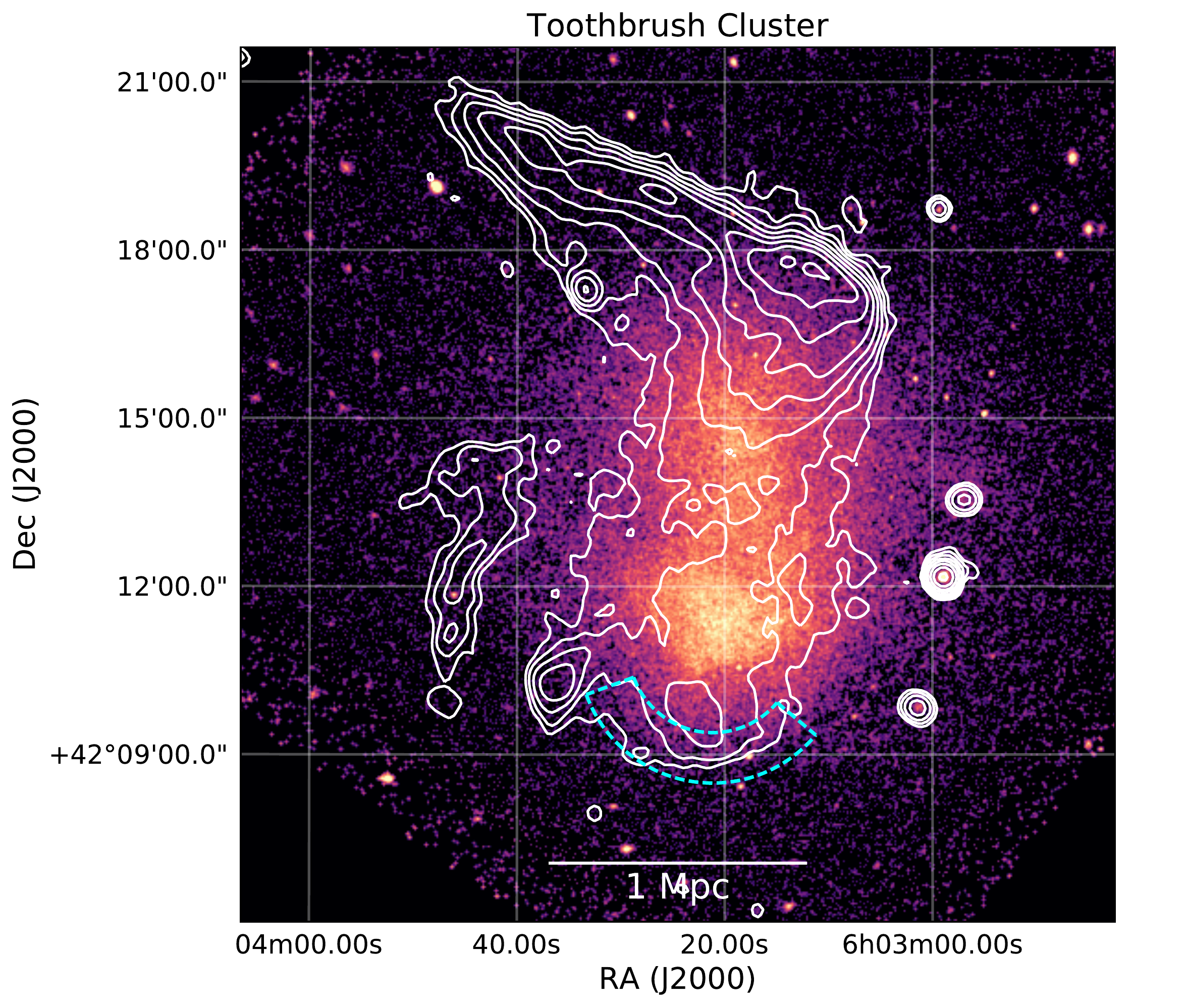}
\caption{Radio halo-shock edges in Abell 520 \citep[\textit{left};][]{2018ApJ...856..162W} and the Toothbrush Cluster \citep[\textit{right};[]{2016ApJ...818..204V}. VLA 1.4~GHz and LOFAR 150~MHz contours are overlaid at levels of $[1,2,4,8,\ldots] \times 5 \sigma_{\rm rms}$ (where $\sigma_{\rm rms}$ is the map noise)
 for the left and right panel images, respectively. The halo-shock edges are indicated by the cyan colored dashed regions.}
\label{fig:haloedges}
\end{figure*}

In a handful of clusters the radio halo emission seems to be bounded by cluster shock fronts  \citep{2005ApJ...627..733M,2011MNRAS.412....2B,2010arXiv1010.3660M,2013A&A...554A.140P,2014AA...561A..52V,2014MNRAS.440.2901S,2016ApJ...818..204V}. Two of these examples of ``halo-shock edges'' are shown in Figure~\ref{fig:haloedges}. The nature of these sharp edges is still unclear.

It is possible that some of the ``halo'' emission near these shocks comes from CR electrons compressed at the shock. Alternatively, these edges are cluster radio shocks where electrons are (re-) accelerated. When these electrons move further downstream they will be re-accelerated again, but now by turbulence generated by the merger. Then, depending on the observing frequency, magnetic field strength (which sets the cooling time), and timescale for the turbulent cascade and re-acceleration, the radio shock and halo emission might blend forming these apparent halo-shock edges.

On the other hand, so far no polarized emission has been observed at these halo-shock edges \citep{2014MNRAS.440.2901S} which would indicate compression. Also, no clear strong downstream spectral gradients due to electron energy losses have been found so far \citep[e.g.,][]{2016ApJ...818..204V,2018ApJ...852...65R,2018arXiv181109713H}. If the synchrotron emission purely comes from a second order Fermi process at these edges, it would imply that there is sufficient post-shock MHD turbulence immediately after the shock \citep[see for example][]{2015ApJ...815..116F}. However, if this turbulence is generated by the shock
passage downstream there might be insufficient time for this turbulence to decay to the smaller scales  that  are  relevant for  particle  acceleration.  To fully understand the nature of halo-shock edges, future high-resolution spectral and polarimetric observations will be crucial. 

\subsection{Mini-halos}
\begin{figure*}[htbp]
\centering
\includegraphics[width=0.49\textwidth]{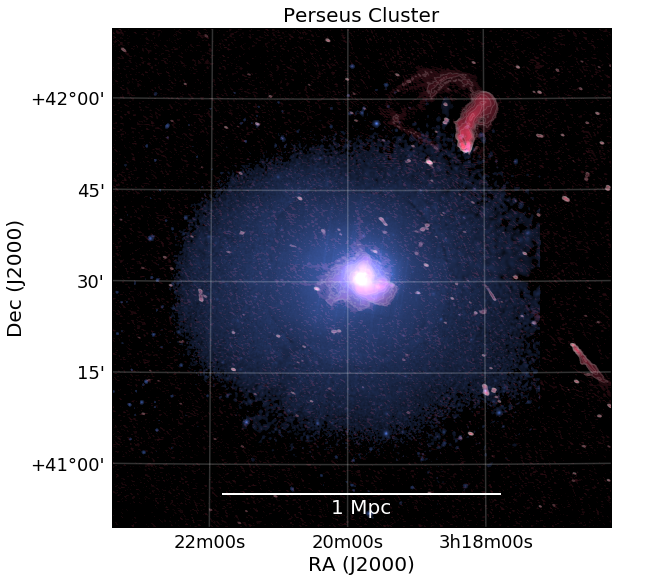}
\includegraphics[width=0.49\textwidth]{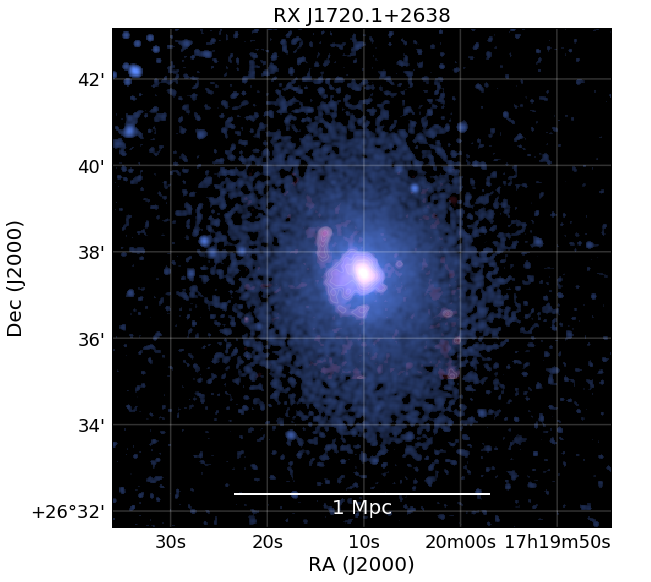}
\caption{Examples of clusters hosting radio mini-halos, see also Figure~\ref{fig:perseusMH}. The radio emission is shown in red and the X-ray emission in blue. Perseus cluster: VLA 230--470~MHz and XMM-Newton 0.4--1.3~keV \citep{2017MNRAS.469.3872G}. RX\,J1720.1+2638: GMRT 617~MHz and Chandra 0.5--2.0~keV \citep{2014ApJ...795...73G,2017ApJ...843...76A}.}
\label{fig:minihalos}
\end{figure*}

\begin{figure*}[htbp]
\centering
\includegraphics[width=0.49\textwidth]{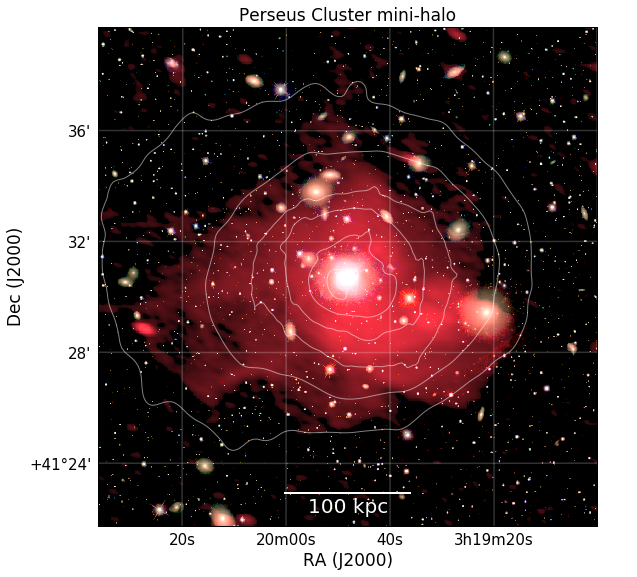}
\includegraphics[width=0.49\textwidth]{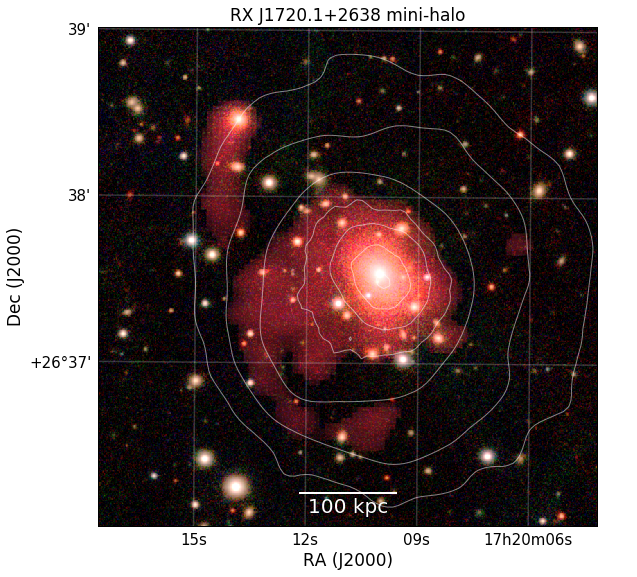}
\caption{Radio-optical overlays of the mini-halos in the Perseus cluster (\textit{left}) and RX\,J1720.1+2638 (\textit{right}). Both mini-halos display clear substructure. X-ray surface brightness contours are shown in white. The X-ray and radio data are the same as listed in Figure~\ref{fig:minihalos}.
The optical  images are taken from SDSS \citep[Perseus; \textit{gri} bands,][]{2018ApJS..235...42A} and Pan-STARRS \citep[RX\,J1720.1+2638; \textit{grz} bands,][]{2016arXiv161205560C}.}
\label{fig:perseusMH}   
\end{figure*}

Radio mini-halos have sizes of $\sim$100--500~kpc and are found in relaxed cool core clusters, with the radio emission surrounding the central radio loud BCG \citep[for a recent overview of mini-halos see][]{2015aska.confE..76G}. The sizes of mini-halos are comparable to that of the central cluster cooling regions. The prototypical mini-halo is the one found in the Perseus cluster \citep{1975AA....45..223M,1982Natur.299..597N,1990MNRAS.246..477P,1992ApJ...388L..49B,sijbring_phd,1998AA...331..901S}, see Figures~\ref{fig:minihalos} and \ref{fig:perseusMH}. Although smaller than radio halos, radio mini-halos also require in-situ acceleration given the short lifetime of synchrotron emitting electrons. The radio emission from mini-halos does therefore not directly originate from the central ANG, unlike the radio lobes that coincide with X-ray cavities in the ICM.  

Radio mini-halos have 1.4~GHz radio powers in the range of $10^{23}-10^{25}$~W~Hz$^{-1}$. The most luminous mini-halos known are located in the clusters PKS\,0745--191 \citep{1991MNRAS.250..737B} and RX\,J1347.5--1145 \citep{2007AA...470L..25G}, although the classification of the radio emission in PKS\,0745--191 as a mini-halo is uncertain \citep{2004AA...417....1G,2007AA...463..937V}.
The most distant mini-halo is found in the Phoenix Cluster \citep{2014ApJ...786L..17V}, although very recently a possible mini-halo in ACT-CL\,J0022.2--0036 at $z=0.8050$ has been reported by \cite{2018arXiv180609579K}.

Compared to giant radio halos, the synchrotron volume emissivities of mini-halos are generally higher \citep{2008A&A...486L..31C,2009AA...499..679M}. \cite{2009AA...499..679M} fitted exponential azimuthal surface brightness profiles (see Equation~\ref{eq:expradialprofile}) and showed that mini-halos have  smaller e-folding radii ($r_e$) compared to giant halos, as expected from their smaller sizes with the emission being mostly confined to the X-ray cooling region.

Since the mini-halo emission surround the central radio galaxy, whose lobes often have excavated cavities in the X-ray emitting gas, the separation between AGN lobes and mini-halos can be difficult, in particular in the absence of high-resolution images. Radio emission that directly surrounds the central AGN (less than a few dozens of kpc), does not necessarily require in-situ re-acceleration. This emission has also been classified as `core-halo' sources. The separation between core-halo sources, amorphous lobe-like structures, and mini-halos is often not clear \citep{1991MNRAS.250..737B,2008ApJ...675L...9M}.
In addition, the central radio galaxies are sometimes very bright, requiring high-dynamic range imaging to bring out the low-surface brightness mini-halos. The classification as a mini-halo is also difficult without X-ray data \citep[e.g.,][]{2009MNRAS.399..601B}. Because of these observational limitations, there is currently a rather strong observational selection bias. For that reason many fainter radio mini-halos could be missing since they fall below the detection limit of current telescopes. Despite these observational difficulties the number of known mini-halo has steadily been increasing \citep{2006AA...448..853G,2012ApJ...753...47D,2011AA...525L..10G,2014ApJ...781....9G,2017ApJ...841...71G}. In Table~\ref{tab:clusterlist} we list the currently known radio mini-halos and candidates. 

An example of a source that is difficult to classify is the one found in the central parts of the cluster Abell\,2626. This source was initially named as a mini-halo by \cite{2004AA...417....1G}. More detailed studies \citep{2013MNRAS.436L..84G,2017AA...604A..21I,2017MNRAS.466L..19K} reveal a complex ``kite-like'' radio structure, complicating the interpretation and classification. The cluster RX\,J1347.5--1145 presents another interesting case. It was found to host a luminous radio mini-halo \citep{2007AA...470L..25G} with an elongation to the south-east. This elongation seems to correspond to a region of shock heated gas induced by a merger event, also detected in the SZ \citep{2001PASJ...53...57K,2004PASJ...56...17K,2010ApJ...716..739M,2011ApJ...734...10K,2012ApJ...751...95J}. This suggests that the south-east emission is not directly related to the central mini-halo, but rather is a separate source \citep{2011AA...534L..12F} which could be classified as a cluster radio shock.

Few detailed high-quality resolved images of mini-halos exist. This makes it hard to study the morphology of mini-halos in detail. Interestingly, \cite{2008ApJ...675L...9M} found that mini-halos are often confined by the cold fronts of cool core clusters (but see Section~\ref{sec:halounification}). The most detailed morphological information is available for the Perseus cluster mini-halo.  \cite{2017MNRAS.469.3872G} presented 230--470~MHz images which revealed  filamentary structures in this mini-halo, extending in various directions (Figure~\ref{fig:perseusMH}). Hints of these structures are already visible at 1.4~GHz \citep{1989ESOC...32..107S}. 
These structures could be related to variations in the ICM magnetic field strength, localized sites of particle re-acceleration, or a non-uniform distribution of fossil electrons. The Perseus cluster mini-halo emission also follows some of the structures observed in X-ray images. Most of the mini-halo emission is contained within a cold front. However, some faint emission extends (``leaks'') beyond the cold front.
Similarly, the  RX\,J1720.1+2638 mini-halo also displays  substructure suggesting that when observed at high resolution and signal-to-noise mini-halos are not fully diffuse.  
 
Spectral indices of radio mini-halos are similar to giant radio halos, although few detailed studies exist. The integrated spectrum for the Perseus mini-halo is consistent with a power-law shape \citep{sijbring_phd}. A hint of spectral steepening above 1.4~GHz is found for RX\,J1532.9+3021 \citep{2013ApJ...777..163H,2014ApJ...781....9G}. An indication of radial spectral steepening for the Ophiuchus cluster \citep{2009AA...499..371G,2009MNRAS.396.2237P} was reported by \cite{2010AA...514A..76M}. The most detailed spectral study so far has been carried out on RX\,J1720.1+2638 \citep{2008ApJ...675L...9M,2014ApJ...795...73G}. This mini-halo shows a spiral-shaped tail, with spectral steepening along the tail. Possible steepening of the integrated spectrum for RX\,J1720.1+2638 at high frequencies has also been reported \citep{2014ApJ...795...73G}. So far no targeted polarization studies of mini-halos have been performed.

\subsubsection{Statistics}
\cite{2014ApJ...781....9G} found no clear correlation between the mini-halo radio power and cluster mass, unlike giant radio halos. However, \cite{2008A&A...486L..31C,2013AA...557A..99K,2015aska.confE..76G} did report evidence for a correlation between radio power and X-ray luminosity. The slope of the correlation was found to be similar to that of giant radio halos \citep{2015aska.confE..76G}. Larger samples are required to obtain better statistics and confirm the found correlations, or lack thereof.

\cite{2017ApJ...841...71G} determined the occurrence of radio mini halos in a sample of 58 clusters with M$_{500} > 6 \times 10^{14}$~M$_\odot$. They found that 80\% of the cool core clusters hosted mini-halos. Therefore, mini-halos are common phenomenon in such systems. No mini-halos were found in non-cool core systems. In addition, tentative evidence was found for a drop in the occurrence rate for lower cluster masses. \cite{2013AA...557A..99K} found a mini-halo occurrence rate of about 50\% in the Extended GMRT Radio Halo Survey ($L_{\rm{X, 0.1-2.4 keV}} > 5 \times 10^{44}$~erg~s$^{-1}$, $0.2<z<0.4$), also indicating mini-halos are rather common.

\subsubsection{Origin of radio mini-halos}
Similar to giant radio halos, hadronic  \citep[e.g.,][]{2004A&A...413...17P} or turbulent re-acceleration models \citep{2002A&A...386..456G} been invoked to explain the presence of the CR synchrotron emitting electrons. 
Unlike giant radio halos, where the turbulence is induced by major cluster mergers, mini-halos would trace turbulence in the cluster cores generated by gas sloshing \citep{2013ApJ...762...78Z,2015ApJ...801..146Z}. The central AGN is a likely candidate for the source of the fossil electrons that are re-accelerated  \citep[e.g.,][]{2007ApJ...663L..61F}. The confinement of mini-halos by cold fronts \citep{2008ApJ...675L...9M} support a scenario where turbulence induced by gas sloshing motions re-accelerates  particles. Simulations by \cite{2013MNRAS.428..599F,2013ApJ...762...78Z,2015ApJ...801..146Z} provided further support for this scenario, reproducing some of the observed morphology, where the emission is bounded by cold fronts.

The radio spectral properties of mini-halos provide another discriminator for the origin of the CR electrons. If the electrons are re-accelerated by magnetohydrodynamical turbulence, the integrated spectra of mini-halos should display a spectral break caused by a cutoff in the electron energy distribution. Due to the limited number of spectral studies available, no clear conclusion can be drawn on the general occurrence of spectral breaks in mini-halo spectra.

\subsubsection{Unification}
\label{sec:halounification}
Despite of the their differences, it is possible that mini-halos and giant halos in clusters are physically related to each other.
For example, cluster merger events could transport CR from cluster cores to larger-scales where they are re-accelerated again \citep[see][]{2014IJMPD..2330007B}. This could lead to ``intermediate'' cases where mini-halos could evolve into giant radio halos and vice-versa. This could either be a transition been turbulent re-acceleration due to core sloshing and merger induced turbulent re-acceleration. Or alternatively, a transition between hadronic mini-halos and merger induced turbulent re-acceleration \citep{2014MNRAS.438..124Z}. Recent observations have provided evidence for such scenarios, finding (mini-)halos with unusual properties.

\cite{2014MNRAS.444L..44B} discovered a large 1.1~Mpc radio halo in CL1821+643 which contains a strong cool core. If this halo is caused by a merger event, the cluster is in a stage where the merger has not (yet) been able to disrupt the cool core as also noted by \cite{2016MNRAS.459.2940K}. For example, because the merger is an off-axis event, or the merger is still in an early stage.
CL1821+643 could therefore be a transitional object, where a mini-halo is switching off and a giant radio halo is just being formed. Similarly, \cite{2017MNRAS.466..996S,2018arXiv181108410S} reported the presence of a $\sim$1~Mpc radio halo in the semi-relaxed cluster Abell\,2261\footnote{The classification of Abell\,2390 as giant radio halo by \cite{2017MNRAS.466..996S} was not confirmed by \cite{2018arXiv181108410S} which suggested the emission belongs to a double lobe radio galaxy.}, questioning the assumption that giant radio halos only occur in clusters undergoing major mergers.

Another peculiar case is the sloshing, minor-merger cluster Abell\,2142. Early work already hinted at the presence of diffuse emission \cite{1977A&A....60...27H,1999NewA....4..141G,2000NewA....5..335G} in this cluster. This was confirmed by \cite{2013ApJ...779..189F} which showed a 2~Mpc radio halo. \cite{2017AA...603A.125V} found that the radio halo consists of two components. The inner component has a higher surface brightness, with properties similar to that of a mini-halo. The outer larger component has a steeper spectrum. They proposed that the inner component is powered by central sloshing turbulence. The outer component  might probe turbulent re-acceleration induced by a less energetic merger event. Alternatively, the different components are the result from a transition between hadronic and turbulent re-acceleration processes. 

The cluster PSZ1\,G139.61+24.20 ($z = 0.267$) was listed as a candidate mini-halo by \cite{2017ApJ...841...71G}. \cite{2018MNRAS.478.2234S} presented the discovery of steep-spectrum emission extending beyond the cool core region of the cluster with LOFAR. They argued that the emission outside the core is produced by turbulent re-acceleration from a minor merger event that has not disrupted the cool core. If this scenario is correct, it indicates that both a giant radio halo and mini halo could co-exist. A very similar situation has recently been found in the cluster RX\,J1720.1+2638. Here, \cite{2018arXiv181108410S} discovered extended faint diffuse steep spectrum emission beyond the cold front and mini-halo region \citep{2008ApJ...675L...9M,2014ApJ...795...73G}.

\subsubsection{Gamma-ray upper limits}
\label{sec:minihalogamma}
The most important gamma-ray limits on mini-halos come from the observations of the Major Atmospheric Gamma Imaging Cherenkov (MAGIC) telescopes of Perseus \citep{2010ApJ...710..634A,2012A&A...541A..99A,2016A&A...589A..33A}, and from the combined likelihood analysis of HIFLUGCS clusters with the \emph{Fermi}-LAT satellite data \citep{2014ApJ...787...18A}. As is the case of Coma for merging clusters, Perseus is the most constraining object when it comes to mini-halos because of its high mass, closeness, and mini-halo brightness. Perseus host two gamma-ray bright AGNs - the central radio galaxy NGC~1275 and IC~310 - detected both by \emph{Fermi} \citep{2009ApJ...699...31A} and by MAGIC \citep{2012A&A...541A..99A,2016A&A...589A..33A}. The poor angular resolution of \emph{Fermi} at low ($<$10~GeV) energies makes it difficult to target the possible diffuse gamma-ray emission in Perseus, and  makes the MAGIC Perseus observations the most constraining for relaxed cool core clusters hosting mini-halos.

Differently from the case of the Coma radio halo, the gamma-ray upper limits on Perseus do not yet allow to exclude the hadronic origin of its mini-halo. The CR energy density in Perseus is constrained to be below about 1--10\% of the thermal energy density, with the exact number depending on the assumptions made regarding the CR-spectral and spatial distribution, e.g., the steeper the spectrum and/or the flatter spatial (radial) distribution, the looser the constraints become. This strong dependence of the constraints on the CR content in clusters on the proton spectral and spatial distributions should be kept in mind when quoting these limits.

Assuming the mini-halo emission is hadronic, the gamma-ray upper limit can be turned in a lower limit on the magnetic field needed to generate the radio emission with secondary electrons. This is similar to what has been done for the Coma radio halo where the magnetic field needed for the hadronic interpretation would be extremely high with an energy density of 1/3 or more of the thermal energy density \citep{2017MNRAS.472.1506B}. In the case of Perseus, current gamma-ray limits imply central magnetic fields above  $\sim5$~$\mu$Gauss, still well below the $\sim25$~$\mu$Gauss inferred from Faraday rotation measurements \citep{2006MNRAS.368.1500T}.

\subsection{Future Gamma-ray studies}

{Future gamma-ray observations of clusters of galaxies will be fundamental for this field as only thanks to gamma-rays the exact amount of CR protons can be directly studied and the degeneracy between secondary and primary sources of electrons in radio-halo models can be addressed. In particular, future observations of the Perseus cluster -- as envisioned in the key science projects of the in-construction Cherenkov Telescope Array \citep[CTA;][]{2017arXiv170907997C} -- will allow to eventually test the hadronic interpretation of mini-halos, and, more importantly, to limit the CR energy density to below about 2\% independently from the assumptions on the CR-proton spectral and spatial distribution. Such low limits will also allow to test the role of AGNs, particularly, the protons confinement in AGN bubbles and how protons are transported from the central AGNs to cluster peripheries. Paramount for an ``order-of-magnitude'' jump in constraining power, also for gamma-ray observations of cluster radio shocks, and hopefully to aim for several detections, will be the satellites proposed as successors of \emph{Fermi} \citep{2016CRPhy..17..663K,2016APS..APRH11004M}. Note, however, that if nature is ``kind'' and the electrons generating the radio halo of Coma are re-accelerated secondaries, continued \emph{Fermi} observations could reach a detection in the near future \citep{2017MNRAS.472.1506B}.}

\subsection{Upcoming large cluster samples}
{With new deep low-frequency radio surveys covering a significant fraction of the sky \citep[such as LoTSS,][]{2017A&A...598A.104S,2018arXiv181107926S} many new radio (mini-)halos are expected to be discovered. In particular those with steep radio spectra. New surveys are also planned at GHz frequencies \citep{2011PASA...28..215N,2013PASA...30...20N} which should also uncover additional diffuse cluster radio sources \citep{2012A&A...548A.100C}.}

{With the improved statistics offered by larger samples, the properties and occurrence rates as a function of cluster mass, dynamical state, and other global cluster properties can be investigated in detail. These samples should also contain a  population of ultra-steep spectrum radio halos that are predicted in the framework of the turbulent re-acceleration model \citep{2010A&A...509A..68C}. Furthermore, large samples might  shed more light on (i) the  possible connections between halos and mini-halos and (ii) the evolution of diffuse cluster radio sources over cosmic time, from $z\sim$1 to the present epoch. For example, changes in the occurrence rate are expected due to the increase of Inverse Compton losses with redshift,  change of the cluster merger rate, evolution of cluster magnetic fields.}

\newpage

\section{Cluster radio shocks (relics) and revived fossil plasma sources}
\label{sec:relics}

Apart from radio halos, we broadly divide diffuse cluster sources into cluster radio shocks and revived fossil plasma sources. The distinction between radio shocks and fossil plasma sources is not always straightforward, since it requires the detection of shocks via SZ or X-ray measurements and the availability of radio spectra. Our adopted classification is similar to that of \cite{2004rcfg.proc..335K}\footnote{We do not consider dying radio galaxies here that have not interacted with the ICM, see \cite{2011A&A...526A.148M}.} who defined radio {\sl Gischt} and {\sl Phoenix}. Given that there is now compelling evidence that Gischt trace shock waves \citep[e.g.,][]{2010ApJ...715.1143F}, we propose to simply call these {\sl cluster radio shocks}. This still leaves open the questions of the underlying (re-)acceleration mechanism that produces the synchrotron emitting CR at these shocks.

Radio shocks and fossil sources are detected in clusters covering a wide range in mass, unlike radio halos which are almost exclusively found in massive systems. Some examples of radio shocks and revived fossil sources in lower mass clusters are discussed in \cite{2003AJ....125.1095S,2017MNRAS.472..940K,2017AA...597A..15D,2018MNRAS.477.3461B,2018MNRAS.477..957D}. Phoenices and other revived AGN fossil sources (such as GreEt) are characterized by their steep radio spectra and presence of high frequency spectral breaks. These sources will be discussed in Section~\ref{sec:fossil}.

Similar to giant radio halos and mini-halos, there are ``hybrid'' or ``intermediate'' sources which share some properties between these two categories. For example AGN fossil plasma that is re-accelerated at a large cluster merger shock \citep[e.g., in Abell\,3411-3412;][]{2017NatAs...1E...5V}.
\subsection{Cluster radio shocks (relics)}
\begin{figure*}[htbp]
\centering
\includegraphics[width=0.49\textwidth]{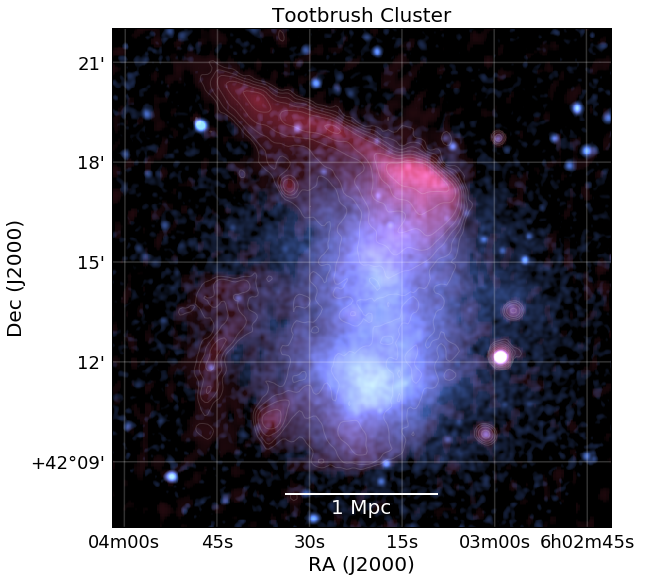}
\includegraphics[width=0.49\textwidth]{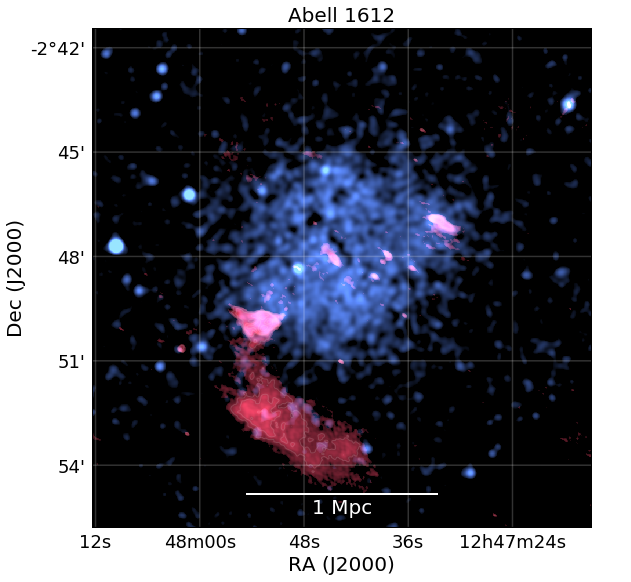}
\includegraphics[width=0.49\textwidth]{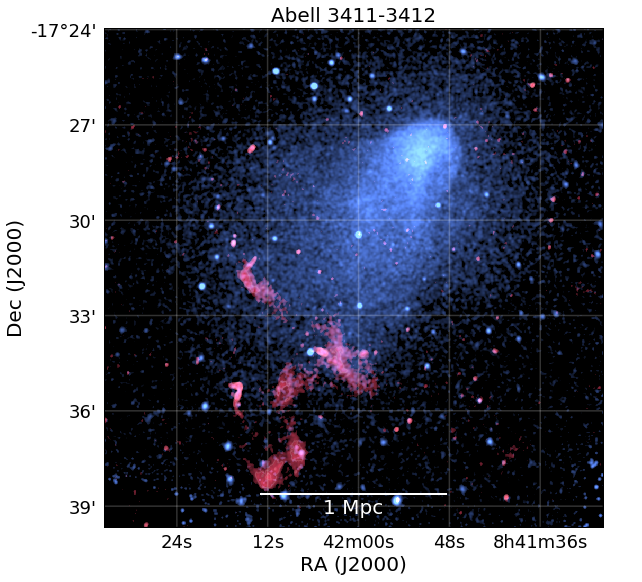}
\includegraphics[width=0.49\textwidth]{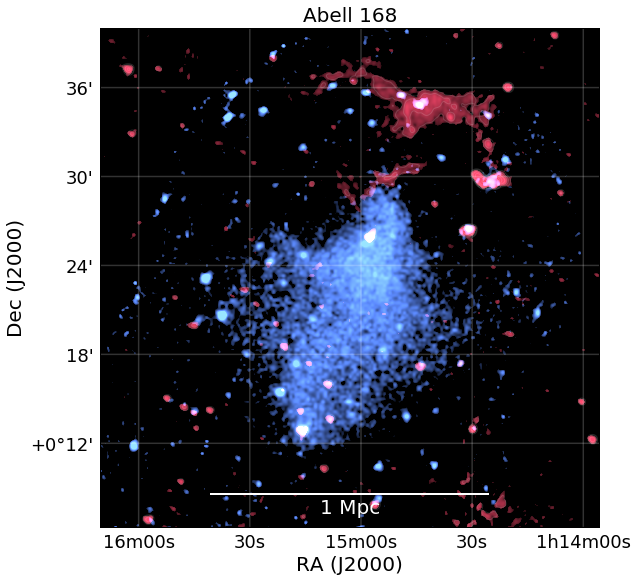}
\caption{Examples of cluster radio shocks. The radio emission is shown in red and the X-ray emission in blue. Toothbrush Cluster: LOFAR 120--180~MHz and Chandra 0.5--2.0~keV \citep{2016ApJ...818..204V}.  Abell\,1612: GMRT 610~MHz and Chandra 0.5--2.0~keV \citep{2011AA...533A..35V,2017AA...600A..18K}. Abell\,3411-3412: GMRT 610~MHz and Chandra 0.5--2.0~keV \citep{2017NatAs...1E...5V}. Abell\,168: GMRT~323~MHz and XMM-Newton 0.4--2.3~keV \citep{2018MNRAS.477..957D}. Additional examples of cluster radio shocks can be found in Figure~\ref{fig:haloexamples} (Abell\,2744, the Bullet cluster, Abell\,2256, the Coma cluster, and MACS\,J0717.5+3745)}.
\label{fig:radioshocks}
\end{figure*}

Cluster radio shocks are mostly found in the outskirts of galaxy clusters, see Figure~\ref{fig:radioshocks}. Unlike radio halos, they have elongated shapes. In addition, radio shocks are strongly polarized at frequencies $\gtrsim 1$~GHz, with polarization fractions of $\gtrsim 20\%$ \citep{1998A&A...332..395E}, see Figures~\ref{fig:cizaspix} and ~\ref{fig:a2744pol}. 

The first identified cluster radio shock was the source 1253+275 in the Coma cluster \citep{1979ApJ...233..453J,1981AA...100..323B}. This radio source has been studied in considerable detail early on by \cite{1985AA...150..302G,1991AA...252..528G}. Recently, evidence for a shock at this location has also been obtained \citep{2013MNRAS.433.1701O,2013PASJ...65...89A}, see also Section~\ref{sec:radioxrayshock}. A couple of other cluster radio shocks that were studied after the discovery of 1253+275 were the ones found in Abell\,2256  \citep[e.g.,][]{1976AA....52..107B,1994ApJ...436..654R} and Abell\,3667 \citep[e.g.,][]{1997MNRAS.290..577R}. The number of detected radio shocks increased significantly with the availability of the NVSS and WENSS surveys \citep{1999NewA....4..141G,2001ApJ...548..639K}. A list of cluster radio shocks in given in Table~\ref{tab:clusterlist}. 
The most powerful cluster radio shock is found in MACSJ0717.5+3745 \citep{2009AA...503..707B,2009A&A...505..991V}. Interestingly, this cluster also hosts the most powerful radio halo. The most distant radio shocks are located in ``El Gordo'' at $z=0.87$ \citep{2012ApJ...748....7M,2014ApJ...786...49L,2016MNRAS.463.1534B}.

\begin{figure}[htbp]
\centering
\includegraphics[width=0.45\textwidth]{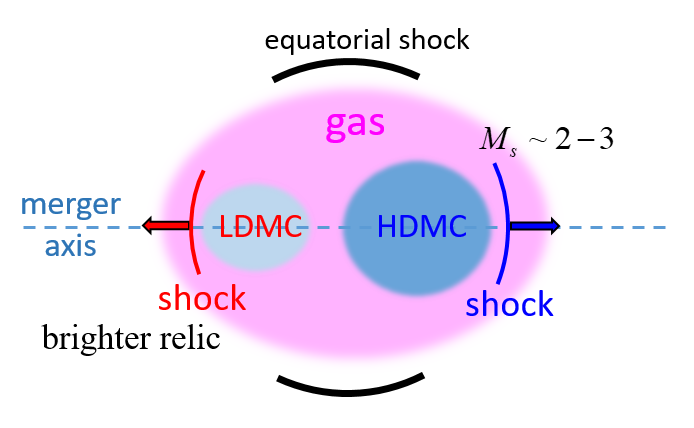}
\caption{Schematic picture of an idealized binary cluster merger about 1~Gyr after core passage. Equatorial shocks expand outwards in the equatorial plane perpendicular to the merger axis, while merger shocks lunch in the opposite directions along the merger axis. The shock-kinetic-energy-weighted Mach number range is $\left< \mathcal{M}_{s} \right>_{\phi} \simeq 2 - 3$. Typically, the shock ahead of lighter DM core has the higher shock kinetic energy flux and becomes the brighter radio shock.}
\label{fig:binarymerger}   
\end{figure}

In an idealized binary merger, `equatorial' shocks form first and move outwards in the equatorial plane, see Figure~\ref{fig:binarymerger}). 
After the dark matter core passage, two `merger' shocks lunch into the opposite directions along the merger axis, which can explain the formation of cluster double radio shocks in observed merging clusters \citep[e.g.,][]{2011MNRAS.418..230V,2011AA...528A..38V,2017ApJ...841...46M}, see also Section~\ref{sec:doublerelics}. \cite{2012MNRAS.421.1868V}  investigated why cluster radio shocks are mostly found in the periphery of clusters using simulations. They showed that the radial distribution of observed radio shocks can be explained by the radial trend of dissipated kinetic energy in shocks, which increases with cluster centric distance up until half of the virial radius. Analyzing the properties of shocks associated with synthetic merging clusters in structure formation simulations, \cite{2018ApJ...857...26H}  found that the CR production peaks at $\sim$1~Gyr after the core passage, with the shock-kinetic-energy-weighted Mach number $\left< \mathcal{M}_s \right>_{\phi} \simeq 2 - 3$ and the CR-flux-weighted Mach number $\left< \mathcal{M}_s \right>_{\rm CR} \simeq 3 -4$. Simulations by \cite{2011ApJ...735...96S,2012MNRAS.421.1868V,2012MNRAS.426...40B,2013ApJ...765...21S,2016MNRAS.459...70V,2017MNRAS.470..240N,2017MNRAS.464.4448W} also produce large-scale radio shock morphologies that provide a reasonable match to what is found in observations.

Some examples of studies showing the connection between radio shocks and cluster mergers using optical spectroscopy and imaging are \cite{2007A&A...469..861B,2010A&A...521A..78B,2009A&A...503..357B,2013MNRAS.434..772B,2015ApJ...805..143D,2017ApJ...838..110G,2016ApJ...831..110G,2017ApJ...841....7B}. This connection is also corroborated by weak lensing studies that reveal
multiple mass peaks in some radio shock hosting clusters \citep[e.g.,][]{2015PASJ...67..114O,2016ApJ...817..179J,2015ApJ...802...46J}. The most comprehensive analysis of a sample of 29~radio shock hosting clusters was performed by \cite{2017arXiv171101347G,2018arXiv180610619G}. They found that the merger axes of radio shock hosting clusters are generally in or near the plane of the sky. This indicates that there are selection biases for finding cluster radio shocks based on the viewing angle. Due to this selection effect, many radio shocks with less favorable orientations are probably missing in current samples.

Cluster radio shocks seems to be less common than radio halos or mini-halos, the occurrence of radio shocks was found to be about about $5\% \pm 3\%$ by \cite{2015AA...579A..92K}. However, unlike radio halos or mini-halos, the merger axis orientation probably plays an important role in detecting these sources, as mentioned.

Some giant cluster radio shocks such as the Sausage and the Toothbrush are thought to be associated with major mergers with a subclump mass ratio $\lesssim 3$ \citep{2015PASJ...67..114O,2015ApJ...802...46J,2016ApJ...817..179J}, while the cluster ZwCl\,0008.8+5215 with a double radio shock and PLCK\,G287.0+32.9 with multiple radio shocks are merging systems with a mass ratio $\gtrsim 5$ \citep{2017ApJ...838..110G,2017ApJ...851...46F}.

In a few clusters the  emission from the cluster radio shocks is attached or overlaps with that of the radio halo \citep[e.g.,][]{2009ApJ...699.1288D,2016ApJ...818..204V}. The nature of these ``bridges'' between halos and cluster radio shocks is still unclear. In some cases, the radio halo emission covers the entire region between double radio shocks \citep{2012MNRAS.426...40B,2017MNRAS.471.1107H,2018ApJ...865...24D}. One possibility is that we observe a transition from first order Fermi (re-)acceleration to second order re-acceleration by turbulence that develops in the post shock region.

\subsubsection{Morphology and sizes}

\begin{figure}[htbp]
\centering
\includegraphics[width=0.49\textwidth]{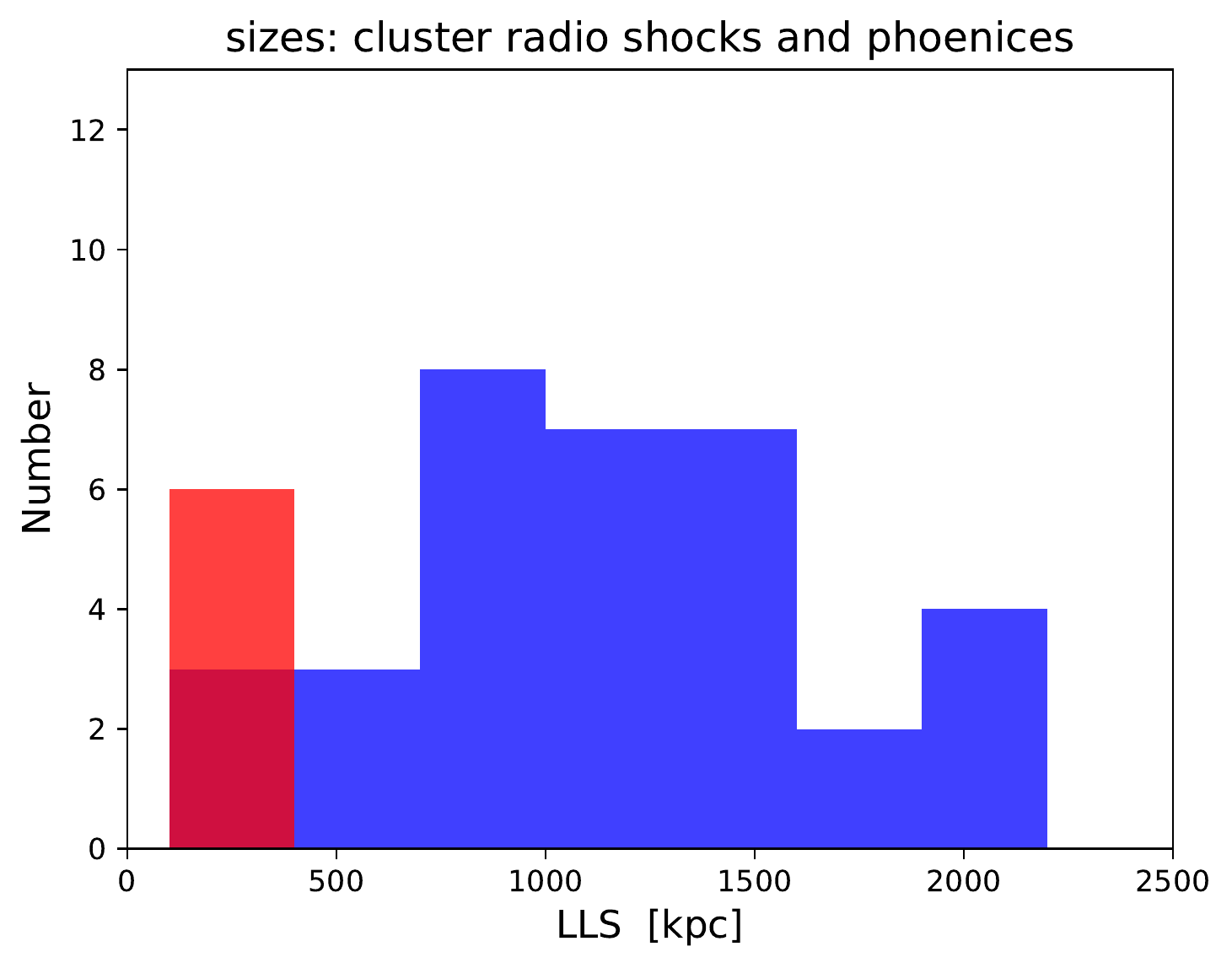}
\caption{Histogram showing the largest linear sizes (LLS) of cluster radio shocks and phoenices. Radio phoenices are shown in red and cluster radio shocks are shown in blue. Sources  and largest angular sizes (LAS) were taken from \url{http://galaxyclusters.com}.}
\label{fig:relicsLLSpower}
\end{figure}

\begin{figure*}[htbp]
\centering
\includegraphics[width=1.0\textwidth]{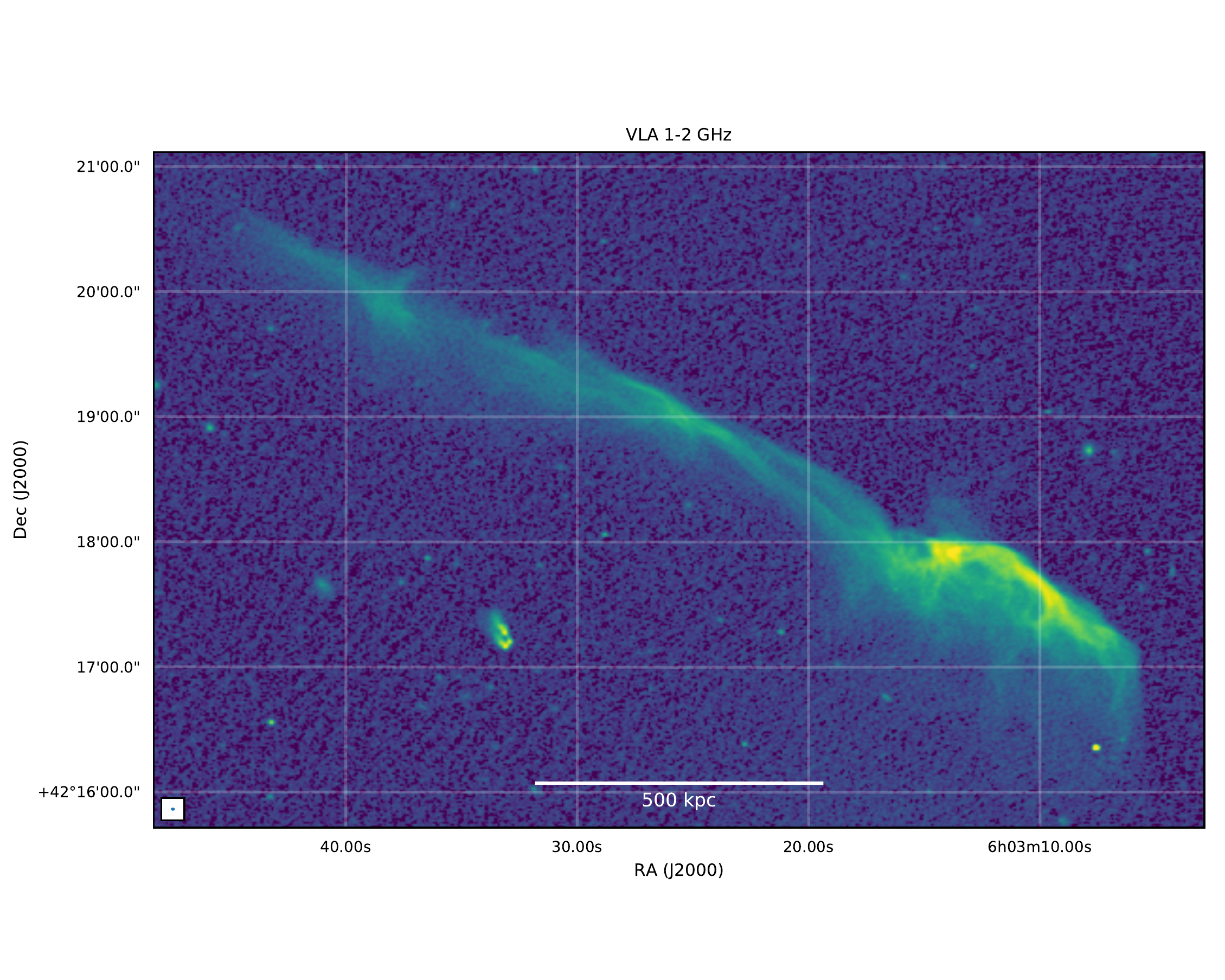}\vspace{-1.4cm}
\includegraphics[width=1.0\textwidth]{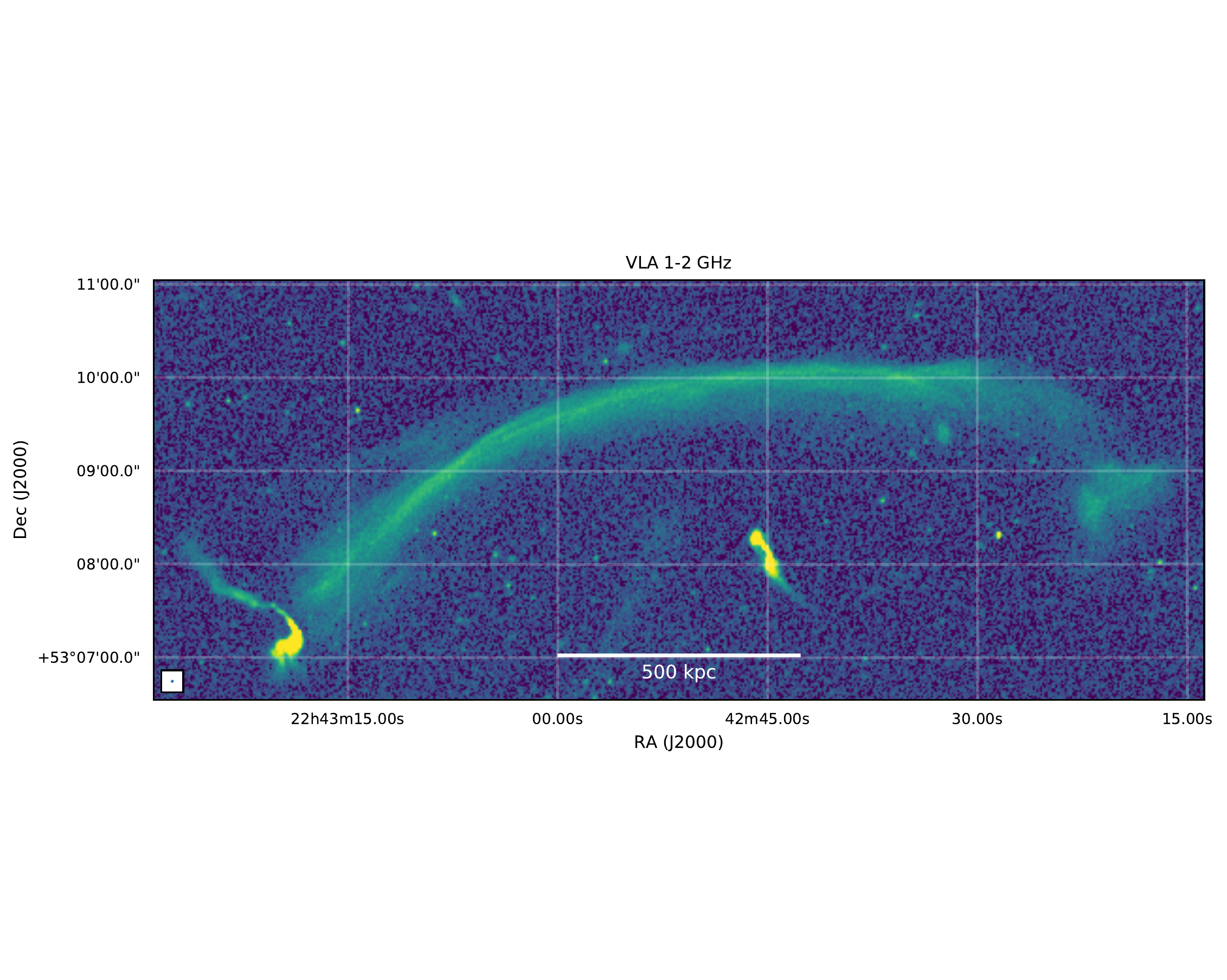}
\vspace{-3.5cm}
\caption{VLA 1--2~GHz high-resolution ($\sim$2$^{\prime\prime}$) images of the Toothbrush \citep[\textit{top panel};][]{2018ApJ...852...65R} and Sausage Cluster \citep[\textit{bottom panel};][]{2018ApJ...865...24D} radio shocks. Both images show the radio shocks consist of multiple filamentary substructures.}
\label{fig:TB}
\end{figure*}

Cluster radio shocks typically have elongated shapes, examples are the sources found in the Coma cluster \citep{1991AA...252..528G}, CIZA\,J2242.8+5301 \citep{2010Sci...330..347V}, Abell\,3667 \citep{1997MNRAS.290..577R,2003PhDT.........3J}, Abell\,115 \citep{2001AA...376..803G}, and Abell\,168 \citep{2018MNRAS.477..957D}. These elongated shapes are expected for sources that trace shock waves in the cluster outskirts and are seen close to edge-on. Examples of radio shocks that are less elongated are found in Abell\,2256 \citep[e.g.,][]{2006AJ....131.2900C} and ZwCl\,2341.1+0000 \citep{2002NewA....7..249B,2009AA...506.1083V}. Cluster radio shocks have sizes that roughly range between 0.5 to 2~Mpc, see Figure~\ref{fig:relicsLLSpower}.
Most large radio shocks that are found in the cluster outskirts show asymmetric transverse brightness profiles, with a sharp edge on the side away from the cluster center. On the side of the cluster center, the emission fades more gradually, see Figures~\ref{fig:cizaspix} and ~\ref{fig:TB}.

\begin{figure*}[htbp]
\centering
\includegraphics[width=1.0\textwidth]{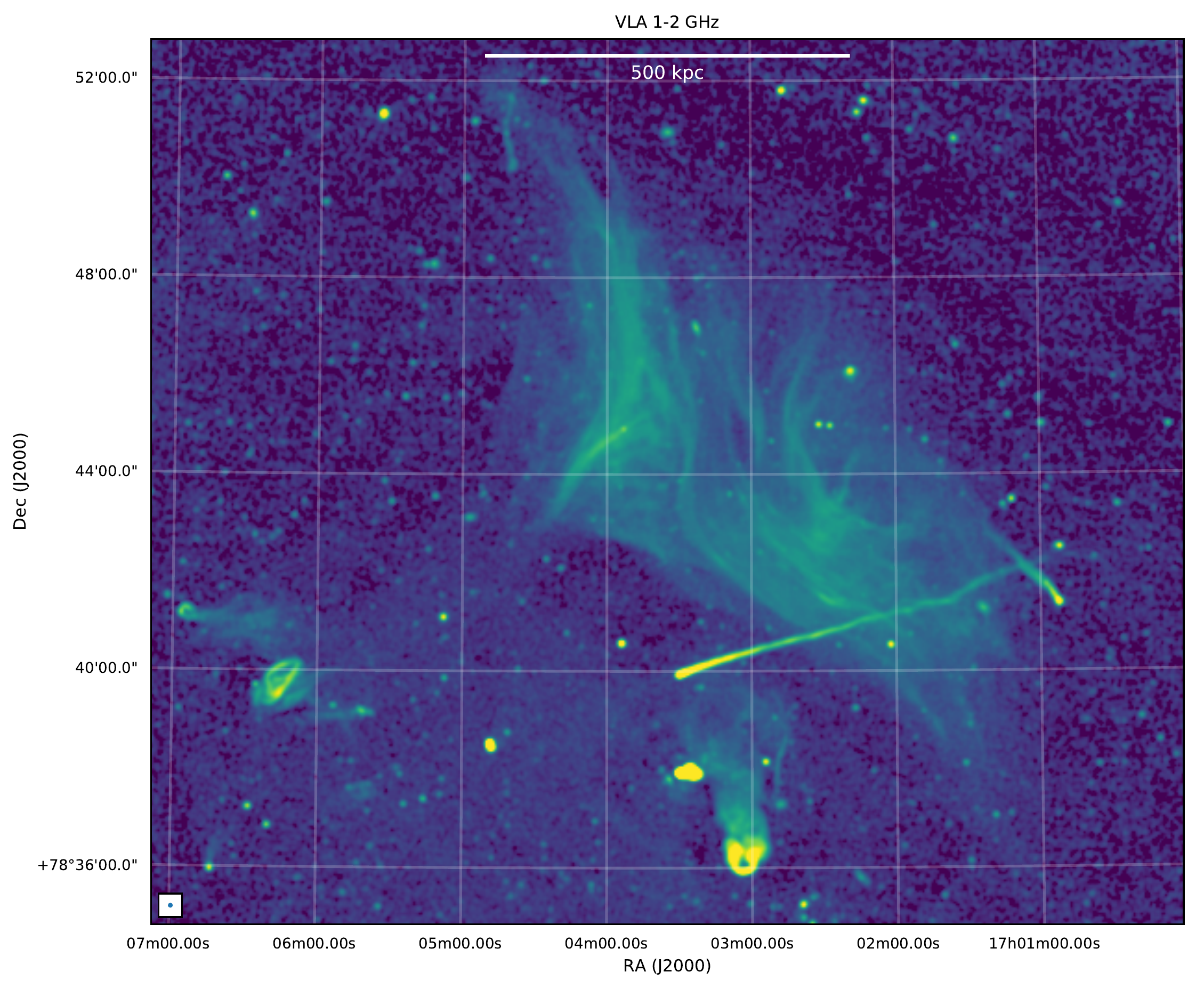}
\caption{VLA 1--2~GHz image with a resolution of $6{\prime\prime}$ of the radio shock region in Abell\,2256  \citep{2014ApJ...794...24O}. This image shows that the radio shock consists of a number filamentary structures. Several prominent tailed radio galaxies are also visible.}
\label{fig:A2256}
\end{figure*}

Deep high-resolution observations of large elongated radio shocks have also revealed a significant amount of filamentary substructures, 
see Figures~\ref{fig:A2256} and \ref{fig:TB}. Large radio shocks that display these filamentary structures are found in Abell\,2256, CIZA\,J2242.8+5301, Toothbrush, MACS\,J0717.5+3745, Abell\,3376, and Abell\,3667. The nature of the filamentary structures is not fully understood.  One possibility is that they trace changes in the magnetic field. Alternatively, they reflect the complex shape of the shock surfaces. The filamentary morphology of cluster radio shocks seems to be ubiquitous because all radio shocks that have been studied with good signal to noise and at high resolution display them.

\subsubsection{Cluster double radio shocks}
\label{sec:doublerelics}

\begin{figure*}[htbp]
\centering
\includegraphics[width=0.49\textwidth]{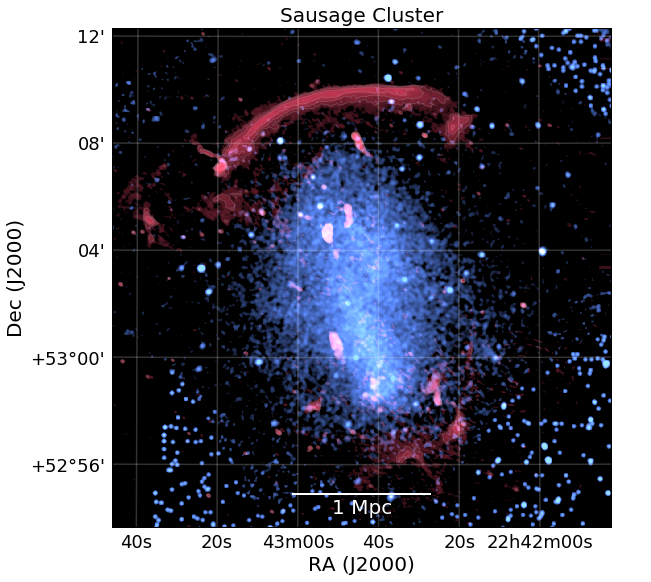}
\includegraphics[width=0.49\textwidth]{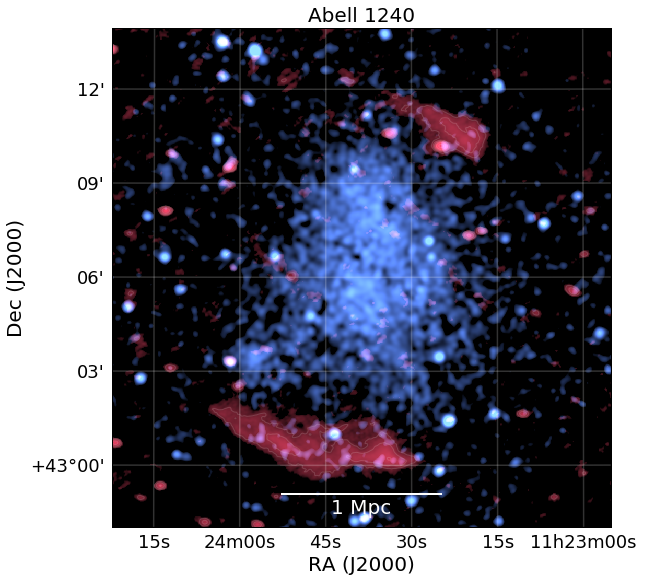}
\includegraphics[width=0.49\textwidth]{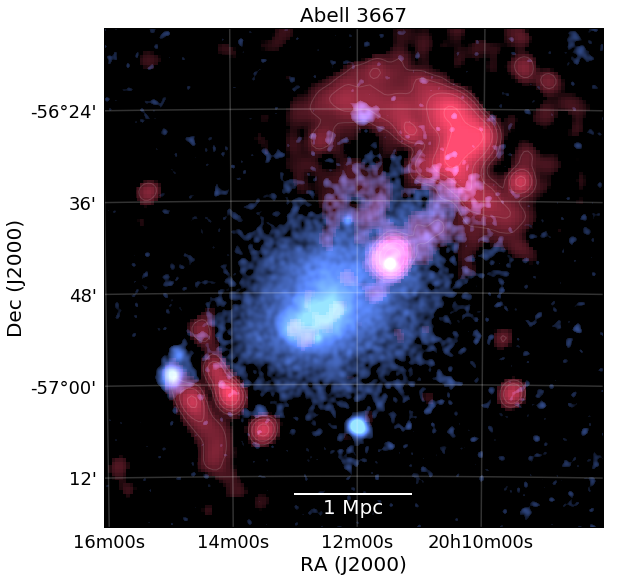}
\includegraphics[width=0.49\textwidth]{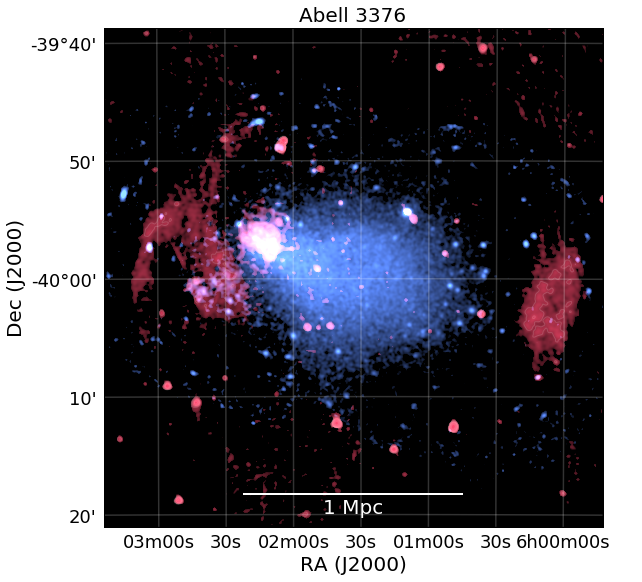}
\caption{Examples of cluster double radio shocks. The radio emission is shown in red and the X-ray emission in blue. Sausage Cluster: GMRT 610~MHz and Chandra 0.5--2.0~keV \citep{2010Sci...330..347V,2014MNRAS.440.3416O}.  Abell\,1240: LOFAR 120--168~MHz and Chandra 0.5--2.0~keV \citep{2018MNRAS.478.2218H}. Abell\,3667: MWA 170--231~MHz and ROSAT PSPC \citep{2017MNRAS.464.1146H,1999A&A...349..389V}. Abell\,3376: GMRT 317~MHz and XMM-Newton 0.3--2.0~keV  \citep{2012MNRAS.426.1204K,2018AA...618A..74U}.}
\label{fig:doubleradioshocks}
\end{figure*}

A particular interesting class of cluster radio shocks are so-called ``double shocks''. Here two large elongated convex radio shocks are found diametrically with respect to the cluster center, see Figure~\ref{fig:binarymerger}. The radio shocks are oriented perpendicular with the respect to the elongated ICM distribution (and merger axis) of the cluster, see Figures~\ref{fig:doubleradioshocks} and~\ref{fig:spixpsz}. Double radio shocks are an important subclass of radio shocks as the cluster merger scenario can be relatively well constrained. In addition, these system seems to be observed close to edge-on. Note that we reserve the classification of a double radio shock for a pair of shock waves that were generated at the same time during core passage. So the presence of two radio shocks in a cluster alone is not a sufficient condition to classify it as a double radio shocks.

About a dozen well-defined double radio shock systems are known, see Table~\ref{tab:clusterlist}. The first cluster double radio shock was found in Abell\,3667 \citep{1997MNRAS.290..577R}. It was realized by \cite{1999ApJ...518..603R,1999ptep.proc..292J} that these radio sources could have resulted from particles accelerated at shocks from a binary merger event. The presence of a shock in the ICM at the location of the northwestern radio source in Abell\,3667 was confirmed via X-ray observations by \cite{2010ApJ...715.1143F}. 
The second double radio shock system was discovered by \cite{2006Sci...314..791B} in Abell\,3376. Other well studied  cluster double radio shocks are the ones in CIZA\,J2242.8+5301 \citep{2010Sci...330..347V}, ZwCl\,0000.8+5215 \citep{2011AA...528A..38V}, MACS\,J1752.0+4440 \citep{2012MNRAS.425L..36V,2012MNRAS.426...40B}, PSZ1\,G108.18-11.53 \citep{2015MNRAS.453.3483D}, and Abell\,1240 \citep{2001ApJ...548..639K,2009AA...494..429B}.

\subsubsection{Radio spectra}
The integrated radio spectra of cluster radio shocks display power-law shapes (but see Section~\ref{sec:relicshighfreq}), with spectral indices ranging from about $-1.0$ to $-1.5$  \citep[e.g.,][]{2012MNRAS.426...40B,2012A&ARv..20...54F,2014MNRAS.444.3130D}. 
One notable exception of a flatter integrated spectrum, with good data available, is Abell\,2256 where the  spectral index is about $-0.8$ \citep{2008AA...489...69B,2012AA...543A..43V,2015AA...575A..45T}. This flat spectral index is difficult to reconcile with particle acceleration models and electron energy losses, see \cite{2012AA...543A..43V} for a discussion. Another  example appeared to be ZwCl\,2341.1+0000 \citep{2009AA...506.1083V} but more recent observations indicate that the spectral index is within the normally observed range \citep{2010AA...511L...5G,2017ApJ...841....7B}.

\begin{figure*}[htbp]
\centering
\includegraphics[width=1.0\textwidth]{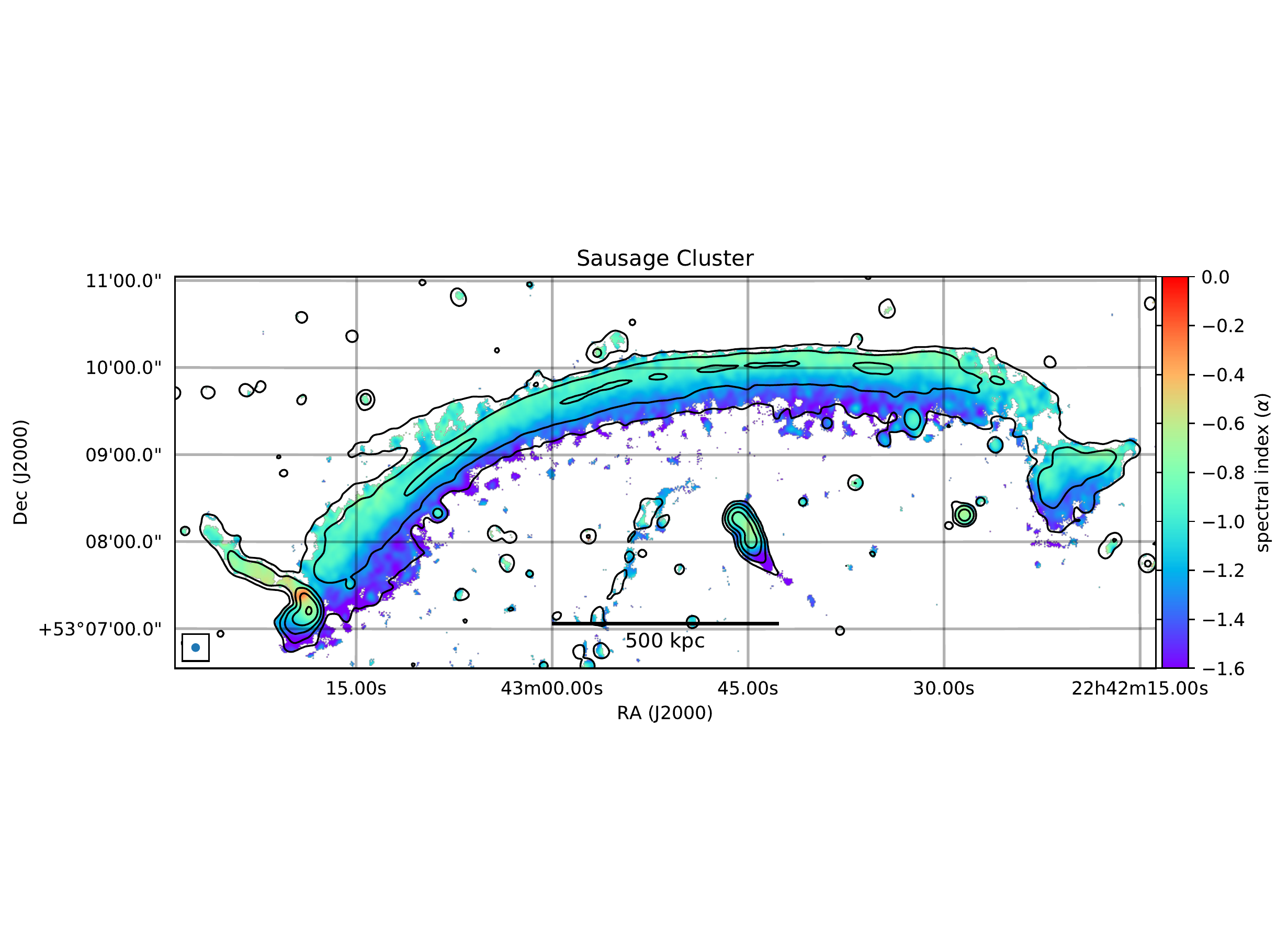}\vspace{-2cm}
\includegraphics[width=1.0\textwidth]{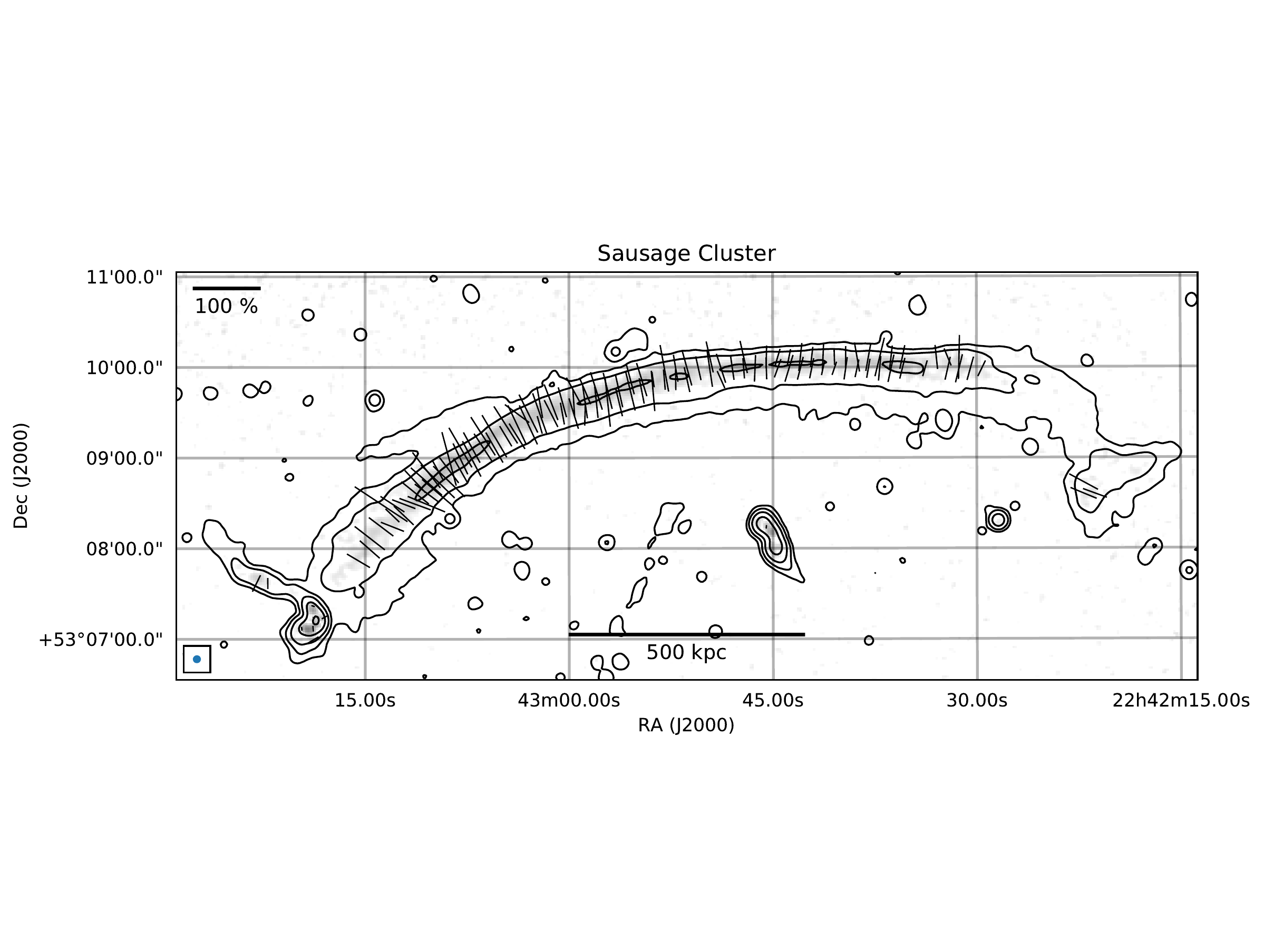}
\vspace{-3cm}
\caption{\textit{Top panel:} Spectral index distribution across the northern cluster radio shock in CIZA\,J2242.8+5301 between 0.15 and 3.0~GHz at 5$^{\prime\prime}$ resolution \citep{2018ApJ...865...24D}. Black contours are from a 1--4 GHz continuum image. Contours are drawn at levels of $[1,4,16,\ldots] \times 5 \sigma_{\rm rms}$, where $\sigma_{\rm rms}$ is the map noise.
\textit{Bottom panel:} Polarized intensity image at 3~GHz (Di Gennaro et al. in prep). Overlaid are the polarization electric field vectors corrected for Faraday Rotation. Black contours are the same as for the top panel.}
\label{fig:cizaspix}   
\end{figure*}

\begin{figure*}[htbp]
\centering
\includegraphics[width=0.48\textwidth]{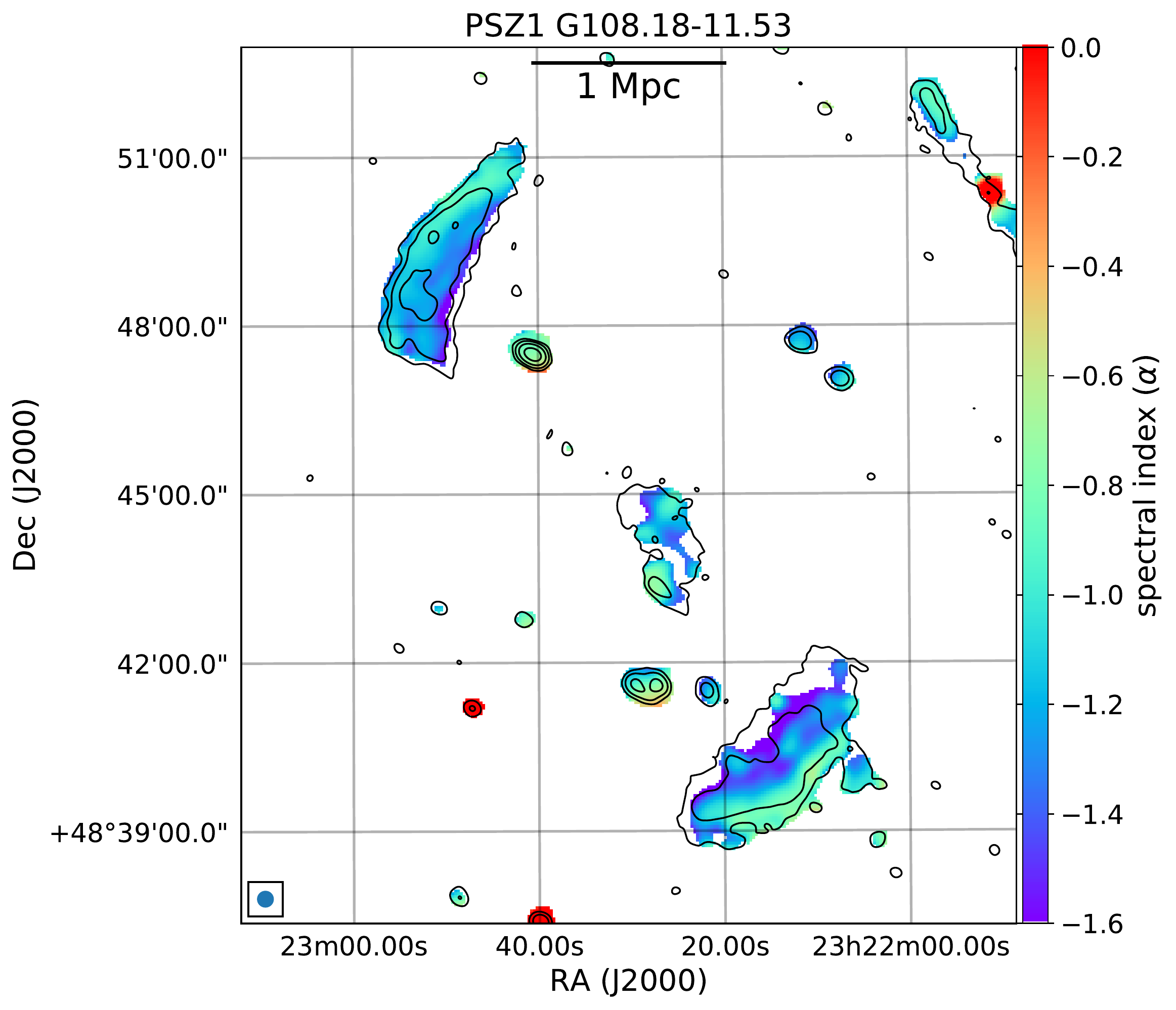}
\includegraphics[width=0.48\textwidth]{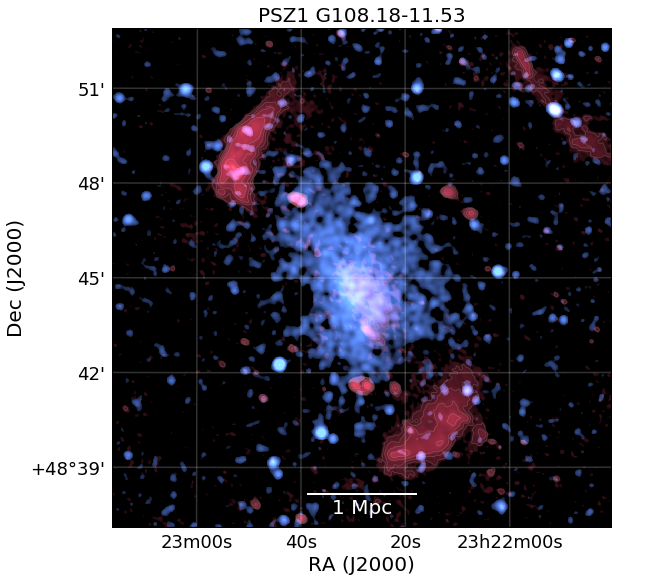}
\caption{{\it Left panel:} Spectral index map for the double radio shock in PSZ1\,G108.18--11.53 between 323 and 1380~MHz from \cite{2015MNRAS.453.3483D}. For both radio shocks, the spectral index steepens in the direction towards the cluster center. The 323~MHz radio contours are overlaid in black at levels of $[1,4,16,\ldots] \times 4 \sigma_{\rm rms}$, where $\sigma_{\rm rms}$ is the map noise. {\it Right panel:} Combined radio (red, GMRT 323~MHz) and X-ray (blue, Chandra 0.5--2.0~keV) image of PSZ1\,G108.18--11.53.}
\label{fig:spixpsz}   
\end{figure*}

Cluster radio shocks often show a clear spectral index gradient across their width, see Figures~\ref{fig:cizaspix} and~\ref{fig:spixpsz}. The region with the flattest spectral index is located on the side away from the cluster center. Towards the cluster center the spectral index steepens. This steepening is thought to be caused by synchrotron and IC losses in the shock downstream region. The majority of well-studied cluster radio shocks, both single shocks (see Figure~\ref{fig:A2744spixhalo}) and double shocks, show this behavior.

\begin{figure}[htbp]
\centering
\includegraphics[width=0.5\textwidth]{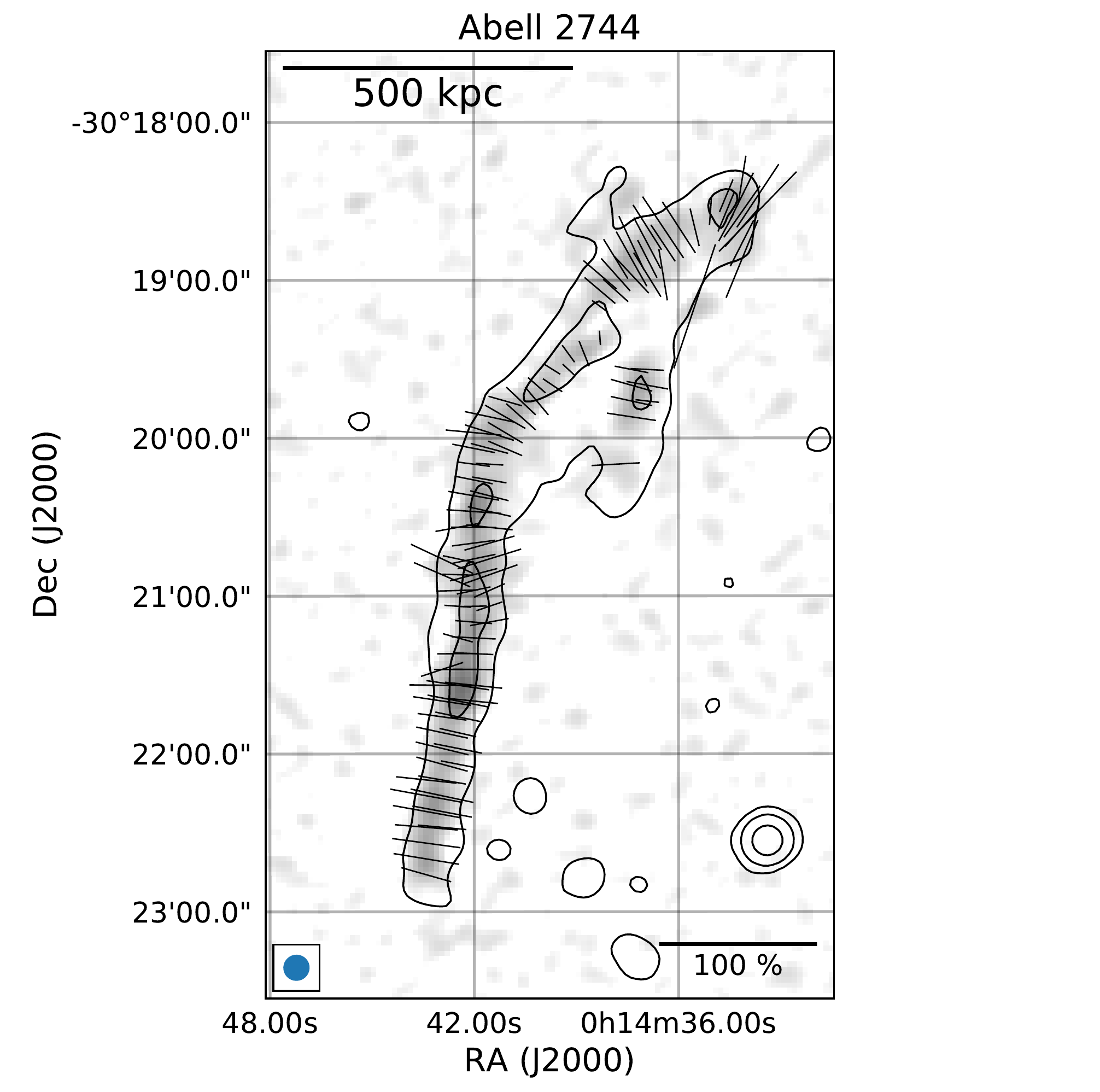}
\caption{Polarized intensity image at 3~GHz of the cluster radio shock in Abell\,2744 from \cite{2017ApJ...845...81P}, see also Figure~\ref{fig:A2744spixhalo}. Overlaid are the polarization electric field vectors corrected for galactic Faraday Rotation. The black contours come from the Stokes~I image and are drawn at levels of $[1,4,16,\ldots] \times 4 \sigma_{\rm rms}$, where $\sigma_{\rm rms}$ is the map noise.}
\label{fig:a2744pol}   
\end{figure}

\subsubsection{Polarization}
\label{sec:relicpolarization}
Cluster radio shocks are amongst the most polarized sources in the extragalactic sky. Very elongated radio shocks usually show the highest polarization fraction, which is expected if they trace edge-on shock waves \citep{1998A&A...332..395E}. For example, CIZA\,J2242.8+5301 shows polarization fractions of $\sim$50\% or more at GHz frequencies for some parts of the radio shock \citep{2010Sci...330..347V}, see Figure~\ref{fig:cizaspix}.

For large cluster radio shocks the intrinsic polarization angles, corrected for the effect of Faraday Rotation, are found to be well aligned. The polarization magnetic field vectors are oriented within the plane of the radio shock \citep[e.g.,][see also Figures~\ref{fig:cizaspix} and~\ref{fig:a2744pol}]{2009AA...494..429B,2010Sci...330..347V,2012MNRAS.426...40B,2017ApJ...845...81P}. 
Only a few Faraday rotation studies have been performed so far of radio shocks. They indicate that for radio shocks projected at large cluster centric radii the Faraday Rotation is mostly caused by the galactic foreground. Faraday Rotation caused by the cluster can be seen for (parts of) radio shocks at smaller cluster centric radii  \citep{2009AA...503..707B,2011AA...525A.104P,2012AA...546A.124V,2014ApJ...794...24O}. From the limited studies available, it seems that large cluster radio shocks strongly depolarize at frequencies $\lesssim 1$~GHz \citep{2008AA...489...69B,2011AA...525A.104P,2015PASJ...67..110O}. Therefore, high-frequency observations (above $\gtrsim 2$~GHz) are best suited to probe the intrinsic polarization properties of radio shocks. For example, the fractional linear polarization in for main `Sausage' and `Toothbrush' radio shocks is on average 40\% at 5--10~GHz, reaching 70\% in localized areas  \citep{2017AA...600A..18K,2017MNRAS.472.3605L}.

\subsubsection{Comparison between radio and X-ray  observations of ICM shocks}
\label{sec:radioxrayshock}

\begin{figure}[htbp]
\includegraphics[width=0.49\textwidth]{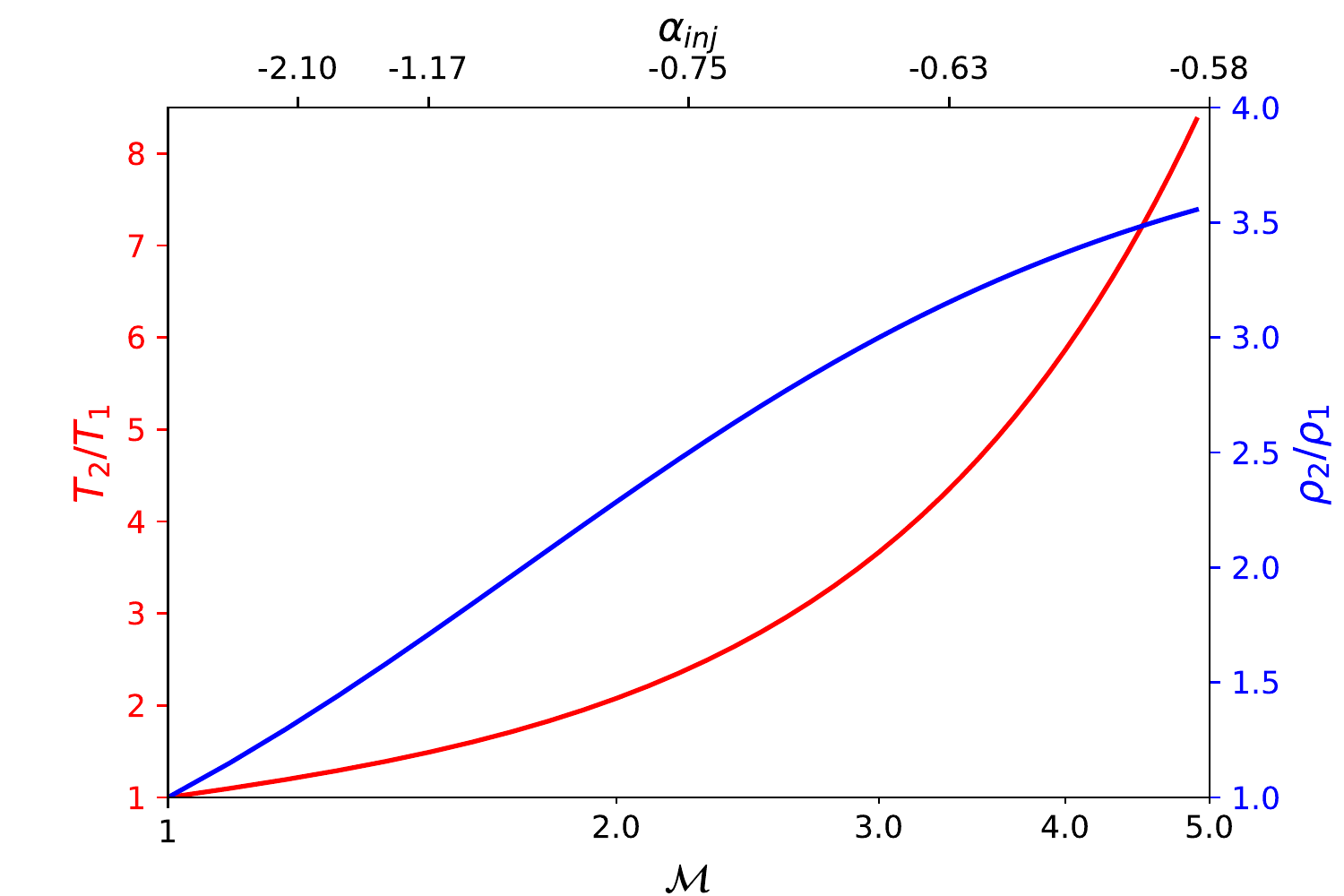}
\caption{The ratio of pre- and post-shock ICM properties (red: temperature, blue density) as a function of shock strength ($\cal M$).
The upper horizontal axis represents the injection spectral index adopting a diffusive shock acceleration (DSA) model under the test particle assumption: one-dimensional planar geometry, constant injection, etc., see \cite{2015JKAS...48....9K,2015JKAS...48..155K,2015ApJ...809..186K} for details.}
\label{fig:Mach_est}   
\end{figure}

Because of their shapes, locations, and spectral and polarimetric properties, cluster radio shocks are considered to trace particles accelerated at shocks. These shocks can be generated by cluster merger activity or accretion flows from surrounding large-scale structures \citep[e.g.,][]{1998A&A...332..395E}. If this assumption is correct, shock waves should coexist at the location of radio shocks. From X-ray observations, the intensity of shock structure can be estimated from the Rankine-Hugoniot jump condition \citep{1959flme.book.....L}. Assuming a ratio of specific heats as $\gamma=5/3$, we have
\begin{equation}
\frac{T_2}{T_1} = \frac{5{\cal M}^4+14{\cal M}^2-3}{16{\cal M}^2} 
\end{equation} 
\begin{equation}
\frac{\rho_2}{\rho_1} =  C=\frac{4{\cal M}^2}{{\cal M}^2+3} \mbox{ ,} 
\end{equation}
where the subscripts 1 and 2 refer to  the pre- and post-shock ICM density ($\rho$) or temperature ($T$), respectively.
The ratios of ICM properties as a function of the shock strength ($\cal M$) are shown in Figure~\ref{fig:Mach_est}.
On the other hand,
based on the assumption of simple DSA theory, the Mach number can be also estimated from the radio injection spectral index ($\alpha_{\rm inj}$) via
\begin{equation}
{{\cal M}_{\rm radio}=\sqrt{ \frac{2\alpha_{\rm inj}-3}{2\alpha_{\rm inj}+1}}} \mbox{ .}
\end{equation} 

In principle, both X-ray and radio approaches are  independent methods to characterize the shock strength, meaning shock strengths inferred from these different wavelength regimes should match each other, if underlaying assumptions are correct. Therefore, the comparison of the shock properties inferred from X-ray and radio data is an important tool to investigate shock related ICM physics. 
Until recently, observational information of radio shocks at X-ray wavelengths were limited because radio shocks are typically located in the cluster periphery, where the ICM X-ray emission is very faint. This makes it challenging to characterize the X-ray shock properties. 

The first detection of a shock wave, co-located with a cluster radio shock (relic), was in the nearby merging cluster Abell\,3667 using \textit{XMM-Newton} observations. \cite{2010ApJ...715.1143F} found a sharp X-ray surface brightness discontinuity at the outer edge of the radio shock, and a significant drop in the ICM temperature at the same location. These discontinuities are consistent with a $\mathcal{M}\sim$2 shock. These results have been confirmed by \cite{2012PASJ...64...67A,2016arXiv160607433S}.

The \textit{Suzaku} satellite, being in a low orbit within Earth's magnetopause, provided  a significantly lower and stable particle background compared to previous X-ray observatories (\textit{XMM-Newton} and \textit{Chandra}).  The low and well calibrated instrumental background of \textit{Suzaku} made it well-suited to study the faint cluster periphery. The first systematic \textit{Suzaku} investigation of cluster radio shocks was performed by  \cite{2013PASJ...65...16A}.  Since the first detection of the shock wave associated with Abell\,3667, there are about 20 X-ray detected shocks corresponding to radio shocks. An overview of radio shocks with X-ray detections is shown in Table~\ref{tab:relicshock}.

At radio wavelength, there are also observational challenges to derive shock properties. One particular difficulty is to measure $\alpha_{\rm inj}$. The integrated spectral index of a radio shock reflects a balance between acceleration and energy losses. As a result, the index of the integrated spectrum is 0.5 steeper compared to $\alpha_{\rm inj}$. This relation ($\alpha_{\rm int}=\alpha_{\rm inj}+0.5$ \citep{1962SvA.....6..317K}) is however somewhat simplistic,  since the shock properties do evolve over time, see also Section~\ref{sec:dsamodel}.  Alternatively, with spatially resolved spectral index maps one can obtain more reliable measurements of $\alpha_{\rm inj}$, avoiding some of the problems with energy losses. Here one needs to measure the spectral index as close as possible to the shock  location. However, even in this case some mixing of different electron energy populations will occur, depending on the spatial resolution, shape of the shock surface, and projection effects.

\begin{figure}[htbp]
\includegraphics[width=0.49\textwidth]{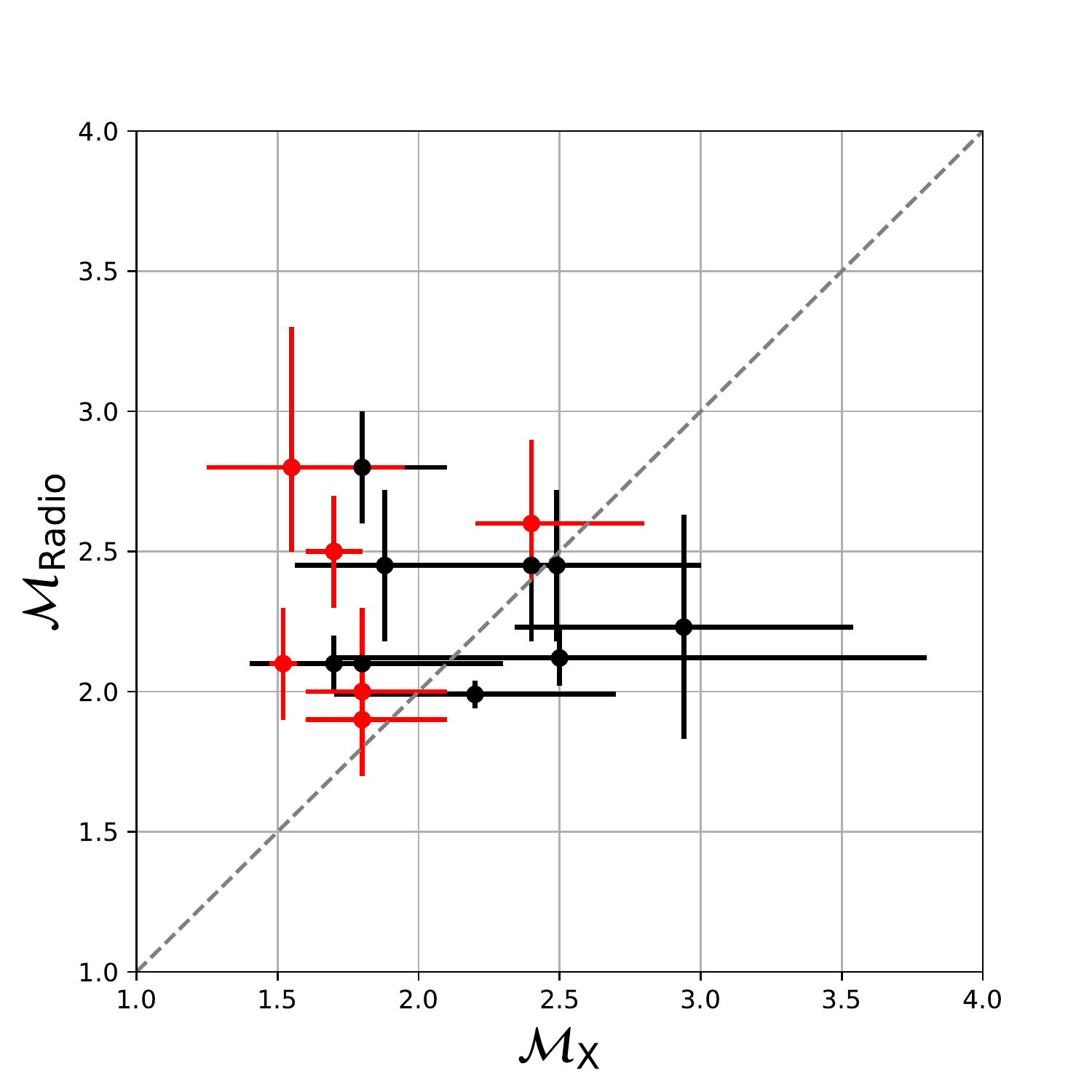}
\caption{Mach numbers for cluster radio shocks derived from the radio spectral index ($M_{\rm radio}$) plotted against the Mach number derived  from the ICM temperature jump ($M_{\rm X}$). The error bars show the statistical uncertainties at the one sigma level. Note that some radio derived Mach numbers were estimated from the integrated spectral index (black) rather than spatially resolved injection spectral index \citep[red: for details see][]{2014MNRAS.445.1213S,2016ApJ...818..204V,2017MNRAS.471.1107H}.}
\label{fig:Mach_comp}   
\end{figure}

For the northern radio shock in CIZA\,J2242.8+5301 a number of detailed comparison between the radio and X-ray derived Mach numbers have been performed. \cite{2010Sci...330..347V} reported a radio injection spectral index of $-0.60\pm0.05$ resulting in $\mathcal{M} = 4.6^{+1.3}_{-0.9}$ (68\% confidence range).
In the X-rays, \cite{2013PASJ...65...16A,2015AA...582A..87A} reported a temperature increase across the radio shock with an amplitude of a factor $\sim$3 resulting giving ${\cal M}_{\rm}=2.7^{+0.7}_{-0.4}$ (including systematics due to the background estimation). This kind of tension, ${\cal M}_{\rm radio}> {\cal M}_{\rm X}$, has been found for other radio shocks, see  Figure~\ref{fig:Mach_comp}.  If this discrepancy is indeed real, this may point to problems in the DSA scenario for shocks in clusters. To explain the observational results, several solutions have been proposed.

For example, it is possible that the X-ray derived Mach numbers are somewhat  underestimated due to unfavorable viewing angles and the complexity of the shock surface. In addition, the shock acceleration efficiency is a thought to be a strong function of shock Mach number \citep{2007MNRAS.375...77H}. Therefore the CR-energy-weighted Mach number is expected to be higher than the kinetic-energy-weighted Mach number \citep{2018ApJ...857...26H}. Thus radio measured Mach numbers will be biased towards parts of the shock with the highest Mach numbers. Difficulties and possible biases with radio based measurements are discussed in \cite{2014MNRAS.445.1213S,2016ApJ...818..204V,2017MNRAS.471.1107H}. The re-acceleration of fossil plasma has also been invoked, see Section~\ref{sec:dsamodel}. \cite{2017AA...600A.100A} investigated possible systematic errors associated with X-ray observations. We refer the reader to Sect 4.3. in their paper for more details.

\begin{table*}
\label{tab:relicshock}
\caption{{A table of clusters which show evidence for shock waves in X-ray observations and that coincide with the location of cluster radio shocks (Akamatsu et al. in prep.)}}
\renewcommand\tabcolsep{4pt}
\begin{center}
\begin{tabular}{lcccll}\hline
Name	& $T$ jump &  $\rho$ jump & Spec index		& X-ray ref		& Radio ref \\ \hline
Coma SW& \CheckmarkBold& 	& \CheckmarkBold	&  \cite{2013PASJ...65...89A} & \cite{1991AA...252..528G} \\
&&&& \cite{2013MNRAS.433.1701O}  &\cite{2003AA...397...53T}\\
Abell\,115	&\CheckmarkBold&\CheckmarkBold	&\CheckmarkBold &  \cite{2016MNRAS.460L..84B}& \cite{2001AA...376..803G}\\
Abell\,754 &\CheckmarkBold&\CheckmarkBold	&\CheckmarkBold&  \cite{2003AstL...29..425K} &  \cite{2001ApJ...559..785K}\\
&&&& \cite{2011ApJ...728...82M} &  \cite{2009ApJ...699.1883K}\\
Abell\,1240 &&\CheckmarkBold	&	\CheckmarkBold &  \cite{2018MNRAS.478.2218H} &  \cite{2001ApJ...548..639K} \\
&&&&& \cite{2009AA...494..429B} \\
&&&&&  \cite{2018MNRAS.478.2218H} \\
Abell\,3667 NW&\CheckmarkBold&\CheckmarkBold&\CheckmarkBold& \cite{2010ApJ...715.1143F}&  \cite{1997MNRAS.290..577R}\\
&&&&  \cite{2012PASJ...64...49A}& \cite{2003PhDT.........3J} \\
&&&& \cite{2016arXiv160607433S} & \cite{2014MNRAS.445..330H}	\\
&&&&& \cite{2017arXiv170604930J} \\                
Abell\,3667	SE&		\CheckmarkBold&\CheckmarkBold&\CheckmarkBold &  \cite{2013PASJ...65...16A} & \cite{2003PhDT.........3J} \\
&&&	&  \cite{2018MNRAS.tmp.1381S}& \cite{2014MNRAS.445..330H}\\
&&&&&  \cite{2015MNRAS.447.1895R} \\
Abell\,3376 W &	\CheckmarkBold &\CheckmarkBold&\CheckmarkBold&  \cite{2012PASJ...64...67A} 	  &  \cite{2006Sci...314..791B}\\
&&&& \cite{2018AA...618A..74U}& 	\cite{2012MNRAS.426.1204K}\\
&&&&& \cite{2015MNRAS.451.4207G}\\
Abell\,3376 E&	\CheckmarkBold&\CheckmarkBold&\CheckmarkBold	 & \cite{2018AA...618A..74U} 	   & \cite{2012MNRAS.426.1204K}\\
Abell\,2255&	\CheckmarkBold&\CheckmarkBold&\CheckmarkBold	&\cite{2017AA...600A.100A}  &\cite{2009AA...507..639P}\\

Abell\,2256 &	\CheckmarkBold&&\CheckmarkBold &  \cite{2015AA...575A..45T}  & \cite{1994ApJ...436..654R}\\
&&&&&\cite{2006AJ....131.2900C} \\
&&&&&\cite{2012AA...543A..43V} \\
Abell\,2744 & \CheckmarkBold	  &\CheckmarkBold		   &\CheckmarkBold		 & \cite{2016MNRAS.461.1302E}  & \\
&&&&  \cite{2017PASJ...69...39H} & \cite{2017ApJ...845...81P}\\
Sausage N & \CheckmarkBold	  & &\CheckmarkBold	   &  \cite{2013PASJ...65...16A} 	  &  \cite{2010Sci...330..347V}\\
&&&&  \cite{2013MNRAS.429.2617O}&  \cite{2013AA...555A.110S,2014MNRAS.441L..41S,2016MNRAS.455.2402S}\\
&&&&  \cite{2014MNRAS.440.3416O}& \cite{2017MNRAS.471.1107H}	\\
&&&& \cite{2015AA...582A..87A}& Loi et al. 2017 \cite{2017MNRAS.472.3605L}\\
&&&&&  \cite{2017AA...600A..18K} \\ 
&&&&& \cite{2018ApJ...865...24D} \\ 
Sausage S & \CheckmarkBold	  & 						    	     &\CheckmarkBold	    &  \cite{2015AA...582A..87A}   &   \cite{2017MNRAS.471.1107H}\\
&&&&&  \cite{2013AA...555A.110S} \\
&&&&&  \cite{2018ApJ...865...24D} \\
Toothbrush	&	\CheckmarkBold	  &\CheckmarkBold		  &\CheckmarkBold		 &  \cite{2013MNRAS.433..812O}  & \cite{2018ApJ...852...65R}\\
&&&&   \cite{2015PASJ...67..113I} &  \cite{2012AA...546A.124V} \\
&&&&  \cite{2016ApJ...818..204V} &  \cite{2016ApJ...818..204V}\\
&&&&& \cite{2016MNRAS.455.2402S} \\
El Gordo	&\CheckmarkBold&\CheckmarkBold&\CheckmarkBold&  \cite{2016MNRAS.463.1534B}	& \cite{2014ApJ...786...49L}   	 \\
&&&&  \cite{2016ApJ...829L..23B}\\
ZwCl\,2341.1+0000  &   &\CheckmarkBold						  &\CheckmarkBold		&  \cite{2014MNRAS.443.2463O} &  \cite{2002NewA....7..249B}	\\
&&&&&  \cite{2009AA...506.1083V} \\
&&&&&  \cite{2010AA...511L...5G} \\
&&&&&  \cite{2017ApJ...841....7B} \\
Bullet reverse & \CheckmarkBold& \CheckmarkBold& \CheckmarkBold&  \cite{2015MNRAS.449.1486S} &  \cite{2015MNRAS.449.1486S}\\
Abell\,2146	&	\CheckmarkBold	&\CheckmarkBold	&&  \cite{2010MNRAS.406.1721R,2011MNRAS.417L...1R,2012MNRAS.423..236R}	& \cite{2018MNRAS.475.2743H}	\\
Abell\,521	&	\CheckmarkBold	  &\CheckmarkBold	&\CheckmarkBold & \cite{2013ApJ...764...82B}&  \cite{2008Natur.455..944B}	\\
&&&&& \cite{2008AA...486..347G} \\
RXJ1314.4-2515 &\CheckmarkBold	  &\CheckmarkBold	&	& \cite{2011MmSAI..82..495M}	& \cite{2005AA...444..157F} \\
&&&&& \cite{2007AA...463..937V} \\
\hline
\end{tabular}
\end{center}
\end{table*}

\begin{figure*}[htbp]
\includegraphics[width=0.49\textwidth]{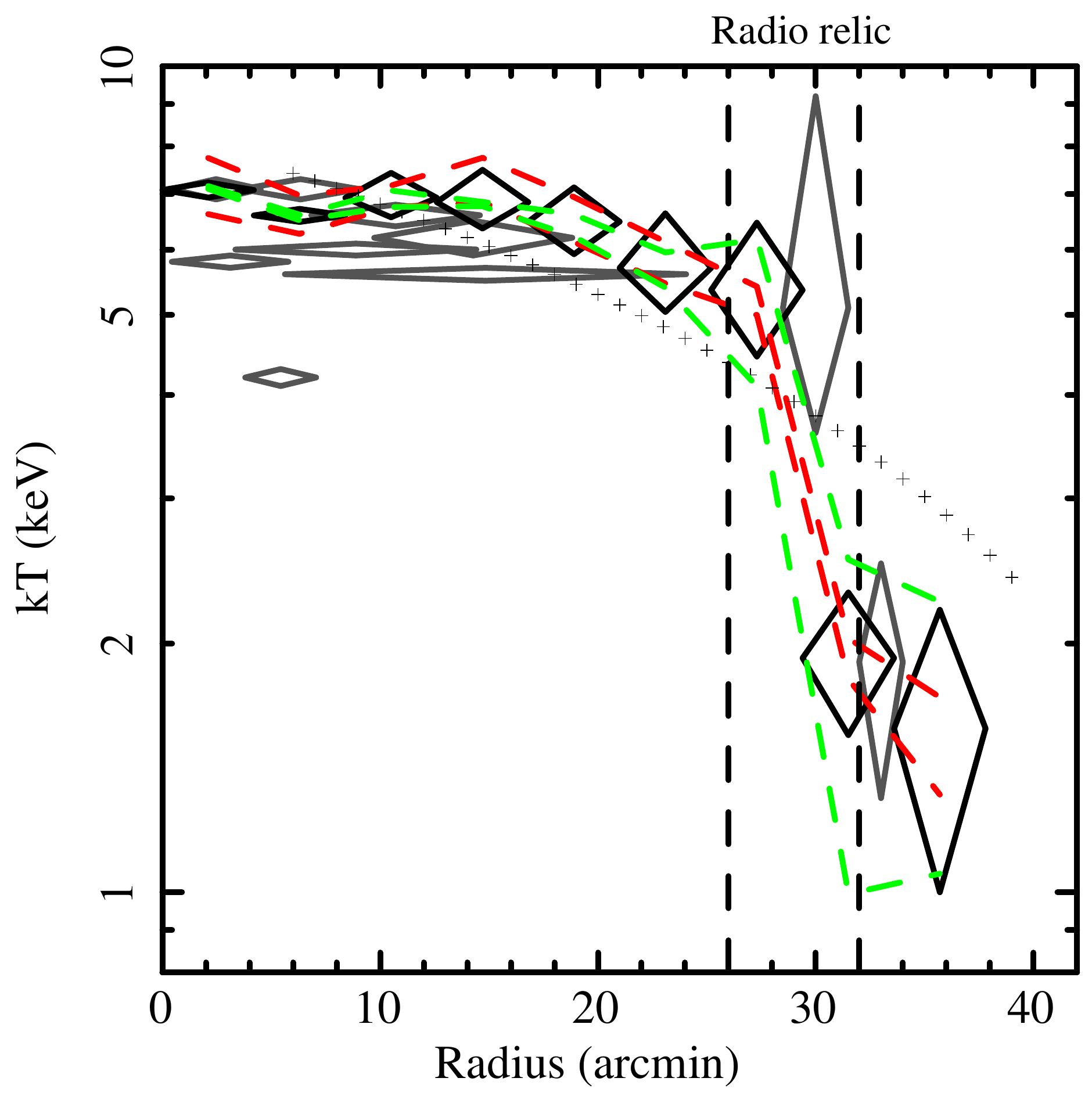}
\includegraphics[width=0.49\textwidth]{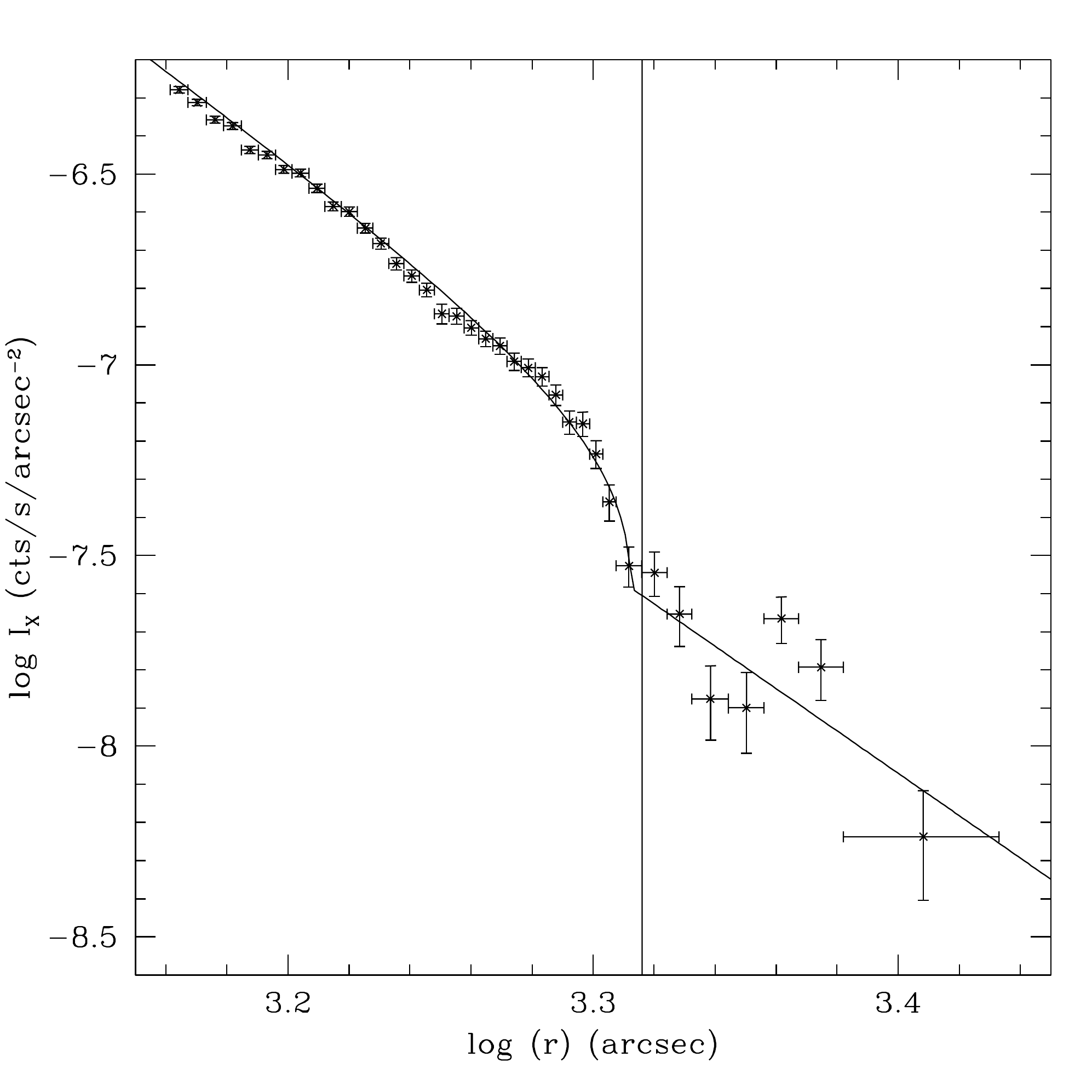}
\caption{
ICM temperature (\textit{left}) and surface brightness (0.5--2.0~keV, \textit{right}) profiles of Abell\,3667 adopted from \cite{2012PASJ...64...67A} and \cite{2016arXiv160607433S}, respectively.
For the ICM temperature profile, the black and gray diamonds represent \textit{Suzaku} and \textit{XMM-Newton} \citep{2010ApJ...715.1143F} best-fit values with 90\% confidence range. The black dashed vertical lines show the approximate radial boundaries of the northwest radio shock. Two (green and red) dashed lines show the systematic uncertainties of the best-fit values due to changes of the optical blocking filter contaminants and the non X-ray background level. The crosses show an average profile given by \cite{2007A&A...461...71P} for Abell\,3667. For the surface brightness profile, the data points are shown with $1\sigma$ uncertainties. The model fit is shown with a solid line. An abrupt drop of the surface brightness (i.e., density) is present near the outer edge of the radio shock, which is indicated by the vertical line.}
\label{fig:A3667_Xray}   
\end{figure*}

{Future X-ray satellites, such as \emph{Athena} \citep{2013arXiv1306.2307N}, will provide precise measurements of cluster merger shocks. This will shed further light on the apparent discrepancy between the Mach numbers derived from radio and X-ray observations. With the improved collecting area with respect to current satellites, the shock properties in faint cluster outskirt can also be determined.}

\subsubsection{SZ observations}
\label{sec:SZhigh}
The thermal ICM electrons in galaxy clusters interact with CMB photons through inverse Compton scattering, resulting in the so-called SZ effect \citep{1970Ap&SS...7....3S}. The SZ effect provides a complementary way of studying the ICM and, because of its redshift-independent nature, is particularly powerful at high-redshift where the X-ray surface brightness suffers from significant cosmological dimming. Low-resolution studies measuring the bulk SZ signal have been very successful at selecting large samples of both relaxed and disturbed clusters up to $z\sim$1.5 \citep[e.g.,][]{2016A&A...594A..27P,2015ApJS..216...27B}. 

In the last decade, efforts at the very highest radio frequencies (above 90~GHz) have focused on measuring SZ at high spatial resolution with the aim of detecting small scale features in the ICM, such as shocks in merging clusters. Great strides have been made possible by the introduction of high-resolution, large field-of-view instruments such as MUSTANG-2\footnote{\url{http://www.gb.nrao.edu/mustang/}} installed on the 100-m Green Bank Telescope ($9^{\prime\prime}$ resolution at 90~GHz) and NIKA/NIKA2\footnote{\url{http://ipag.osug.fr/nika2/Welcome.html}} on the 30-m IRAM telescope (reaching $10^{\prime\prime}$--$20^{\prime\prime}$ resolution at 150 and 260~GHz). The power of these instruments has already been demonstrated through high-resolution SZ images showing substructure in merging clusters and in the cores of relaxed clusters \citep{2017A&A...598A.115A,2018A&A...612A..39R,2018A&A...614A.118A}. For the nearby Coma cluster the Planck satellite has provided resolved SZ images \citep{2013A&A...554A.140P}, including the likely detection of two $\mathcal{M} \sim 2$ shocks in the cluster periphery.

Following pioneering work detecting a weak shock in MACS0744+3927 \citep{2011ApJ...734...10K}, more recent observational work with the Atacama Large Millimeter/submillimeter Array (ALMA)\footnote{\url{http://www.eso.org/public/teles-instr/alma/receiver-bands/}} has enabled a direct detection and measurement of a cluster merger shock in `El Gordo' \citep{2016ApJ...829L..23B}. These observations demonstrate great potential for future SZ determinations of shock properties (particularly the Mach number), especially at large cluster-centric distances and high-redshift, where X-ray measurements of the ICM properties become challenging.

\subsubsection{Gamma-rays from cluster radio shocks}
\label{sec:gammaradioshock}
Apart from (re-)accelerating electrons, shocks should also accelerate protons. For DSA, the number of accelerated protons should be much larger than electrons. Similar to the secondary model for radio halos, these CR protons should collide with the thermal ICM and produce gamma-rays via hadronic interactions.

It has been noted by \cite{2014MNRAS.437.2291V,2015MNRAS.451.2198V,2016MNRAS.459...70V} that the expected gamma-ray emission for DSA shock acceleration at radio shocks is in tension with gamma-ray upper limits for some  clusters. This indicates that the relative acceleration efficiency of electrons and protons is at odds with predictions from DSA. Adding the re-acceleration of fossil particles to this prediction does not change this conclusion. One possible explanation for the lack CR protons is that the magnetic field at radio shocks is predominantly perpendicular to the shock normal. Work by \cite{2014ApJ...783...91C} indicates that the acceleration efficiency of protons is strongly suppressed at such shocks. Simulations by \cite{2017MNRAS.464.4448W} indicate this could reduce the tension with the low gamma-ray upper limits.

Recently, claims of gamma-ray emission from the virial shocks around the Coma cluster \citep{2017ApJ...845...24K}, as well as from a stacking of other clusters \citep{2017arXiv170505376R}, have been put forward. We underline, however, that so far these claims have been not been confirmed \citep{2012ApJ...757..123A,2014MNRAS.440..663Z,2014MNRAS.441.2309P,2016ApJ...819..149A}.

\subsubsection{High-frequency studies of radio shocks}
\label{sec:relicshighfreq}

Owing to their their steep spectra, radio shocks have been classically observed at relatively low frequencies ($<2$ GHz). In this Section we review the current state-of-the-art high-frequency observations of radio shocks, by focusing on observations above 5~GHz. High-frequency observations pose particular challenges: (i) radio shocks have steep-spectra making them very faint at high-frequencies; (ii) radio interferometers typically have small fields of view at high frequency and thus have difficulty in detecting extended diffuse sources. Until 2014, the highest frequency detection of a radio shocks were in the clusters Abell\,521 and MACS\,J0717.5+3745 at 5~GHz \citep{2008AA...486..347G,2009AA...503..707B}. The interest in high-frequency observations of cluster radio shocks, and the number of detections, has grown over the past few years. This interest has been motivated by the study of the injected electrons and their aging mechanism \citep[as discussed for example by][]{2016JKAS...49..145K,2016MNRAS.462.2014D,2016ApJ...823...13K}.

\begin{figure}[htbp]
\centering
\includegraphics[width=0.5\textwidth]{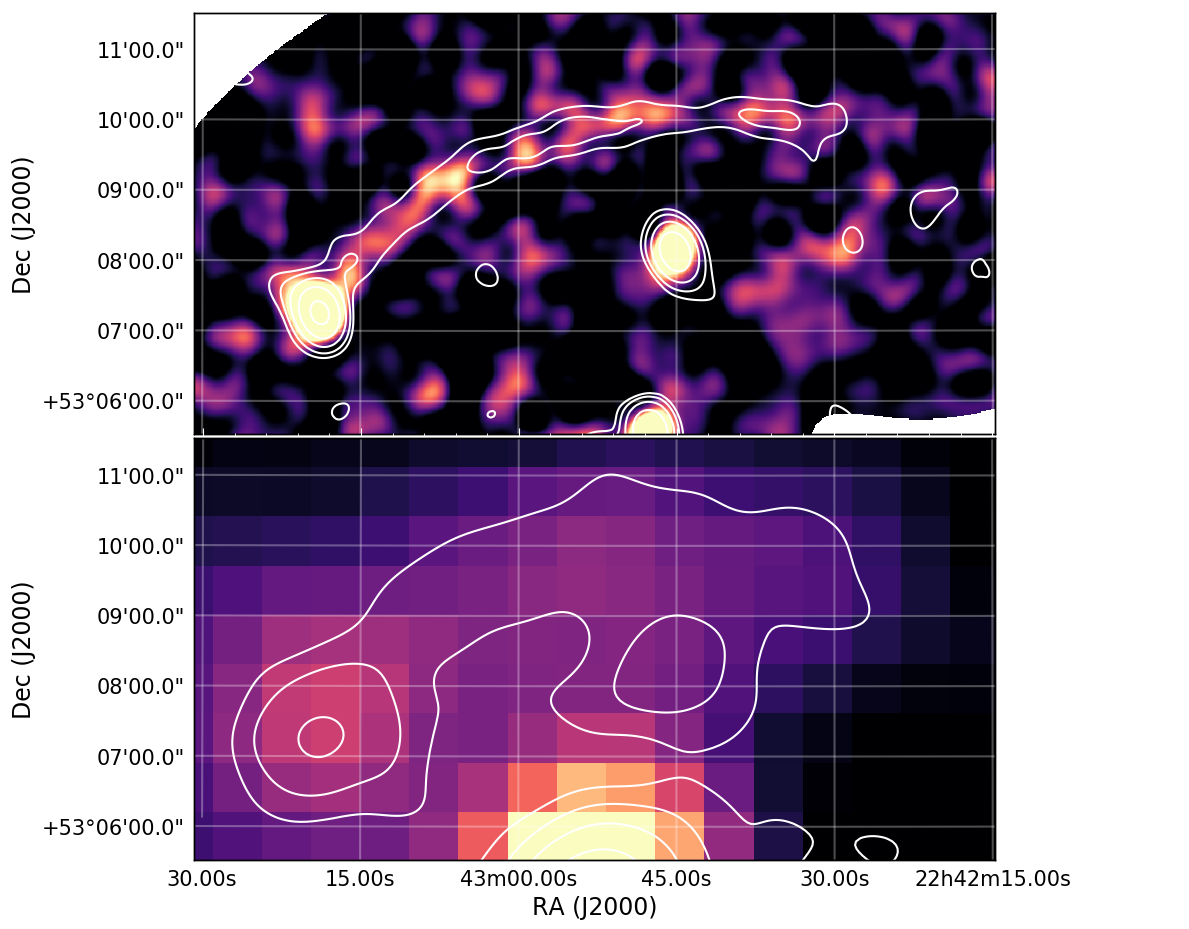}
\caption{High-frequency view of the main radio shock in the `Sausage' cluster. The \textit{top} panel shows interferometric images at $\sim$30$^{\prime\prime}$ resolution taken at 16~GHz with AMI (contours) and 30~GHz with CARMA \citep[background image;][]{2014MNRAS.441L..41S,2016MNRAS.455.2402S}. The \textit{bottom} panel shows single-dish measurements from Effelsberg at 6.6~GHz with $174^{\prime\prime}$  resolution and 8.35~GHz at $90^{\prime\prime}$ resolution \citep{2016MNRAS.455.2402S,2017AA...600A..18K,2017MNRAS.472.3605L}.}
\label{fig:sausageHF}   
\end{figure}

Instruments that helped make progress at high frequencies include interferometers, such as the {Arcminute} Microkelvin Imager (AMI, 16~GHz), the Combined Array for Research in Millimeter-wave Astronomy (CARMA, 30~GHz) and the VLA (4--10~GHz), and single dish antennas such as Effelsberg (up to 10~GHz) and the Sardinia Radio Telescope (SRT, up to 19~GHz), see Figure~\ref{fig:sausageHF}.

 At the moment of writing, six clusters benefit from radio shock detections above 5~GHz: the main radio shocks in the `Sausage' and the `Toothbrush' clusters \citep[both up to 30~GHz,][]{2014MNRAS.441L..41S,2016MNRAS.455.2402S,2017AA...600A..18K,2017MNRAS.472.3605L}, Abell\,2256 \citep[at 5~GHz,][]{2015AA...575A..45T} , the `Bullet' cluster radio shock \citep[5.5 and 9 GHz,][]{2016Ap&SS.361..255M}, ZwCl\,0008.8+5215 and Abell\,1612 \citep[at 5 and 8~GHz,][]{2017AA...600A..18K}. In combination with low frequency measurements, integrated cluster radio shock spectra spanning over 3 orders of magnitude in frequency have been produced, for example, covering the range from 74/150~MHz to 30~GHz, as is the case for the `Sausage' and the `Toothbrush' radio shocks \citep{2012AA...546A.124V,2016MNRAS.455.2402S}.

Interferometric observations from 150~MHz to 30~GHz have revealed a possible steepening of the integrated radio shock spectra beyond 2--5~GHz \citep{2014MNRAS.441L..41S,2016MNRAS.455.2402S,2015AA...575A..45T}, which challenges the radio shock formation model involving DSA acceleration at planar shocks. However, studies combining high-frequency single-dish observations with low-frequency interferometric observations \citep{2017AA...600A..18K,2017MNRAS.472.3605L} do not corroborate this finding (for more details on the caveats of both methods, see below). The mismatch between observations and theory has sparked a discussion as to what is causing the decrement in the flux density of cluster radio shocks at high frequencies (see also Section~\ref{sec:dsamodel}). One possibility is that the decrement is not intrinsic to the CR electron distribution at the shock, but is caused by  the SZ effect. At 10--30~GHz, the SZ effect is expected to result in a decrement in flux density. Even though the radio shocks are typically located $1-1.5$~Mpc away from the cluster center, authors have argued that the  sharp pressure discontinuity from the shock could explain $\sim 20-50\%$ of the decrement (for more typical examples such as the `Sausage', `Toothbrush', or Coma cluster), even up to 100\% at the highest frequencies for extreme cases, such as the `El Gordo' or Abell 2256 clusters \citep[depending on the shock geometry,][]{2015MNRAS.447.2497E,2016A&A...591A.142B}.

Various alternatives to the simple shock acceleration model have also been proposed. By contrast to acceleration at time invariant shocks, which results in power-law integrated spectra, curved spectra could be a natural result of spherically-expanding ICM shocks \citep{2015JKAS...48..155K,2015JKAS...48....9K}. The simple radio shock formation model assumes that the associated shock wave injects thermal electrons. A scenario where the shock predominantly injects non-thermal fossil electrons, pre-accelerated by previous AGN activity, could also reproduce the observed curved radio spectra \citep{2015ApJ...809..186K}. The downstream steepening, as well as the steepening of the integrated spectrum, can be recovered if there is non-uniform magnetic field in the downstream area of the shock \citep{2016MNRAS.462.2014D} or if the electrons, after shock acceleration, are further re-accelerated by turbulence \citep{2015ApJ...815..116F,2017JKAS...50...93K}. Tailored DSA simulations aimed at reproducing the observed parameters of radio shocks with good spectral coverage are now also becoming available \citep{2015ApJ...809..186K,2016JKAS...49...83K,2017ApJ...840...42K}.

\paragraph{Limitations and caveats:}\label{sec:caveats}

The combination of low-frequency and high-frequency data to produce wide-frequency spectra can be complicated. Different approaches have been taken in the literature to achieve this: (i) using common baselines at all frequencies trying to ensure recovery of the same spatial scales \citep{2014MNRAS.441L..41S,2016MNRAS.455.2402S} or (ii) combining low-frequency datasets with the best available coverage at short baselines with single dish measurements \citep{2015AA...575A..45T,2017AA...600A..18K,2017MNRAS.472.3605L}. However, both these approaches come with caveats. Unlike low-frequencies, at high-frequencies, current interferometers do not have very good short-baseline coverage and therefore have trouble recovering extended emission. When using common baseline coverage, the data at high frequency can become too noisy and the spectral behavior of the extended flux is lost. Using low-frequency interferometric data together with single dish measurements has the intrinsic problem of resolving out flux in the interferometric data. In addition, the removal of flux from compact sources that contaminate the single dish measurements is not always straightforward. Current cluster radio shock observations at $10-90$ GHz are limited by their resolution. Interferometric observations in the literature can push down to half arcmin resolution at 10--30~GHz \citep{2014MNRAS.441L..41S,2016MNRAS.455.2402S}. By contrast, single dish measurements have the advantage of measuring the total power, but the resolution they can achieve is relatively poor. The largest single dish telescopes (such as the 100-m Effelsberg) can reach $20^{\prime\prime}$ resolution at the highest frequency, but can only achieve one to a few arcmin resolution at 10--30~GHz \citep{2015AA...575A..45T,2016MNRAS.455.2402S,2017AA...600A..18K,2017MNRAS.472.3605L}.

{\paragraph{Future prospects:} The number of radio shock detections above 5~GHz is expected to steadily rise in the following years with observations coming from current instruments, such as the VLA and single-dish telescopes. A number of new facilities are coming or will shortly come online, which will have a significant impact on the study of radio shocks at high-frequencies. Particularly, instruments mounted on large single dish telescopes, such as MUSTANG-2 and NIKA2, will enable SZ studies at high resolution and thus pave the way for joint SZ and X-ray studies of  shocks.} 

{Upcoming interferometers will enable the study of the diffuse synchrotron emission from radio shocks at never-before achieved resolution. The low-bands of the ALMA, will provide $5^{\prime\prime}$--$15^{\prime\prime}$ resolution over the 35--50~GHz (Band 1) and $65-90$~GHz (Band 2) range in its most compact configuration. Particularly interesting will be the combination of the ALMA 12-m array with the Atacama Compact Array (a compact configuration of 7-m dishes), which is expected to provide a good compromise in terms of mapping of large scale structures and resolution. In the 2020s, the Square Kilometre Array (SKA)\footnote{\url{https://www.skatelescope.org/}} will have observing capabilities up to 10~GHz providing exquisite low surface brightness sensitivity at high resolution (up to 2~milliarcsec at 10~GHz).}

\subsubsection{Scaling relations}
Similar to radio halos, a correlation is found  between cluster X-ray luminosity  and radio power of cluster radio shocks \citep{2012A&ARv..20...54F}. This correlation likely reflects an underlying correlation between mass an radio power, with $P \propto M^{2.8}$ \citep{2014MNRAS.444.3130D}. 
In addition, there is a correlation between the largest linear size (LLS) and distance from the cluster center of the radio shock \citep{2009AA...508...75V,2012MNRAS.426...40B,2014MNRAS.444.3130D}. This is in line with the prediction that in the periphery of clusters the shock surfaces are larger. There is no clear evidence for a correlation between  LLS and radio spectral index. Previously, the existence such a correlation has been reported by \cite{2009AA...508...75V}. However, this LSS--$\alpha$ correlation was produced by the radio phoenices present in the \cite{2009AA...508...75V} sample, because radio phoenices generally have smaller LLS and steeper spectra than radio shocks. \cite{2012MNRAS.420.2006N,2012MNRAS.423.2325A,2017MNRAS.470..240N} investigated whether simulations can reproduce the luminosity function, shapes, and LLS distribution of radio shocks. They found reasonable agreement with the properties of radio shocks detectable in the NVSS survey.

\subsection{Cosmic ray acceleration modeling at cluster shocks}
\label{sec:dsamodel}
The acceleration of CR electrons at ICM shocks depends critically on the injection of background electrons into the Fermi-I process
and the self-generation of plasma waves that scatter electrons both upstream and downstream of the shock.
Background electrons need to be pre-accelerated above the injection momentum, $p_{\rm inj} \sim 130-200~ p_{\rm th,e}$ ($p_{\rm th,e}=\sqrt{2m_{\rm e} k T_2}$), in order to cross the shock transition layer whose width is of the order of gyro radii of thermal protons.
Particle-in-cell (PIC) simulations of low Mach number shocks in high beta plasma by \cite{2014ApJ...794..153G,2014ApJ...797...47G} 
demonstrated that incoming electrons are specularly reflected at the shock ramp by magnetic mirrors and gain energy via multiple cycles of shock drift acceleration, resulting in a suprathermal power-law population of electrons. Necessary scattering waves are self-excited by the firehorse instability. This process is most efficient at quasi-perpendicular shocks where the mean background magnetic field is nearly perpendicular to the shock flow direction. However, the full Fermi-I acceleration that involves scattering of electrons in both upstream and downstream regions of the shock has yet to be studied by PIC simulations.

The merger-shock DSA models for cluster radio shocks have to adopt a set of shock parameters including the pre-shock temperature, $kT_1$, sonic Mach number, $\mathcal{M}_s$, post-shock magnetic field strength profile, $B_2(r)$, and optionally a turbulent acceleration timescale, $\tau_{\rm acc}$,
and assume a specific viewing geometry often parameterized with extension angles, $\psi$'s.
In addition, in the re-acceleration model, one assumes a fossil electron population with energy spectrum, $N_{\rm fossil}= N_{e}(r) E^{-p}\exp[-(E/E_c)^2]$, in a large volume over $\sim$1~Mpc scale. The power-law slope, $p$, and the energy cutoff, $E_c$, can be adjusted to reproduce radio observations.

In particular, DSA models have been successful in reproducing some of observed properties of giant radio shocks such as the thin elongated morphologies, radio flux ($S_{\nu}$) and spectral index ($\alpha_{\nu}$) profiles, and integrated radio spectra ($J_{\nu}$) \citep{2012ApJ...756...97K}.
In the case where the radio-inferred Mach number, $\mathcal{M}_{\rm radio}=[(2\alpha_{\rm sh}-3)/(2\alpha_{\rm sh}+1)]^{1/2}\approx 3-4$, is greater than the X-ray-inferred Mach number, $\mathcal{M}_{\rm X}\approx 1.5-3$, the re-acceleration of fossil electrons with a flat spectrum could explain the observed discrepancy \citep{2013MNRAS.435.1061P,2016ApJ...823...13K}. On the other hand, \cite{2017NatAs...1E.163Z,2018MNRAS.478.4922Z} suggested superdiffusive shock acceleration (SSA) as an alternative explanation for the Mach number discrepancy. SSA is based on superdiffusive transport of energetic particles due to a non-Gaussian (L\'evy) random walk. It may lead to CR energy spectra flatter than expected from DSA with normal diffusion.

In addition, the Fermi-II acceleration by post-shock turbulence via transit-time-damping resonance has been invoked to explain the broad downstream steepening of the radio spectrum behind the observed radio shocks  \citep{2007MNRAS.378..245B,2017ApJ...840...42K,2017JKAS...50...93K}. Thus, the model parameters need to be fine-tuned by comparing theoretical predictions against observations of radio shocks, especially,
$S_{\nu}(R)$, $\alpha_{\nu}(R)$, and $J_{\nu}$, at the least.

\begin{figure}[htbp]
\centerline{\includegraphics[width=0.55\textwidth]{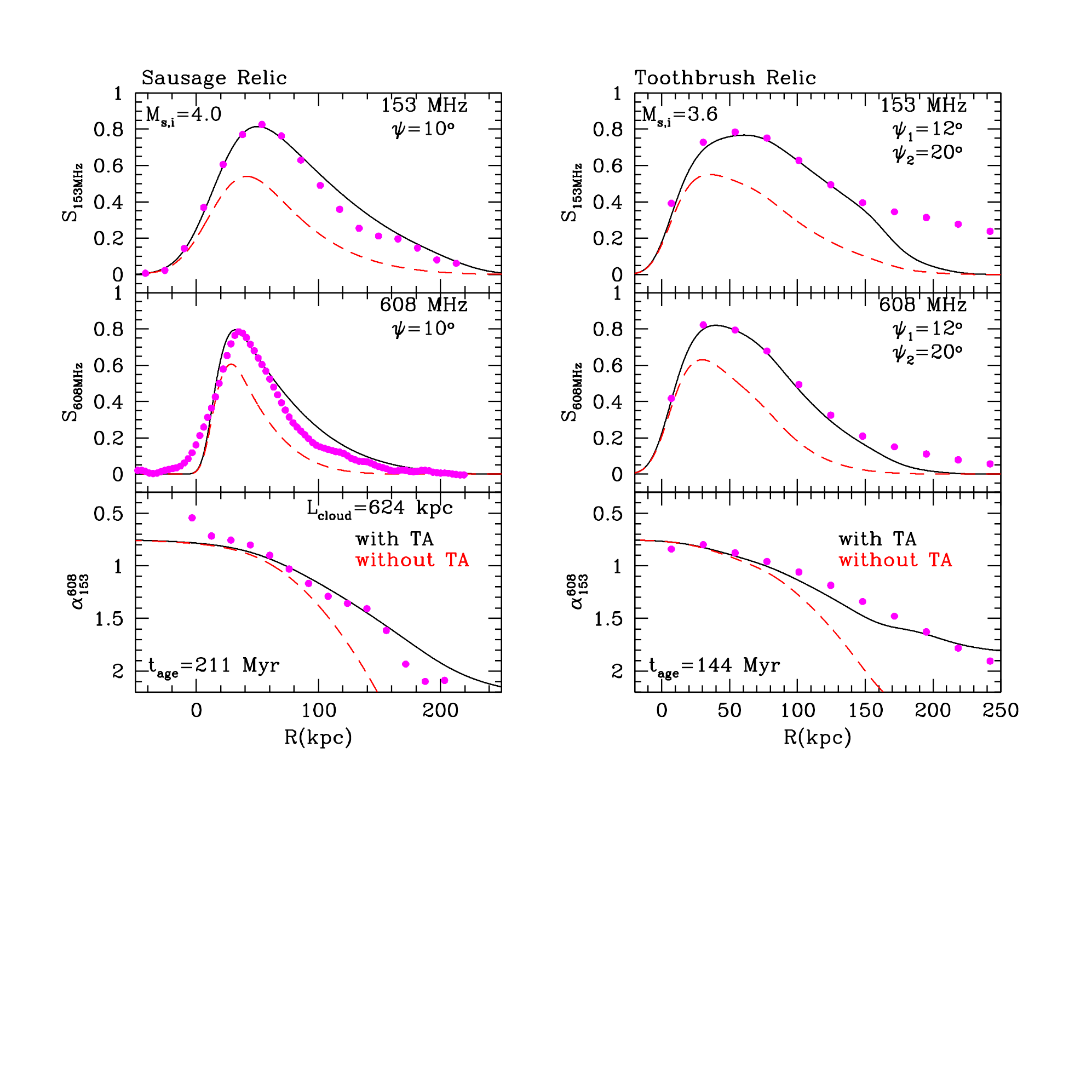}}
\vspace{-3.0cm}
\caption{Radio flux density, $S_{\nu}$, at 150 MHz (top panels) and at 610 MHz (middle panels)
in arbitrary units, and the spectral index, $|\alpha_{150}^{610}|$, between the two frequencies (bottom panels),
plotted as a function of the projected distance behind the shock, $R$ (kpc). The red dashed line is for the model that includes turbulent re-acceleration (TA) in the shock downstream region. The magenta dots are the observational data of the Sausage \citep{2016MNRAS.455.2402S} and the Toothbrush radio shock \citep{2016ApJ...818..204V}.}
\label{fig:SausageTooth_SA}   
\end{figure}

\begin{figure}[htbp]
\centerline{\includegraphics[width=0.55\textwidth]{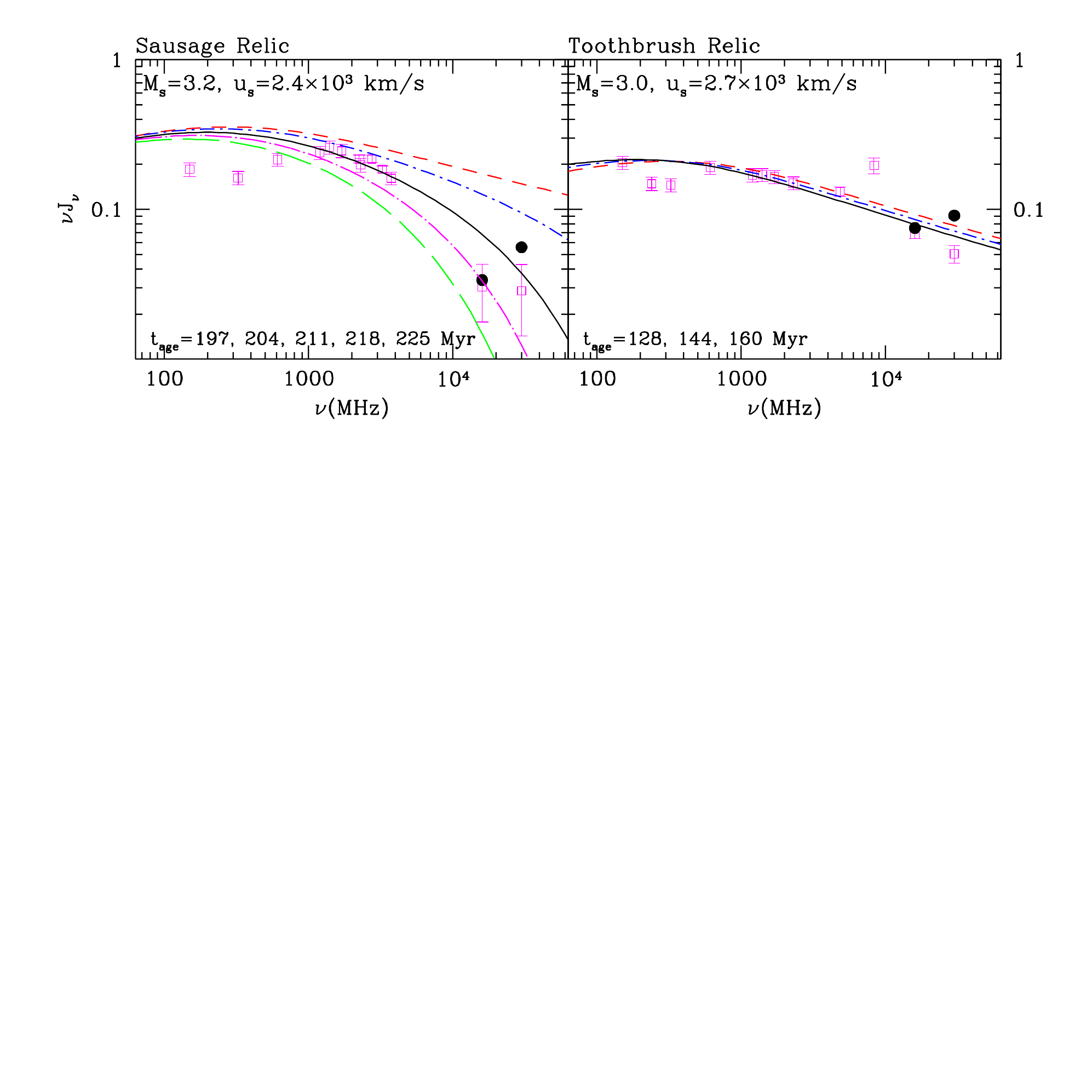}}
\vspace{-5.8cm}
\caption{Time evolution of volume-integrated radio spectrum are shown in chronological order by the red, blue, black, magenta, and green lines for the same two models as shown in Figure~\ref{fig:SausageTooth_SA}.
The open magenta squares and the error bars are the observational data from \cite{2016MNRAS.455.2402S}.
The solid black circles are the data points which could represent the SZ corrected
fluxes \citep{2016A&A...591A.142B}.}
\label{fig:SausageTooth_IS}   
\end{figure}

Figures \ref{fig:SausageTooth_SA} and \ref{fig:SausageTooth_IS} demonstrate that such shock re-acceleration models could reproduce the radio observations of the Sausage and the Toothbrush radio shocks.
In Figure \ref{fig:SausageTooth_SA}, the shock Mach number is $\mathcal{M}_{\rm s}\approx 3.2$ at $t_{\rm age}=211$~Myr for the Sausage radio shock and 
$\mathcal{M}_{\rm s}\approx 3.0$ at $t_{\rm age}=144$~Myr for the Toothbrush radio shock.
The turbulence acceleration time scale $\tau_{\rm acc}\approx 10^8$~yrs is adopted, and the DSA models with (black lines) and without (red lines) post-shock turbulent acceleration are compared.
Note that the model shock Mach number $\mathcal{M}_s$ is slightly higher than $\mathcal{M}_{\rm X}\approx 2.7$ for the Sausage \citep{2015AA...582A..87A}, while $\mathcal{M}_s$ is much higher than $\mathcal{M}_{\rm X}\approx 1.2-1.5$ for the Toothbrush \citep{2016ApJ...818..204V}.
Although alternative models with $\mathcal{M}_s$ closer to $\mathcal{M}_{\rm X}$ could be adopted to explain both radio and X-ray observations, fossil electrons with hard spectra ($s \approx 1-2\alpha_{\rm obs}$) should be present over a large volume in the ICM. But that seems unrealistic because of the fast cooling time scales of GeV electrons.
For the Sausage, the merger scenario itself also remains unclear. Numerical models have trouble to both produce the X-ray properties (like post-shock temperature) and low Mach numbers \citep{2017MNRAS.471.4587D}.

As mentioned in Section~\ref{sec:relicshighfreq},
the steepening above 2~GHz in the integrated spectrum of the Sausage radio shock has yet to be understood \citep{2016MNRAS.455.2402S}.
Re-acceleration of electrons by post-shock Alfv\'enic turbulence \citep{2015ApJ...815..116F} and magnetic field amplification behind the shock \citep{2016MNRAS.462.2014D} have been suggested to explain such steepening at high frequencies. In the model shown in Figure \ref{fig:SausageTooth_IS}, the shock sweeps through a finite region of fossil electrons, resulting in continuous softening of $J_{\nu}$ in time \citep{2016ApJ...823...13K,2017JKAS...50...93K}. The spectrum at 211~Myr (black solid line) shows the best match with the observed radio data.

\subsection{Fossil plasma and CRe re-energization}
\label{sec:fossil}

\begin{figure*}[htbp]
\centering
\includegraphics[width=0.49\textwidth]{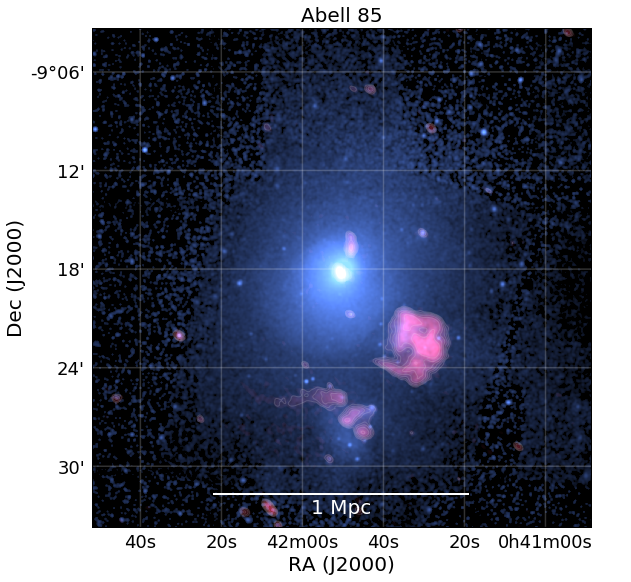}
\includegraphics[width=0.49\textwidth]{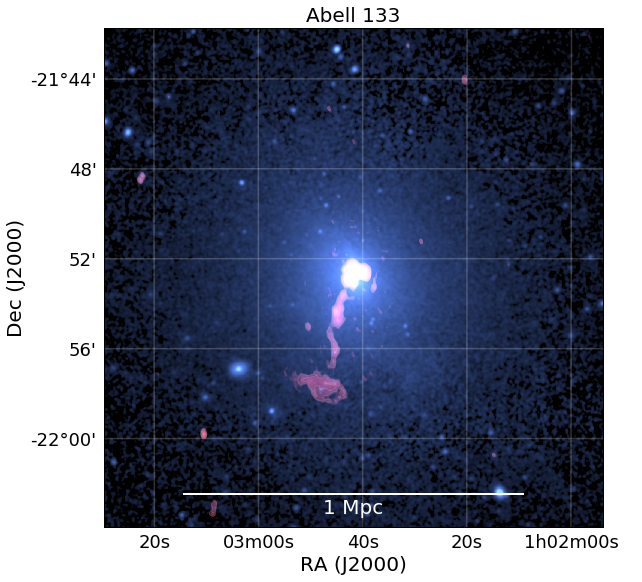}
\includegraphics[width=0.49\textwidth]{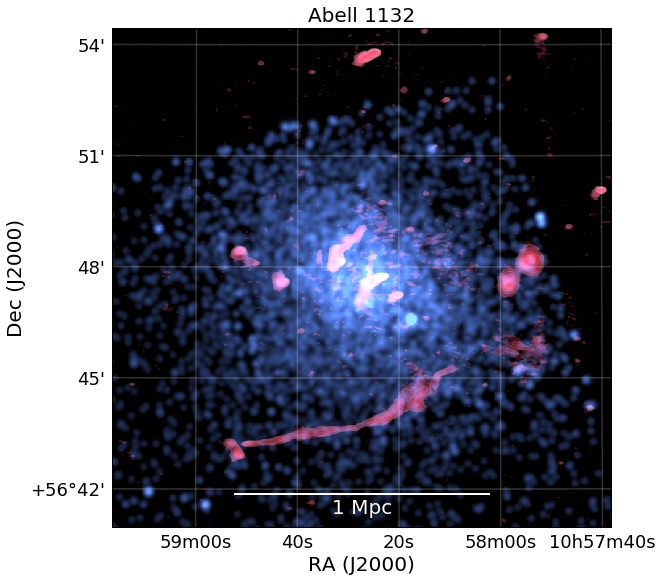}
\includegraphics[width=0.49\textwidth]{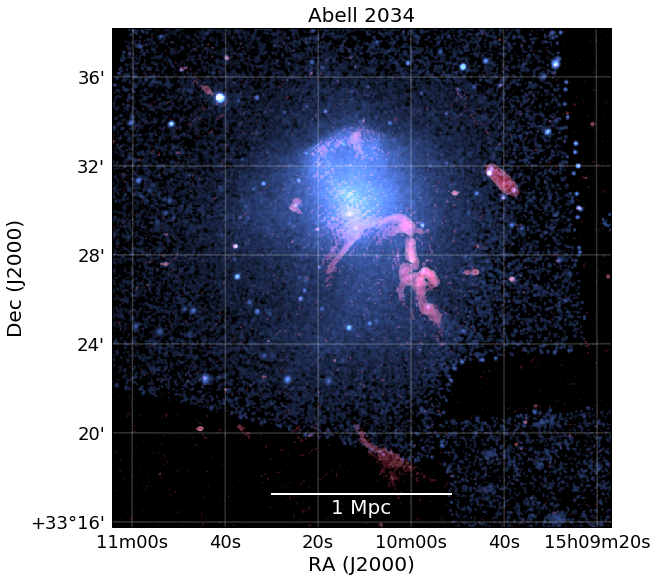}
\caption{Examples of radio phoenices and fossil plasma sources. The radio emission is shown in red and the X-ray emission in blue.  Abell 85: GMRT 148~MHz and Chandra 0.5--2.0~keV \citep{2017ApJ...843...76A}. Abell\,133: GMRT 325~MHz and Chandra 0.5--2.0~keV \citep{2017ApJ...843...76A}. Abell\,1132: LOFAR 144~MHz and Chandra 0.5--2.0~keV \citep{2018MNRAS.473.3536W}. MKW3S: GMRT 322~MHz and Chandra 0.5--2.0~keV \citep{2017ApJ...843...76A}. Abell\,2034: LOFAR 118--166~MHz and Chandra 0.5--2.0~keV \citep{2016MNRAS.459..277S}.}
\label{fig:phoenix}
\end{figure*}

The study of mildly relativistic AGN fossil plasma throughout clusters is and important topic since, as discussed, old populations of relativistic electrons have been invoked as seed particles for the formation of radio halos and cluster radio shocks. They also retrace past AGN activity and constitute a source of non-thermal pressure in the ICM. Examples of radio phoenices and revived fossil plasma sources are shown in Figure~\ref{fig:phoenix}.

\subsubsection{Radio phoenices and revived fossil plasma}

The currently favored scenario is that  phoenices trace old radio plasma from past episodes of AGN activity. When a shock compresses this old  plasma, the resulting increase in the momentum of the relativistic electrons and the magnetic field strength can produce a source characterized by a steep and curved radio spectrum \citep{2001A&A...366...26E}. Simulations also predict that these sources should often have complex morphologies \citep{2002MNRAS.331.1011E}. It should be noted that so far  direct observational evidence for a connection between shocks waves and phoenices is still missing. 
Therefore, the formation scenario for these revived fossil plasma sources remains somewhat uncertain.

\begin{figure*}[htbp]
\centering
\includegraphics[width=0.49\textwidth]{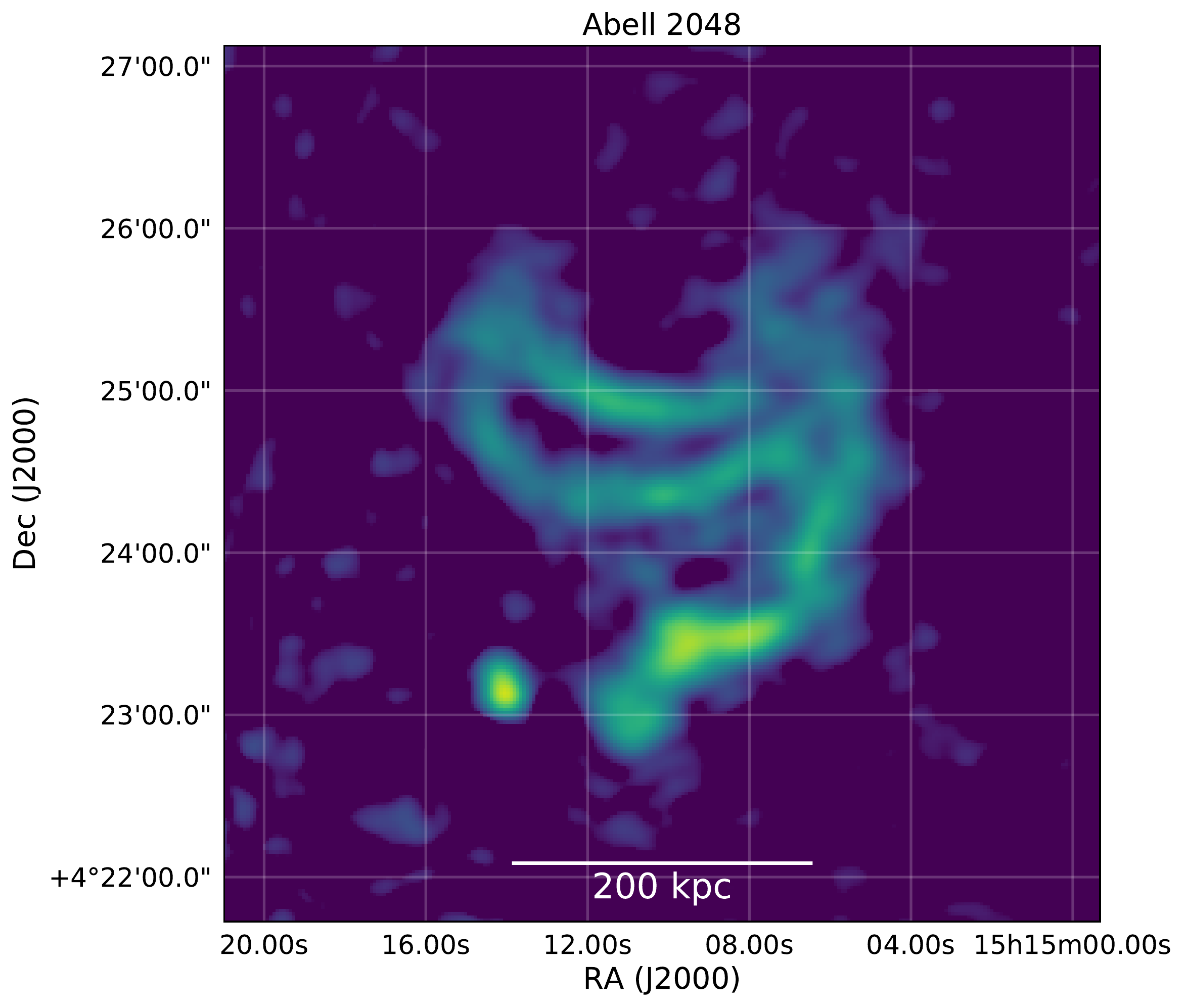}
\includegraphics[width=0.49\textwidth]{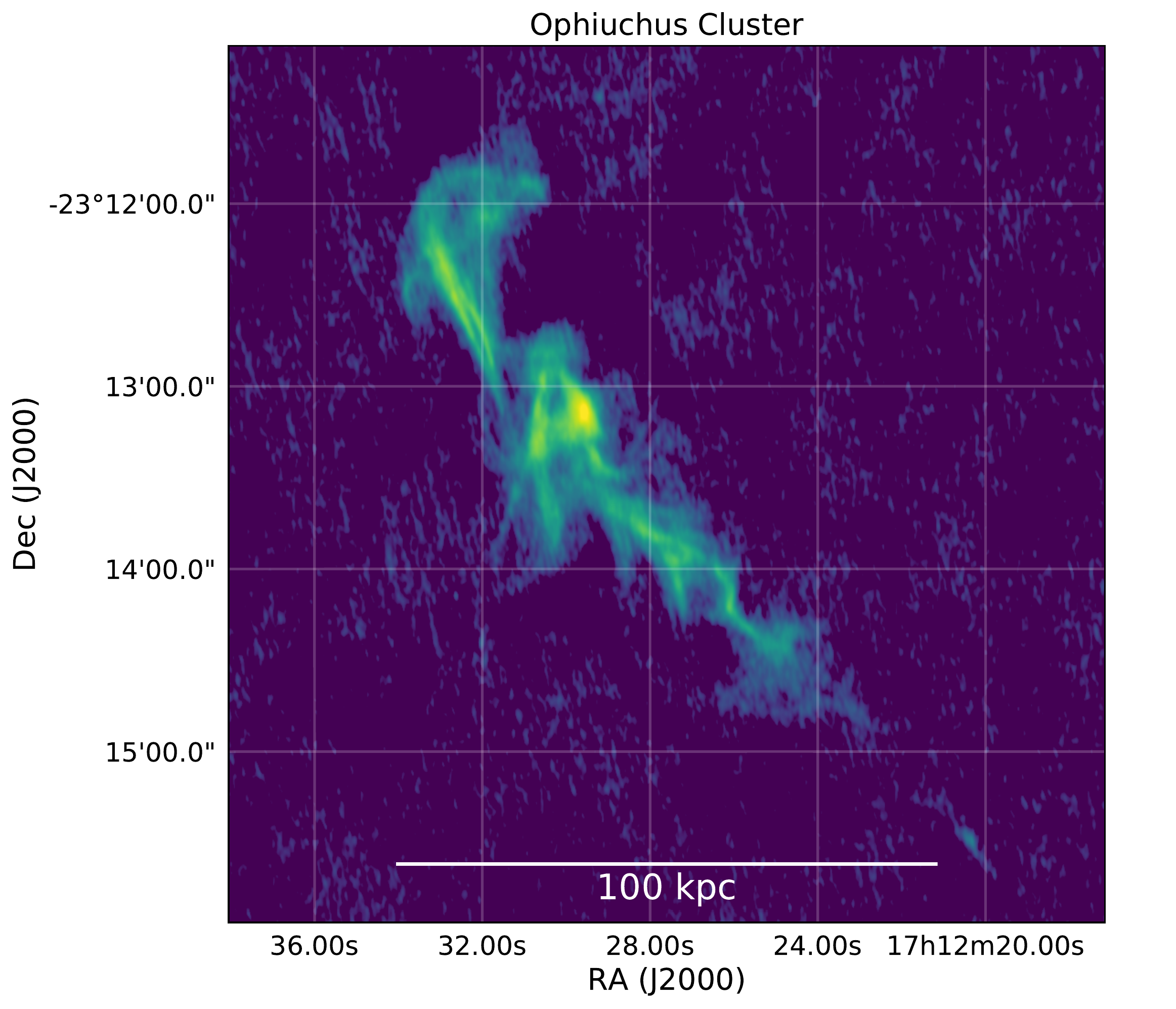}
\includegraphics[width=0.49\textwidth]{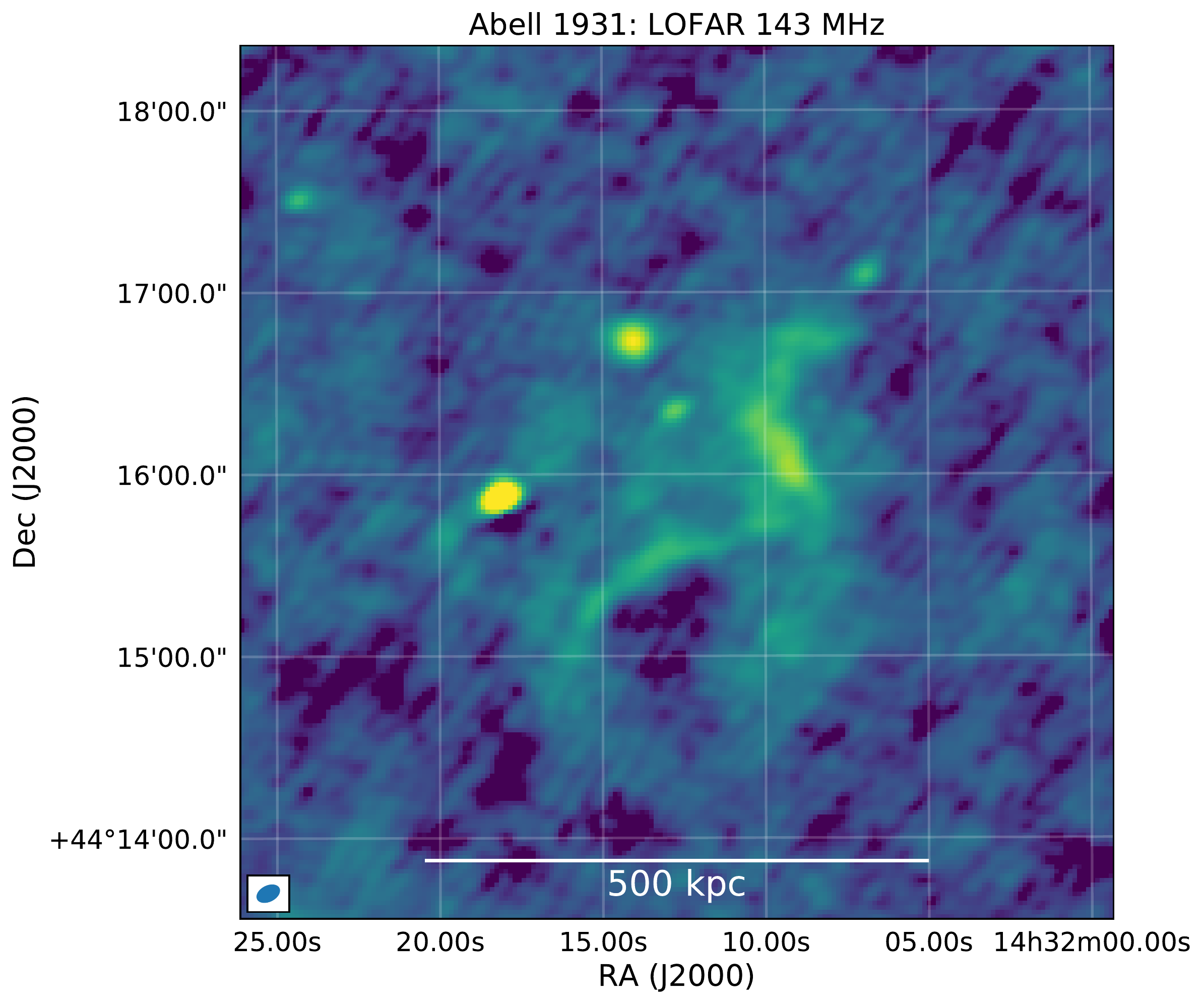}
\includegraphics[width=0.49\textwidth]{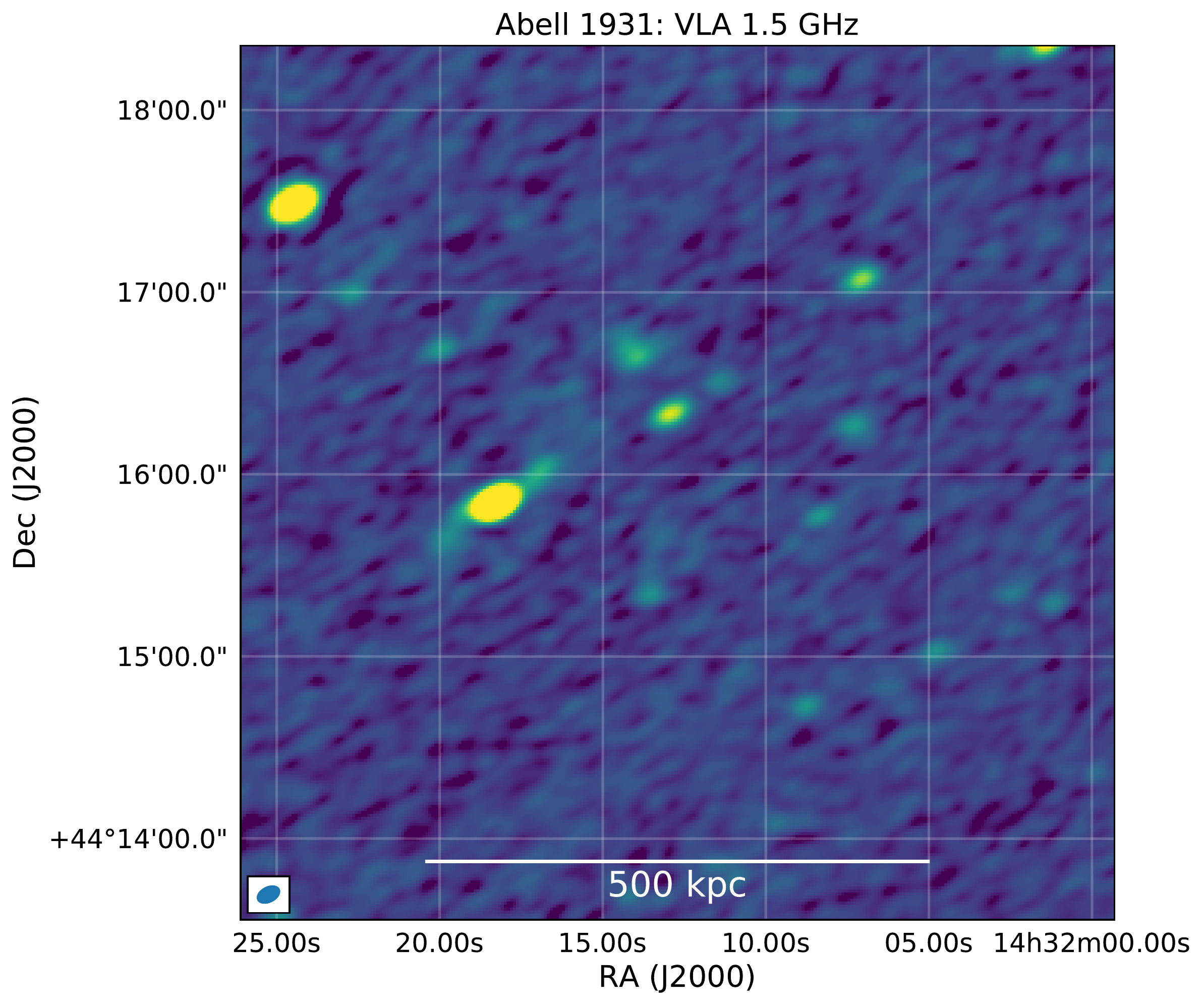}
\caption{\textit{Top panels:} Examples of radio phoenix sources in Abell\,2048 \citep[$z=0.097$; GMRT 325~MHz,][]{2011AA...527A.114V} and the Ophiuchus cluster \citep[$z=0.028$, VLA 1.5~GHz,][]{2016MNRAS.460.2752W}. \textit{Bottom panels:} Radio images of Abell 1931 at 143~MHz (\textit{left panel}) and 1.5~GHz (\textit{right panel}). The elongated source visible in the 143~MHz image is characterized by a very steep radio spectrum, making it invisible at GHz frequencies \citep{2018MNRAS.477.3461B}.}
\label{fig:phoenixexamples}
\end{figure*}

Compared to cluster radio shocks, revived fossil plasma sources and phoenices are on average found at smaller cluster centric distances \citep{2012A&ARv..20...54F}, have smaller sizes ($\lesssim$300--400~kpc, see Figure~\ref{fig:relicsLLSpower}, and have lower radio powers. These revived fossil sources have a range of morphologies, from roundish shapes \citep[e.g., Abell\,1664,][]{1999NewA....4..141G,2001AA...376..803G,2012ApJ...744...46K}) to elongated and filamentary \citep[e.g., Abell\,13, Abell\,85, Abell\,2048, Abell\,4038, Abell\,2443, Abell\,1033\footnote{Not to be confused with the GReET discussed in Section~\ref{sec:greet}.}, Abell\,1914, Abell\,1931, and the Ophiuchus cluster,][]{1983AuJPh..36..101S,2001AJ....122.1172S,2011AA...527A.114V,2015MNRAS.448.2197D,2009AA...508.1269V,2016MNRAS.460.2752W,2010AA...514A..76M,2018MNRAS.477.3461B,2018arXiv181108430M}. The elongated and filamentary morphologies, see Figures~\ref{fig:phoenixexamples}, are the most common \citep[e.g.,][]{1998MNRAS.297L..86S,2001AJ....122.1172S}. Some of these objects are found in cool core clusters such as Abell\,85, Abell\,1664, and Abell\,4038, unlike cluster radio shocks. This indicates that major merger events are not required for their formation.

\begin{figure}[htbp]
\centering
\includegraphics[width=0.49\textwidth]{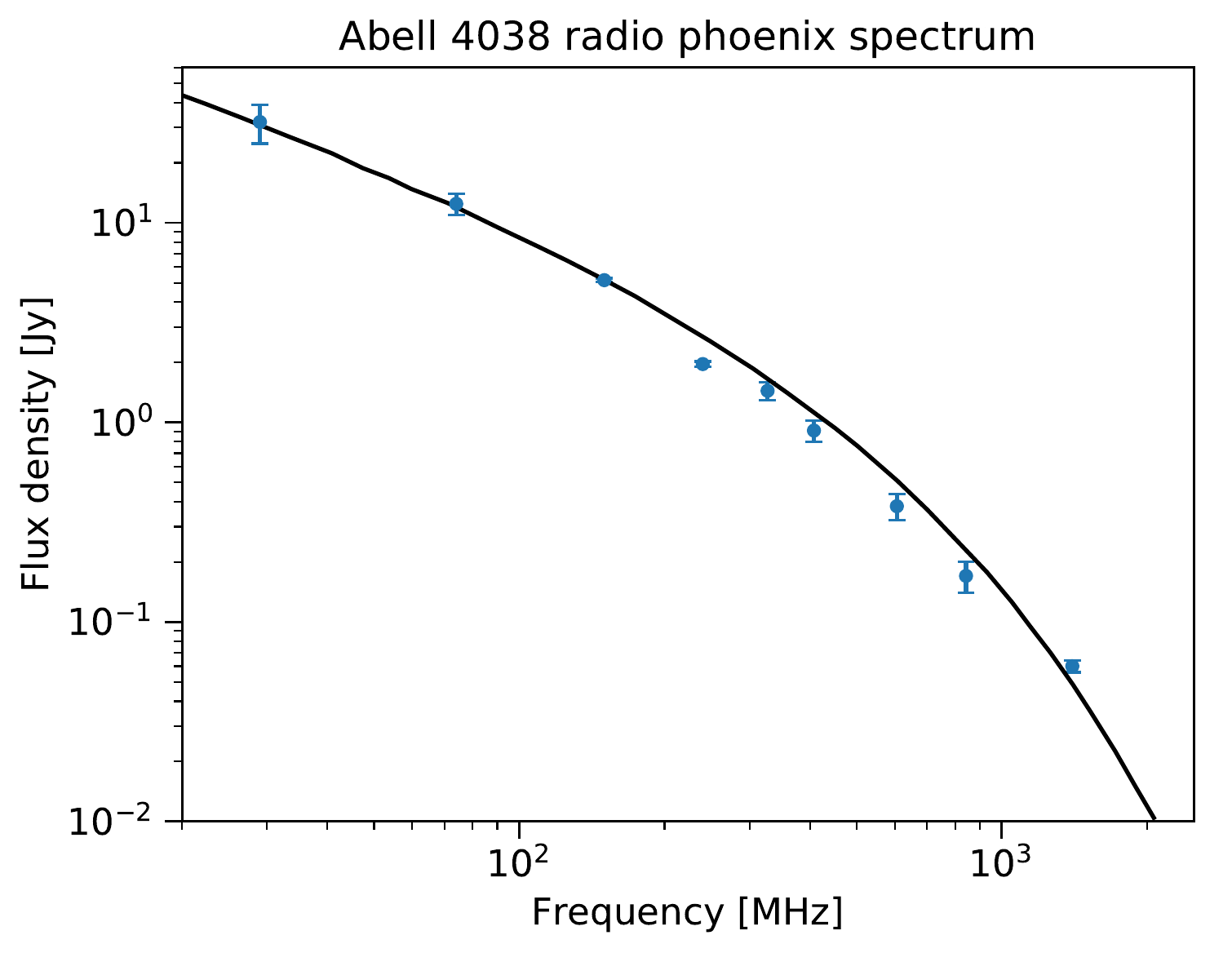}
\caption{Integrated radio spectrum of the radio phoenix in the cluster Abell\,4038 using the data presented in \cite{2012ApJ...744...46K} and references therein. The black line shows an adiabatic compression model fit \citep{2001A&A...366...26E}.}
\label{fig:a4038spectrum}
\end{figure}

Radio phoenices and revived fossil sources have integrated spectra that are typically steeper than $-1.5$. In many instances the spectra are curved \citep{2011AJ....141..149C,2001AJ....122.1172S,2009AA...508...75V}, showing high-frequency spectral steepening, see Figure~\ref{fig:a4038spectrum} for an example. The spectral index distribution across these sources is irregular without clear common trends  \citep{2011AA...527A.114V,2011AJ....141..149C,2012ApJ...744...46K}.

Polarized emission from radio phoenices and revived fossil sources has also been detected. The polarization fractions are generally lower than for cluster radio shocks and show larger variations
\citep[e.g.,][]{2001AJ....122.1172S}. However, it should be remarked that only a few  polarization studies have been performed so far of these sources.

\subsubsection{Re-acceleration and fossil plasma}
As discussed before, DSA shock models proposed for CRe acceleration have found that the acceleration efficiency is often low when electrons are accelerated directly from the thermal pool. This low efficiency is hard to reconcile with the observed brightness and radio spectrum of some cluster radio shocks which suggest a higher acceleration efficiency \citep[e.g.,][]{2011ApJ...734...18K,2014MNRAS.437.2291V}. AGN activity continuously supplies fresh CRs in the ICM creating bright radio galaxies. Due to synchrotron losses, these CRs are visible only for few tens of Myr at Gigahertz frequencies. Although direct observations are prohibitive, a certain amount of CRe with $\gamma \sim 100$ should be present mixed with the ICM \cite{1999ApJ...520..529S,2001ApJ...557..560P,2013MNRAS.435.1061P}. Therefore, CR electrons might be re-accelerated from this seed population in the ICM \citep{1998A&A...332..395E,2005ApJ...627..733M,2011ApJ...734...18K,2012ApJ...756...97K}, mitigating some of the DSA requirements (see also Section~\ref{sec:dsamodel}). {An underlying assumption here is that the jets and lobes of radio galaxies are lepton-dominated \citep[e.g.,][]{2016MNRAS.459...70V}. Otherwise many CR protons would be re-accelerated, possibly causing problems with the \emph{Fermi} gamma-ray upper limits (Section~\ref{sec:gammaradioshock}).}

Instead of fossil ($\gamma \sim 100$) CRe, more energetic CRe from the lobes of a currently active radio galaxy could also re-accelerated \citep{2016ApJ...823...13K}. A few observational pieces of evidence for this scenario were recently reported. In PLCK\,G287.0+32.9, two large radio shocks have been  discovered \citep{2011ApJ...736L...8B,2014ApJ...785....1B}. One of the two radio shocks appears to be connected to the lobes of a radio galaxy. However, no optical counterpart for the radio galaxy could be located and the radio spectral index across the source remains difficult to interpret. In the Bullet cluster (1E\,0657--55.8), a 930~kpc long radio shock is located opposite to the \textit{bullet} direction \citep{2015MNRAS.449.1486S}. In this radio shock, a region of 330~kpc has a much higher surface brightness. This might haven be caused by a pre-existing population of CRe of AGN origin. The best example of CRe of AGN origin re-accelerated by a merger shock comes from Abell\,3411-3412 \citep{2017NatAs...1E...5V,2017NatAs...1E..14J}. In this mering system a morphological connection between a radio galaxy and a radio shock is evident. Both polarization and spectral features are in agreement with particle re-acceleration. Furthermore, X-ray data show the presence of a surface brightness discontinuity at the radio shock's outer edge. However, in the great majority of cases, the presence of a source of CR electrons near the radio shock is missing, leaving unanswered the question: are pre-energized CRe necessary to power all radio shocks? A similar problem is present with radio halos that also require an initial reservoir of mildly energetic CRe to re-energize \citep{2014IJMPD..2330007B}.

With the increase in resolution, sensitivity, and sky coverage of low-frequency telescopes more steep spectrum fossil sources are being discovered. This should shed more light on the connection between diffuse cluster radio sources and AGN fossil plasma in the near future.
It has already become clear that galaxy cluster host sources with such steep spectra that they are completely missed at GHz frequencies. Several of these examples have now been uncovered with LOFAR, such as in Abell\,1033 (see Section~\ref{sec:greet}), Abell\,~1931, and Abell\,2034. Recently, the MWA has also found a significant number of new fossil plasma sources and candidates \citep{2017arXiv170703517D}.

\subsubsection{GReET}
\label{sec:greet}
\begin{figure*}[htbp]
\centering
\includegraphics[width=0.49\textwidth]{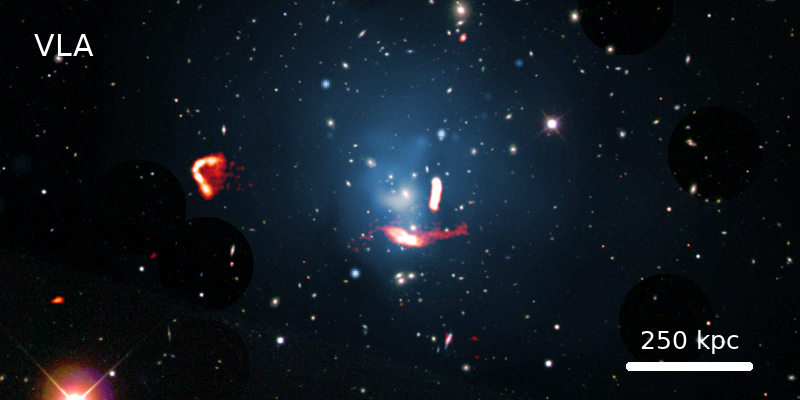}
\includegraphics[width=0.49\textwidth]{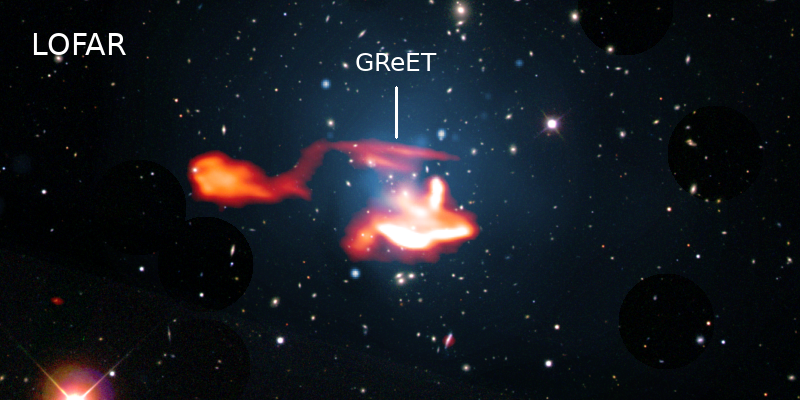}

\caption{Optical image of the galaxy cluster Abell 1033, with over plotted in blue the X-ray emission and in orange the synchrotron radio emission. The left panel shows our view of the galaxy cluster at conventional radio frequencies (VLA at 1.4 GHz). The right panel shows the discovery of the first GReET (gently re-energised radio tail), a new type of radio source visible uniquely at very low frequencies \citep[LOFAR at 140 MHz;][]{2017SciA....3E1634D}. To detect the GReET at 1.4 GHz would require a century of VLA observing time.}\label{fig:abell1033}
\end{figure*}

From observations of extended radio sources in the galaxy cluster Abell\,1033 \citep[see Figure~\ref{fig:abell1033};][]{2017SciA....3E1634D}, the presence of a possible new mechanism to energize old radio plasma was inferred. In this cluster a WAT source fades into a pair of fairly thin filaments within which the emission starts to brighten and the synchrotron spectrum flattens again. This process of re-energisation is so gentle that it barely balances the radiative losses of cosmic rays, with a particle acceleration time-scale comparable to the radiative loss time-scale of the electrons emitting at $<$100~MHz. This source has been labeled ``GReET'' (gently re-energized tail). 

{A proposed physical explanation for the re-energisation mechanism is that Rayleigh-Taylor and Kelvin-Helmholtz instabilities in the tails generate turbulent waves that re-accelerate electrons via second order Fermi mechanisms. The challenge is to understand how the re-acceleration rate is maintained quasi-constant in the tail over a long time-scale. A proposed solution is to assume that turbulence is continuously forced in the tail by the interaction between perturbations in the surrounding medium with the tail itself \citep{2017SciA....3E1634D}. These perturbations are driven in the medium by the cluster dynamics for a time-scale and on spatial-scales that are larger/comparable to that of the GReET.}

If this gentle re-energizing process observed in Abell\,1033 is common in tails of radio galaxies in galaxy clusters, then electrons released by radio galaxies in the ICM could live as long as seen in the case of Abell\,1033 ($>0.5$~Gyr) and they would be able to accumulate in larger quantities and with higher energies. This could produce a seed population of energetic particles for merger-induced re-acceleration mechanisms, such as turbulence and shocks, that were proposed to explain cluster-scale radio sources. Two other possible GReETs are present in ZwCl\,0634.1+4750 \citep{2018AA...609A..61C} and in Abell\,1314 \citep{2018arXiv181107929W}. In both cases a tailed radio galaxy shows an increase in surface brightness along its tail and an unexpected flattening in the spectral index. {Because very few examples of GReETs are known, the precise nature of GReETs and their existence as a distinct class of objects remains to be confirmed.}

\subsubsection{Future prospects}
A vast phenomenology of re-energized plasma of AGN origin has recently been emerging, and it attests to the different mechanisms at play: compression \citep[radio phoenices;][]{2001A&A...366...26E}, Fermi-I shock re-acceleration \citep[cluster radio shocks;][]{2017NatAs...1E...5V}, turbulence \citep[(mini-)halos;][]{2013ApJ...762...78Z} or complex plasma interactions \citep[GReETs;][]{2017SciA....3E1634D}. In most of these cases the re-energization is mild and the radio spectrum is steep, implying that conventional GHz-frequency telescopes overlooked the great majority of these phenomena. 

With the current low-frequency telescopes LOFAR, MWA, and the uGMRT, and the future SKA-low, many more revived fossil plasma sources are to be discovered. This should help to better understand the variety of sources present and their spatial distribution in the ICM. Future low-frequency observations should also reveal more connections with cluster radio shocks and possibly with radio halos. These connections can then be studied in more detail. Particularly interesting will be to push these observations towards to lowest frequencies possible ($\lesssim 10-50$~MHz) as current MWA, LOFAR, and uGMRT observations in the 100--300~MHz range probably only probe the tip of the iceberg.

\section{Diffuse radio emission outside clusters}
\label{sec:radiocosmicweb}

Elongated filaments of galaxies span the regions between clusters. Compared to the intracluster medium, the intergalactic medium of galaxy filaments has a significantly lower density ($\lesssim 10^{-4}$~particles~cm$^{-3}$) and cooler temperature ($10^{5-7}$~K). About half of the Universe’s baryons reside in this WHIM \citep{1999ApJ...514....1C}. Galaxy filaments are expected to be surrounded by strong accretion shocks, where the plasma is first shock-heated \citep{1972A&A....20..189S,2000ApJ...542..608M}. However, studying the WHIM and associated shocks is notoriously difficult due to a lack of sensitive observational tools.

Owing to their high-Mach numbers ($\mathcal{M} \gtrsim 10$), WHIM accretion shocks should accelerate particles \citep{2001ApJ...559...59M,2003ApJ...593..599R,2003ApJ...585..128K}, similar to what happens in clusters. Radio studies of the WHIM would offer a unique diagnostic tool to determine the strength of the WHIM magnetic field and pinpoint the location of the accretion shocks. However, the detection of the very faint radio emission from these shocks around galaxy filaments is difficult. With larger catalogs of polarized sources, deep continuum images, and simulations, some progress has already been made in deriving the properties of magnetic fields beyond clusters in the cosmic web \citep[e.g.,][]{2006ApJ...637...19X,2017MNRAS.468.4246B,2017MNRAS.467.4914V,2015A&A...580A.119V,2017CQGra..34w4001V,2016MNRAS.462..448G}. Deep observations with the SKA and other radio telescopes might have the sensitivity to detect radio emission beyond cluster environments \citep{2012MNRAS.423.2325A,2015A&A...580A.119V} and in low mass systems such as groups. A challenge will be to properly classify such emission, since deep  observations will  also pick up extended low-surface brightness emission associated with (old) radio galaxies.

Despite the expected challenges, some studies have already reported possible extended synchrotron sources from poor clusters and group environments \citep{2017A&A...603A..97N,2009AJ....137.3158B}. 
Also, a candidate for a radio filament was found 5~Mpc away from the cluster Abell\,786 \citep{1991ApJS...76.1055D,1993AJ....105..769H,2000NewA....5..335G}. However, a more recent study suggests that the emission belongs to an old giant radio galaxy \citep{2012ApJ...744...46K}. 
Radio emission surrounding the ZwCl\,2341.1+0000  cluster was reported by \cite{2002NewA....7..249B}. Later studies indicate that the radio radio emission is probably associated with a  cluster merger event \citep{2010AA...511L...5G,2009AA...506.1083V}. Other possible cases of radio emission around clusters associated with accretion shocks (and not from merger events) are the extended radio emission located near MACS\,J0520.7--1328 \citep{2014AA...565A..13M}, Abell\,3444,  \citep{2009AA...507.1257G}, Abell\,2255 \citep{2008AA...481L..91P,2009ApJ...697.1341R},
Abell\,1758N--1758S \citep{2018MNRAS.478..885B}, and MACS\,J0717.5+3745 \citep{2018MNRAS.478.2927B}. Additionally, candidate radio emission connected to large-scale filaments was recently reported based on the SRT observations \citep{2018MNRAS.479..776V}. However, in all the above mentioned cases the nature of these radio sources still needs to be confirmed, requiring additional observations to shed more light on their origin. 


\onecolumn

\tablefirsthead{\toprule \multicolumn{1}{l}{cluster}&\multicolumn{1}{l}{z} & \multicolumn{1}{l}{classification} &\multicolumn{1}{l}{notes} &\multicolumn{1}{l}{references}    \\ \midrule}
\tablehead{%
\multicolumn{2}{c}%
{{\bfseries  Continued from the previous page}} \\
\toprule
\multicolumn{1}{l}{cluster}&\multicolumn{1}{l}{z} & \multicolumn{1}{l}{classification} &\multicolumn{1}{l}{notes} &\multicolumn{1}{l}{references} \\ \midrule}
\tabletail{%
\midrule \multicolumn{2}{r}{{Continued on the next page}} \\ \midrule}
\tablelasttail{%
\\\midrule
\multicolumn{2}{r}{{  }} \\ \bottomrule}

\topcaption{Clusters with diffuse radio emission}
\renewcommand\tabcolsep{3pt}
{\scriptsize 
\begin{supertabular}{lllll}
 Abell\,S753                       & 0.014 & RS/F & & \cite{1987MNRAS.226..979G,2003AJ....125.1095S} \\   
Perseus cluster & 0.018 & MH && \cite{1975AA....45..223M,1982Natur.299..597N} \\ 
& & & & \cite{1990MNRAS.246..477P,1992ApJ...388L..49B,sijbring_phd} \\
& & & & \cite{1998AA...331..901S,2011AA...526A...9B} \\
& & & & \cite{2017MNRAS.469.3872G} \\
Abell\,1367   & 0.022 & RS & &\cite{1978AA....69..355G,1980AJ.....85.1565H,1981AA...100..323B} \\  
& & & & \cite{1983ApJ...270..410G,1987AA...186L...1G} \\
& & & & \cite{2013ApJ...779..189F} \\
Coma cluster$^a$           & 0.023 & H, RS & &\cite{1959Natur.183.1663L,1970MNRAS.151....1W,1979ApJ...233..453J} \\ 
& & & & \cite{1981AA...100..323B,1984AA...133..252A} \\
& & & & \cite{1985AA...150..302G,1987AA...182...21S} \\
& & & & \cite{1991AA...252..528G,1993ApJ...406..399G} \\
& & & & \cite{1997AA...321...55D,2003AA...397...53T,2010PhDT.......259P} \\
& & & & \cite{2011MNRAS.412....2B} \\
Ophiuchus cluster & 0.028 & MH, F & & \cite{2009AA...499..679M,2009AA...499..371G,2010AA...514A..76M} \\ 
& & & & \cite{2016MNRAS.460.2752W} \\
Abell\,4038  & 0.030 & F  && \cite{1984PASAu...5..516S,1994AuJPh..47..145S,1998MNRAS.297L..86S}\\    
& & & & \cite{2001AJ....122.1172S,2012ApJ...744...46K,2018MNRAS.480.5352K} \\
& & & & \cite{2017arXiv170703517D} \\
2A\,0335+096$^b$ & 0.036 & MH & & \cite{1995ApJ...451..125S} \\ 
 Abell\,1314 & 0.034&   F & &\cite{2009ApJ...697.1341R,2018arXiv181107929W} \\ 
 Abell\,548b                       & 0.045 & RS/F && \cite{1999NewA....4..141G,2006AN....327..563G,2006MNRAS.368..544F}  \\       
 & & & & \cite{2017MNRAS.467..936G} \\
Abell\,168 & 0.045 & mRS & & \cite{2018MNRAS.477..957D} \\ 
Abell\,3376                       & 0.046 & dRS & & \cite{2006Sci...314..791B,2012MNRAS.426.1204K,2015MNRAS.451.4207G} \\ 
Abell\,1213 &0.047 & cH & & \cite{2009AA...507.1257G} \\ 
Abell\,3562 & 0.049 & H && \cite{2000MNRAS.314..594V,2003AA...402..913V,2005AA...440..867G} \\ 
Abell\,754  & 0.054 &  H, RS/F &  &\cite{1977Natur.266..239W,1978MNRAS.185P..51M,1980AA....90..283H}  \\ 
& & & & \cite{1999NewA....4..141G,2001ApJ...559..785K} \\
& & & & \cite{2003AA...400..465B,2009ApJ...699.1883K} \\
& & & & \cite{2011ApJ...728...82M} \\
Abell\,85 & 0.055  & F  && \cite{1984PASAu...5..516S,1994AuJPh..47..145S} \\ 
& & & & \cite{2000NewA....5..335G,2017arXiv170703517D} \\
Abell\,2626 & 0.055& U & previously classified as MH & \cite{2004AA...417....1G,2013MNRAS.436L..84G,2017AA...604A..21I}  \\ 
& & & & \cite{2017MNRAS.466L..19K} \\
Abell\,3667   & 0.056&  dRS &  non-confirmed MH  & \cite{1975MmRAS..79....1S,1982MNRAS.198..259G}\\ 
& & & & \cite{1992ApJS...80..137J,1997MNRAS.290..577R} \\
& & & & \cite{2003PhDT.........3J,2015MNRAS.447.1895R} \\
& & & & \cite{2014MNRAS.445..330H,2017arXiv170604930J}\\
Abell\,2319 & 0.056 & H  & & \cite{1978AAS...34..117H,1997NewA....2..501F}\\ 
& & & & \cite{1999NewA....4..141G,2001ApJ...548..639K}\\
& & & & \cite{2013ApJ...779..189F,2015MNRAS.448.2495S} \\
Abell\,133   & 0.057 & cF & &\cite{1984PASAu...5..516S,2001AJ....122.1172S,1994AuJPh..47..145S,2001AJ....122.1172S}\\ 
& & & & \cite{1999NewA....4..141G,2010ApJ...722..825R} \\
& & & & \cite{2017arXiv170703517D} \\
Abell\,2256 & 0.058 & H, RS, mF & & \cite{1976AA....52..107B,1978MNRAS.185..607M} \\
& & & & \cite{1979AA....80..201B,1994ApJ...436..654R,1999JKAS...32...75K} \\
& & & & \cite{1999NewA....4..141G,2001ApJ...548..639K}\\
& & & & \cite{2006AJ....131.2900C,2008AA...489...69B}\\
& & & & \cite{2009AA...508.1269V,2009ApJ...697.1341R} \\ 
& & & & \cite{2010ApJ...718..939K,2012AA...543A..43V}\\
& & & & \cite{2014ApJ...794...24O,2015AA...575A..45T,2015PASJ...67..110O} \\
RXC\,J0225.1--2928   & 0.060  & cRS & & \cite{2016MNRAS.459.2525S} \\     
Abell\,1795 & 0.062 & cMH & &\cite{2014ApJ...781....9G}\\ 
CIZA\,J0649.3+1801    & 0.064 & cRS && \cite{2011AA...533A..35V} \\  
CL\,0217$+$70  & 0.065$^c$ & dRS, H  && \cite{2011ApJ...727L..25B} \\ 
RXC\,J1053.7+5452                   & 0.070 & RS && \cite{2009ApJ...697.1341R,2011AA...533A..35V}\\           
Abell\,399 & 0.072 & H & close to Abell\,401 & \cite{2010AA...509A..86M} \\  
Abell\,2065 & 0.072 &cH & &\cite{2013ApJ...779..189F}\\ 
Abell\,401 &0.074  & H & close to Abell\,399 &  \cite{1980AAS...39..215H,1981AA...100....7R,1999NewA....4..141G}  \\ 
& & & & \cite{2000NewA....5..335G,2003AA...400..465B} \\
& & & & \cite{2010AA...509A..86M} \\
Abell\,2067&0.074 &cR & & \cite{2013ApJ...779..189F}\\ 
ZwCl\,1742.1+3306 & 0.076& cMH &  &\cite{2014ApJ...781....9G} \\ 
Abell\,2029 & 0.077  & MH & &\cite{2009AA...499..679M,2009AA...499..371G} \\ 
Abell\,2061                       & 0.078 & RS, cH &    & \cite{2001ApJ...548..639K,2009ApJ...697.1341R} \\ 
& & & & \cite{2011AA...533A..35V,2013ApJ...779..189F} \\
Abell\,2255 & 0.081 & H, cRS, mF & cUSS  & \cite{1979ApJ...233..453J,1980AAS...39..215H,1995ApJ...446..583B}\\
 & & & & \cite{1997AA...317..432F,1999NewA....4..141G} \\
 & & & & \cite{2001ApJ...548..639K,2005AA...430L...5G} \\
 & & & & \cite{2008AA...481L..91P,2009AA...507..639P} \\
 & & & & \cite{2009ApJ...697.1341R,2011AA...525A.104P} \\
Abell\,2556  & 0.087  & cF & cUSS &  \cite{2017arXiv170703517D} \\
Abell\,478 & 0.088 & MH& &\cite{2014ApJ...781....9G,2018arXiv181108410S} \\  
Abell\,2142 & 0.089 & H, cMH & H-MH & \cite{1999NewA....4..141G,2000NewA....5..335G} \\ 
& & & & \cite{2013ApJ...779..189F,2017AA...603A.125V} \\
Abell\,725                        & 0.092  & cRS && \cite{2001ApJ...548..639K} \\    
Abell\,3365                       & 0.093  & dcRS & at least one RS present  & \cite{2011AA...533A..35V}\\   
Abell\,13      & 0.094 & F &&\cite{1984PASAu...5..516S,1994AuJPh..47..145S} \\ 
& & & & \cite{1999NewA....4..141G} \\
& & & & \cite{2001AJ....122.1172S,2017MNRAS.467..936G} \\
& & & & \cite{2017arXiv170703517D} \\
Abell\,610     & 0.095 &  RS&& \cite{2000NewA....5..335G}\\ 
Abell\,2048    & 0.097 & F & & \cite{2009AA...508...75V,2011AA...527A.114V}\\  
PKS\,0745--191 &0.103 & cMH &  &\cite{1991MNRAS.250..737B,2004AA...417....1G}\\
& & & & \cite{2007AA...463..937V} \\
ZwCl\,0008.8+5215  & 0.103 & dRS & & \cite{2011AA...528A..38V,2017AA...600A..18K}\\
& & & & \cite{2017ApJ...838..110G} \\
Abell\,523 &  0.104 & cH/cRS &    &\cite{2011AA...530L...5G,2011AA...533A..35V} \\ 
& & & & \cite{2016MNRAS.456.2829G} \\
Abell 2798 &  0.105 & RS && \cite{2017arXiv170703517D} \\
CIZA\,J0107.7+5408 & 0.107 &  cDR, cH, cF & diffuse emission present & \cite{2011AA...533A..35V,2016ApJ...823...94R}\\ 
Abell 2751 &  0.107 & RS && \cite{2017arXiv170703517D} \\
Abell S0084 & 0.108 & cH & &  \cite{2017arXiv170703517D} \\
Abell\,2443   & 0.108  & F/RS & complex & \cite{2011AJ....141..149C}\\   
Abell\,2811 &  0.108 & H &  &    \cite{2017arXiv170703517D}\\
Abell\,2034 & 0.113  &  cF, cRS, cH  &  diffuse emission present  & \cite{2001ApJ...548..639K,2009AA...507.1257G} \\ 
& & & & \cite{2011AA...533A..35V,2009ApJ...697.1341R} \\
& & & & \cite{2016MNRAS.459..277S} \\
Abell\,2721 & 0.114 & cRS & &    \cite{2017arXiv170703517D} \\
Abell\,2069 & 0.115 & H & &\cite{2013ApJ...779..189F,2015AA...575A...8D}\\ 

Abell\,2496  &0.122 &cRS &&   \cite{2017arXiv170703517D}  \\
Abell\,1664                       & 0.128   & RS/F &  &\cite{1999NewA....4..141G,2001AA...376..803G} \\  
& & & & \cite{2012ApJ...744...46K} \\
Abell\,1033                       & 0.130  & F, GReET && \cite{2009ApJ...697.1341R,2015MNRAS.448.2197D} \\ 
& & & & \cite{2017SciA....3E1634D} \\
Abell\,1132 &  0.137 & H & cUSS& \cite{2009ApJ...697.1341R,2018MNRAS.473.3536W} \\ 
Abell\,1068 & 0.138  &  cMH  & &\cite{2009AA...499..371G} \\ 
Abell\,22 & 0.142 & cH/cRS & & \cite{2017arXiv170703517D} \\ 
Abell\,1413 & 0.143   & MH & & \cite{2009AA...499..371G,2018arXiv181108410S} \\ 

24P73  & 0.150$^d$   & F && \cite{2009AA...508...75V,2011AA...527A.114V}  \\ 
Abell\,3888 &  0.151 & H & & \cite{2016MNRAS.459.2525S} \\ 
Abell\,2204 &  0.152 & MH&&\cite{2014ApJ...781....9G}\\ 
Abell\,545 &0.154 &H && \cite{1999NewA....4..141G,2003AA...400..465B} \\ 
Abell\,1240                       & 0.159 & dRS &  & \cite{2001ApJ...548..639K,2009AA...494..429B}\\  
& & & & \cite{2018MNRAS.478.2218H} \\
WHL\,J143150.1+133205 & 0.160    & F & & \cite{2009AA...508...75V,2011AA...527A.114V,2011MNRAS.414.1175O}   \\ 
& & & & \cite{2015AA...583A..89S} \\
Abell\,3411--3412                       & 0.162  & H, mRS  & &\cite{2013ApJ...769..101V,2013MNRAS.435..518G}\\ 
& & & & \cite{2017NatAs...1E...5V} \\
RXC\,J1720.1+2637 & 0.164 & MH & H-MH& \cite{2008ApJ...675L...9M,2014ApJ...795...73G} \\ 
& & & & \cite{2018arXiv181108410S}\\
Abell\,2294 &0.169 & cH  &  &\cite{1999dtrp.conf....9O,2009AA...507.1257G} \\ 
Abell\,907 & 0.167 & MH & & \cite{2017ApJ...841...71G} \\ 

Abell\,1914 & 0.171& F, cH &   & \cite{1985AA...148..323R,1999NewA....4..141G}\\ 
& & & & \cite{2001ApJ...548..639K,2003AA...400..465B} \\
& & & & \cite{2009ApJ...697.1341R,2018arXiv181108430M} \\
Abell\,2218 & 0.171 & H & & \cite{1999NewA....4..141G,2000NewA....5..335G}\\ 
& & & & \cite{2001ApJ...548..639K,2009ApJ...697.1341R} \\
Abell\,2073&0.172 &cR & & \cite{2013ApJ...779..189F}\\ 
Abell\,2693 & 0.173 &  cH & & \cite{2017arXiv170703517D} \\
ZwCl\,0634.1+4750 & 0.174 &H &&\cite{2018AA...609A..61C} \\ 
Abell\,2680  & 0.177& cH & &  \cite{2017arXiv170703517D} \\
Abell\,2345   & 0.177  & dRS  & & \cite{1999NewA....4..141G,2009AA...494..429B} \\  
& & & & \cite{2017MNRAS.467..936G} \\
Abell\,1931 &  0.178& cF && \cite{2018MNRAS.477.3461B}  \\
Abell\, 2254 & 0.178 & H & &\cite{1999NewA....4..141G,2001AA...376..803G} \\ 
& & & & \cite{2017MNRAS.467..936G} \\

Abell\,1612                       & 0.179  & RS/F &  & \cite{2011AA...533A..35V,2017AA...600A..18K} \\        

Abell\,665  & 0.182 &H & &\cite{1999NewA....4..141G,2000NewA....5..335G}  \\ 
& & & & \cite{2001ApJ...548..639K,2004AA...423..111F} \\
& & & & \cite{2010AA...514A..71V,2009ApJ...697.1341R} \\
Abell\,1689 & 0.183 & H & & \cite{2011AA...535A..82V} \\ 
CIZA\,J2242.8+5301$^e$  & 0.192  & F, mdRS, H &   & \cite{2010Sci...330..347V,2013AA...555A.110S,2016MNRAS.455.2402S}\\
& & & & \cite{2017AA...600A..18K,2017MNRAS.472.3605L} \\
& & & & \cite{2018ApJ...865...24D} \\
Abell\,115 &  0.197& RS & & \cite{1999NewA....4..141G,2001AA...376..803G} \\  
& & & & \cite{2016MNRAS.460L..84B,2018ApJ...859...44H} \\
Abell\,1451 & 0.199& H &&\cite{2018AA...609A..61C} \\ 

Abell 3527-bis & 0.200& RS && \cite{2017AA...597A..15D} \\ 
Abell\,2163 & 0.203 & H, RS & &\cite{1994AAS...185.5307H,1999NewA....4..141G} \\ 
& & & & \cite{2001AA...373..106F,2004AA...423..111F,2017MNRAS.467..936G} \\
& & & & \cite{2018AA...619A..68T} \\
Abell\,520   &0.203 & H, cmRS & & \cite{1999NewA....4..141G,2001AA...376..803G}  \\ 
& & & & \cite{2003AA...400..465B,2014AA...561A..52V} \\
& & & & \cite{2018ApJ...856..162W,2018arXiv181109713H} \\
Abell\,910   & 0.206  & RS && \cite{2012AA...545A..74G}\\  
Abell\,209 & 0.206 & H&& \cite{1999NewA....4..141G,2006AN....327..563G,2007AA...463..937V}\\ 
& & & & \cite{2009AA...507.1257G,2013AA...551A..24V} \\
RXC\,J1504.1--0248 & 0.215 & MH & & \cite{2011AA...525L..10G}  \\ 
Abell\,773  & 0.217 & H & & \cite{1999NewA....4..141G,2001ApJ...548..639K} \\ 
& & & & \cite{2001AA...376..803G,2009ApJ...697.1341R} \\

PLCK\,G200.9--28.2 & 0.220 & RS && \cite{2017MNRAS.472..940K}\\ 
Abell\,800 &0.222 &  H&& \cite{2012AA...545A..74G} \\ 
RXC\,J1514.9--1523 & 0.223 & H & USS & \cite{2011AA...534A..57G} \\ 
Abell\,2261 & 0.224 & H &  & \cite{2008AA...484..327V,2017MNRAS.466..996S}   \\ 
& & & & \cite{2017ApJ...849...59B,2018arXiv181108410S}\\
{[VRI2012]}\,Toothbrush$^f$ & 0.225 & mRS, cF, H & & \cite{2012AA...546A.124V,2016MNRAS.455.2402S}  \\ 
& & & & \cite{2016ApJ...818..204V,2017AA...600A..18K} \\
& & & & \cite{2018ApJ...852...65R} \\
Abell\,2667 & 0.226 & MH & & \cite{2017ApJ...841...71G} \\ 
Abell\,1682 & 0.226 & cH, cRS, cmFS & diffuse emission present & \cite{2008AA...484..327V,2013AA...551A.141M}\\ 
& & & & \cite{2013AA...551A..24V,2009ApJ...697.1341R} \\
Abell\,2219 & 0.228 & H & & \cite{1999NewA....4..141G,2001ApJ...548..639K}\\ 
& & & & \cite{2003AA...400..465B,2007AA...467..943O}\\
RXC\,J1234.2+0947$^g$ & 0.229  & cH, RS & & \cite{2015AA...579A..92K}  \\ 
Abell\,141    & 0.230 & H & USS &\cite{2017arXiv170703517D}\\
Abell\,2146 & 0.232 & cH, cDRS & diffuse emission present &\cite{2018MNRAS.475.2743H} \\ 
& & & & \cite{2018arXiv181109708H} \\
Abell\,746 & 0.232 & H, RS &&\cite{2011AA...533A..35V} \\ 
Abell\,2390 & 0.233  &\ldots & non-confirmed H& \cite{1999NewA....4..141G,2003AA...400..465B}\\ 
& & & & \cite{2017MNRAS.466..996S,2018arXiv181108410S} \\
Abell\,33 & 0.234 & cRS & &    \cite{2017arXiv170703517D} \\ 
RX\,J2129.6+0005 & 0.235 & MH & & \cite{2015AA...579A..92K}\\ 
Abell\,S780 & 0.236 & MH   & & \cite{2015AA...579A..92K}  \\ 

RXC\,J1314.4--2515  & 0.247 & H, dRS&& \cite{2005AA...444..157F,2007AA...463..937V,2013AA...551A..24V}\\ 
& & & & \cite{2017MNRAS.467..936G} \\
Abell\,521  & 0.248 & H, RS &USS &  \cite{2006AA...446..417F,2007AA...463..937V} \\  
& & & & \cite{2009ApJ...699.1288D,2009AA...507.1257G} \\
& & & & \cite{2008Natur.455..944B,2008AA...486..347G} \\
& & & & \cite{2013AA...551A..24V,2013AA...551A.141M} \\
RXC\,J2351.0--1954 &0.248 & cdRS, cH &&  \cite{2017arXiv170703517D} \\
Abell\,3444 & 0.253 & MH  & &  \cite{2007AA...463..937V,2009AA...507.1257G} \\ 
& & & & \cite{2015AA...579A..92K} \\
Abell\,1835 & 0.253 & MH & & \cite{2009AA...499..679M,2009AA...499..371G} \\  
Abell\,1550 & 0.254 &H & & \cite{2012AA...545A..74G}\\ 
ZwCl~1454.8+2233$^h$ & 0.258 & MH && \cite{2008AA...484..327V} \\ 

CIZAJ1938.3+5409 &0.260 & cH  && \cite{2015MNRAS.454.3391B}  \\ 
PSZ1\,G139.61+24.20& 0.267 &  MH & H-MH & \cite{2017ApJ...841...71G,2018MNRAS.478.2234S}\\  
Abell\,1443 & 0.269  &cH, cRS & & \cite{2015MNRAS.454.3391B} \\ 
PSZ1\,G171.96--40.64 & 0.270 & H & cUSS & \cite{2013ApJ...766...18G} \\ 
ZwCl\,2341.1+0000                   & 0.270   & dRS, cH  && \cite{2002NewA....7..249B,2009AA...506.1083V}   \\ 
& & & & \cite{2010AA...511L...5G,2017ApJ...841....7B} \\

Abell\,1758N$^i$   &0.280 & H  &close to Abell\,1758S  & \cite{1999NewA....4..141G,2006AN....327..563G,2009AA...507.1257G} \\ 
& & & & \cite{2001ApJ...548..639K}\\
& & & & \cite{2009ApJ...697.1341R,2013AA...551A..24V} \\
& & & & \cite{2018MNRAS.478..885B} \\
Abell\,1758S$^j$   &0.280 & H, cRS  & close to Abell\,1758N  & \cite{2018MNRAS.478..885B}\\ 
Abell\,697  & 0.282 & H & USS &  \cite{2001ApJ...548..639K,2008AA...484..327V} \\ 
& & & & \cite{2009AA...507.1257G} \\
& & & & \cite{2009ApJ...697.1341R} \\
& & & & \cite{2010AA...517A..43M,2011AA...533A..35V} \\
& & & & \cite{2013AA...551A..24V,2013AA...551A.141M} \\
ZwCl\,1021.0+0426$^k$ & 0.291  & MH & & \cite{2014ApJ...781....9G,2015AA...579A..92K} \\ 
RXC\,J1501.3+4220 & 0.292 & H & & \cite{2018arXiv181107929W} \\
Bullet cluster$^l$      & 0.296     & H, RS & & \cite{2000ApJ...544..686L,2014MNRAS.440.2901S,2015MNRAS.449.1486S}   \\
Abell\,781  & 0.298 & cRS/cF & non-confirmed H & \cite{2008AA...484..327V,2009ApJ...697.1341R} \\ 
& & & & \cite{2011AA...529A..69G,2011MNRAS.414L..65V,2013AA...551A..24V} \\
& & & & \cite{2018arXiv181107930B} \\

PSZ1 G096.89+24.17$^m$ &0.300 & dRS, cH && \cite{2014MNRAS.444.3130D}\\ 
SPT-CL\,J0245-5302 &  0.300& cdRS & &\cite{2018MNRAS.479..730Z} \\ 
Abell\,2552 & 0.305  &   cH & &\cite{2015AA...579A..92K}\\ 
Abell\,2744 & 0.308 & H, mRS & &\cite{1999NewA....4..141G,2001AA...376..803G} \\ 
& & & & \cite{2003AA...400..465B,2007AA...467..943O} \\
& & & & \cite{2013AA...551A..24V,2017ApJ...845...81P} \\
& & & & \cite{2017MNRAS.467..936G,2017arXiv170703517D}\\
Abell\,1300 &  0.308 & H, RS, cRS & USS  &\cite{2013AA...551A..24V,1999MNRAS.302..571R} \\ 
& & & & \cite{1999NewA....4..141G,2017MNRAS.464.2752P} \\
RXC\,J2003.5--2323 & 0.317 & H & &\cite{2007AA...463..937V,2009AA...505...45G}\\ 
& & & & \cite{2013AA...551A..24V} \\
Abell\,1995 & 0.318 & H && \cite{1999dtrp.conf....9O,2009AA...507.1257G} \\ 

MACS\,J0257.6--2209 &  0.322 & cH & & \cite{2017ApJ...841...71G} \\  
Abell\,1351 & 0.322 & H & & \cite{2009ApJ...704L..54G,2009AA...507.1257G} \\ 
WHL\,J091541.3+251206 & 0.324 &  cF & & \cite{2009AA...508...75V,2011AA...527A.114V}  \\ 
PSZ1\,G094.00+27.41$^n$ & 0.332 & H && \cite{2014MNRAS.444L..44B,2016MNRAS.459.2940K}   \\  
PSZ1\,G108.18--11.53                   & 0.335  & dRS, H  && \cite{2015MNRAS.453.3483D} \\ 
MACS\,J0520.7--1328 & 0.336& cRS & close to 1WGA\,J0521.0--1333 &  \cite{2014AA...565A..13M} \\  

1WGA\,J0521.0--1333 & 0.340 & cRS, cH &diffuse emission present & \cite{2014AA...565A..13M}  \\  
& & & close to MACS\,J0520.7-1328 & \\

RXC\,J1115.8+0129 & 0.345 & cMH && \cite{2016sf2a.conf..367P} \\ 
RBS\,797&0.345&MH&&\cite{2006AA...448..853G,2012ApJ...753...47D}\\ 
MACS\,J1931.8--2634 &0.352 & U & &\cite{2014ApJ...781....9G}\\ 
MACS\,J0308.9+2645 & 0.356& cH & &  \cite{2017MNRAS.464.2752P} \\ 
Abell\,S1121& 0.358 & H & &    \cite{2017arXiv170703517D} \\ 

RXC\,J0256.5+0006$^o$ &0.360 & H &&\cite{2016MNRAS.459.4240K,2018arXiv180609579K} \\ 
RX\,J1532.9+3021 & 0.363 & MH && \cite{2013AA...557A..99K,2013ApJ...777..163H} \\ 
& & & & \cite{2014ApJ...781....9G} \\
MACS\,J1752.0+4440                  & 0.366  & dRS, H  && \cite{2012MNRAS.425L..36V,2012MNRAS.426...40B}   \\  
ZwCl\,1447.2+2619$^p$   & 0.370  & H,  RS && \cite{2009AA...507.1257G,2012AA...545A..74G}\\    
ZwCl\,0847.2+3617$^q$ & 0.373 & cU & & \cite{2009ApJ...697.1341R,2015AA...579A..92K} \\ 

RXC\,J0949.8+1707$^r$ & 0.383 & cH  & & \cite{2008AA...484..327V,2013AA...551A..24V}\\ 
MACS\,J0949.8+1708  & 0.383 & H&  & \cite{2015MNRAS.454.3391B}  \\ 
PLCK\,G287.0+32.9                   & 0.390   & H, RS, F && \cite{2011ApJ...736L...8B,2014ApJ...785....1B}  \\    
& & & & \cite{2017MNRAS.467..936G} \\
PLCKESZ\,G284.99--23.70  & 0.390 &H & & \cite{2016AA...595A.116M,2018AA...611A..94M} \\ 
RX\,J1720.2+3536$^s$ & 0.391 &cMH & &\cite{2017ApJ...841...71G}\footnote{Venturi et al. in prep}  \\ 
GMBCG J357.91841--08.97978$^t$ & 0.394 & cH  && \cite{2017arXiv170703517D} \\
MACS\,J0416.1--2403 & 0.396 & H  &cUSS & \cite{2015ApJ...812..153O,2015sf2a.conf..247P} \\

MACS\,J0159.8--0849 & 0.405 & MH&& \cite{2014ApJ...781....9G,2017ApJ...841...71G}  \\ 
MACSJ0553.4--3342 & 0.407& H && \cite{2012MNRAS.426...40B}\\ 
Abell\,851 & 0.407& cH&& \cite{1999dtrp.conf....9O,2009AA...507.1257G} \\ 
PSZ1 G262.72--40.92  &0.421& U &&\cite{2018AA...611A..94M}\\ 
MACS J0417.5--1154 & 0.443 & H, RS &  & \cite{2011JApA...32..529D,2017MNRAS.464.2752P} \\ 
& & & & \cite{2018ApSS.363..133S} \\
MACS\,J2243.3--0935                  &   0.447  & H, dRS & & \cite{2017MNRAS.464.2752P,2016MNRAS.458.1803C} \\ 
& & & & \cite{2017arXiv170703517D} \\
MACS\,J0329.6--0211 & 0.450  & MH  &&\cite{2014ApJ...781....9G,2017ApJ...841...71G} \\ 
RX\,J1347.5--1145 & 0.452 &  MH, cRS &  & \cite{2007AA...470L..25G,2011AA...534L..12F}\\ 

PLCK\,G004.5-19.5 & 0.516 & H, RS &&\cite{2017AA...607A...4A} \\
ACT-CL J0014.9--0057 &0.533 & cRS &&\cite{2018arXiv180609579K}\\ 
CL\,0016+16$^u$  & 0.541 &H & & \cite{1989AJ.....98.1148M}\\ 
& & & & \cite{2000NewA....5..335G} \\
MACS\,J1149.5+2223                  & 0.544 & RS, cRS, H& USS & \cite{2012MNRAS.426...40B} \\ 
ACT-CL\,J0045.2--0152 &0.545 & U & &\cite{2018arXiv180609579K}\\ 
MACS\,J0717.5+3745                  & 0.546 & H, mRS & & \cite{2009AA...508...75V,2009AA...503..707B}   \\  
& & & & \cite{2013AA...557A.117P,2018MNRAS.478.2927B} \\
& & & & \cite{2018MNRAS.478.2927B} \\
MACS\,J0025.4--1222                & 0.586  &  cdRS && \cite{2017AA...597A..96R}  \\ 
Phoenix cluster & 0.596 & MH & &\cite{2014ApJ...786L..17V}\\ 
ACT-CL\,J0022.2--0036 & 0.805& cMH && \cite{2018arXiv180609579K}\\ 
ACT-CL\,J0102--4915$^v$& 0.870 & H, dRS & &  \cite{2012ApJ...748....7M,2014ApJ...786...49L}  \\
& & & & \cite{2016MNRAS.463.1534B}\\
RXC\,J2351.0--1954 &  \ldots & cH & &\cite{2017arXiv170703517D} \\
\label{tab:clusterlist}
\end{supertabular}%
\\
H = giant radio halo; MH = radio-mini-halo; F = revived fossil plasma source; RS = cluster radio shock (relic); U = unclassified; d = double; m = multiple; c = candidate; USS = ultra-steep spectrum; H-MH = ``intermediate'' or ``hybrid'' halo--mini-halo\\
$^a$Abell\,1656; $^b$RXC\,J0338.6+0958; $^c$uncertain; $^d$uncertain;  $e$Sausage cluster; $^f$RX\,J0603.3+4214; $^g$Z5247; $^h$Z7160, MS\,1455.0+2232; $^i$Abell\,1758a ;$^j$Abell\,1758b; $^k$Z3146; $^l$1E\,0657--5655; $^m$ZwCl 1856.8+6616; $^n$CL\,1821+643; $^o$ACT-CL\,J0256.5+0006; $^p$CL\,14 46$+$26, CL\,1447+26; $^q$ Z1953; $^r$Z2661; $^s$ Z8201; $^t$WHL\,J235151.0--0.085929; $^u$MACS J0018.5+1626, RXC J0018.5+1626; $^v$El Gordo   

\twocolumn

\begin{acknowledgements}
We thank the reviewer for the constructive feedback and suggestions. We thank Francesca Loi, Simona Giacintucci, Annalisa Bonafede, Felipe Andrade-Santos, Marie-Lou Gendron-Marsolais, Frazer Owen, Shea Brown, Lawrence Rudnick, Nobert Werner, Tim Shimwell, and Hans B\"ohringer for sharing their images and data.
We thank Kathrin B\"ockmann for helping to compile the cluster catalogs. 
We thank Amanda Wilber, Federica Savini, and Julius Donnert for feedback on the manuscript. We acknowledge the help of Soumyajit Mandal, Huib Intema, and Dawoon Jung for making some of the displayed GMRT images with SPAM \citep{2009A&A...501.1185I,2014ascl.soft08006I}. 
\def\url#1{\expandafter\string\csname #1\endcsname}
XMM-Newton images from the Perseus cluster, Abell\,2256, and Coma cluster are from Steve Snowden NASA/GSFC. 

RJvW acknowledges support from the ERC Advanced Investigator programme NewClusters 321271 and the VIDI research programme with project number 639.042.729, which is financed by the Netherlands Organisation for Scientific Research (NWO). 
FdG is supported by the VENI research programme with project number 639.041.542, which is financed by the Netherlands Organisation for Scientific Research (NWO).
HA acknowledges support from the VENI research programme, which is financed by NWO.
HK is supported by the National Research Foundation (NRF) of Korea through grants 2016R1A5A1013277 and 2017R1D1A1A09000567.
This research made use of APLpy, an open-source plotting package for Python \citep{2012ascl.soft08017R}. This research made use of Astropy, a community-developed core Python package for Astronomy \citep{2013A&A...558A..33A}. 
\end{acknowledgements}



\bibliographystyle{apju}
\interlinepenalty=10000
\bibliography{ref_filaments}   

\end{document}